\documentclass[twocolumn,showpacs,aps,prx,amsmath,amssymb,superscriptaddress]{revtex4-2}

\usepackage{graphicx}
\usepackage{dcolumn}
\usepackage{bm}
\usepackage{hyperref}
\usepackage{subfigure}
\usepackage{epstopdf}
\usepackage{color,bbm}
\usepackage{ulem}

\newcommand{\beq}{\begin{equation}}
	\newcommand{\eeq}{\end{equation}}
\newcommand{\bqa}{\begin{eqnarray}}
	\newcommand{\eqa}{\end{eqnarray}}
\newcommand{\nn}{\nonumber}

\newcommand{\bra}[1]{ \langle{#1} |}
\newcommand{\ket}[1]{ |{#1} \rangle}

\newcommand{\sch}{Schr\"odinger}

\newcommand{\sq}[1]{\left[ {#1} \right]}

\newcommand{\tr}[1]{{\rm Tr}\sq{ {#1} }}

\definecolor{nblue}{rgb}{0.3,0.3,1.0}
\definecolor{ngreen}{rgb}{0.2,0.7,0.2}
\definecolor{nred}{rgb}{0.9,0.1,0}
\definecolor{nblack}{rgb}{0,0,0}
\definecolor{maroon}{rgb}{0.7,0,0}
\definecolor{golden}{rgb}{0.8,0.6,0.1}

\newcommand{\blu}{\textcolor{nblue}}

\newcommand{\red}{\color{nblack}}

\begin{document}
	
	\title{Quantum Steering: {\red Practical} Challenges and Perspectives}
	
	\author{Yu Xiang}
	\address{State Key Laboratory for Mesoscopic Physics, School of Physics, Frontiers Science Center for Nano-optoelectronics, $\&$ Collaborative Innovation Center of Quantum Matter, Peking University, Beijing 100871, China}
	\address{Collaborative Innovation Center of Extreme Optics, Shanxi University, Taiyuan, Shanxi 030006, China}
	\author{Shuming Cheng}
	\address{The Department of Control Science and Engineering, Tongji University, Shanghai 201804, China}
	\address{Institute for Advanced Study, Tongji University, Shanghai, 200092, China}
	\author{Qihuang Gong}
	\address{State Key Laboratory for Mesoscopic Physics, School of Physics, Frontiers Science Center for Nano-optoelectronics, $\&$ Collaborative Innovation Center of Quantum Matter, Peking University, Beijing 100871, China}
	\address{Collaborative Innovation Center of Extreme Optics, Shanxi University, Taiyuan, Shanxi 030006, China}
	\address{Peking University Yangtze Delta Institute of Optoelectronics, Nantong 226010, Jiangsu, China}
	\author{Zbigniew Ficek}
	\address{Quantum Optics and Engineering Division, Institute of Physics, University of Zielona G\'{o}ra, Prof. Z. Szafrana 4a, 65-516 Zielona G\'{o}ra, Poland}	
	\author{Qiongyi He}
	\email{qiongyihe@pku.edu.cn}
	\address{State Key Laboratory for Mesoscopic Physics, School of Physics, Frontiers Science Center for Nano-optoelectronics, $\&$ Collaborative Innovation Center of Quantum Matter, Peking University, Beijing 100871, China}
	\address{Collaborative Innovation Center of Extreme Optics, Shanxi University, Taiyuan, Shanxi 030006, China}
	\address{Peking University Yangtze Delta Institute of Optoelectronics, Nantong 226010, Jiangsu, China}
	
	\date{\today}
	\begin{abstract}
		Einstein-Rosen-Podolsky (EPR) steering or quantum steering describes the ``spooky-action-at-a-distance'' that one party is able to remotely {\red alter} the states of the other if they share a certain entangled state. Generally, it admits an operational interpretation as the task of verifying entanglement without trust in the steering party's devices, making it lying intermediate between Bell nonlocality and {\red entanglement}. Together with the asymmetrical nature, quantum steering has attracted a considerable interest from theoretical and experimental sides over past decades. In this Perspective, we present a brief overview of the EPR steering with emphasis on recent progress, discuss current challenges, opportunities and propose various future directions. We look to the future which directs research to a larger-scale level beyond massless and microscopic systems to reveal steering of higher dimensionality, and to build up steered networks composed of multiple parties.
	\end{abstract}
	
	\maketitle
	
	\section{Introduction}
	
The concept of {\it steering} a quantum system was first introduced by \sch~\cite{s35} in his response to the Einstein, Podolsky, and Rosen (EPR) paper which objected the complete description of reality provided by quantum mechanics~\cite{epr35}. The objection, called EPR paradox, refers to entanglement between two spatially separated particles that performing local measurements on one particle steers (adjusts) the state of the other distant one. This unusual property called ``spooky action-at-a-distance” implies a violation of local realism in entangled systems and refers to a measurement process that in quantum mechanics we cannot associate objective physical reality with the entangled systems that is independent of the measurement we choose to make. {\red The subject has received its renewed interest after an article by Reid~\cite{mr89}, in which she proposed the practical inequality for testing EPR correlation based on a Heisenberg uncertainty relation of the amplitude and phase quadratures. The EPR paradox was first demonstrated experimentally in a continuous variables (CV) system via nondegenerate parametric amplification}~\cite{op92}. {\red The approach has been applied with considerable success to study EPR steering in variety 
of continuous variable systems}~\cite{rd09,cs16,ucn20,xsh21}.

The study of EPR steering received further stimulus in 2007 when Wiseman {\it  et al.} formalized it for mixed states and provided an operational interpretation via the task of verifying entanglement~\cite{wj07,jwd07}. It provides an approach to certify the existence of entanglement without assumptions of trust devices at the site of steering party. Therefore, steering is often called as a one-sided device-independent (1SDI) scenario for verification of entanglement. This property makes quantum steering a necessary resource for quantum information processing where some of the parties are considered untrusted, such as 1SDI quantum key distribution~\cite{bc12,lct14,gh15,wh16}, quantum secret sharing~\cite{aw15,kx17,wt21}, quantum teleportation~\cite{mr13,hz15}, and randomness generation~\cite{lt14,pc15,sc18,gc19}. Moreover, it is also capable to discriminate sub-channels~\cite{pw15} and enables high metrological precision beyond the standard limit~\cite{yf21}. 

Among three types of nonlocal correlations existing in nature, Bell nonlocality, EPR steering and entanglement, the EPR steering lies in between Bell nonlicality~\cite{bcp14} and  entanglement~\cite{hhhh09} in the sense that not all entangled states exhibit EPR steering and not all steerable states are able to violate Bell inequalities~\cite{qv15}. Unlike the Bell nonlocality and entanglement, 
which are symmetric under the permutation of the parties involved, steering is intrinsically a directional form of correlation that the steerability from subsystem $A$ to subsystem $B$ may not be equal to the opposite direction~\cite{wj08}. {\red For example, in the case of Gaussian measurements the presence of asymmetric noises or losses in the subsystems creates the possibility of asymmetric steering~\cite{hg15,zt15} which, in principle, may achieve the limit of one-way steering, i.e. $A$ can steer $B$ while $B$ cannot steer $A$~\cite{mf10}. The property of one-way steering has been predicted in a number of systems~\cite{o17,mo13,bv14,bh16,bs18,hr13,hf14,tz15,zs19,wc20} and demonstrated in several experimental configurations~\cite{he12,dx17,wx20,wn16,sy16,xy17,tg18}.} The aspect of one-way steering has also generated a great deal of interest in the study of EPR steering in the more complex situation of multipartite systems~\cite{hr13a,js21,cs16,ucn20,cs15}. 

The research work on EPR steering can be divided essentially into two groups. The largest of these groups is comprised of investigations of conceptional aspects and quantification of quantum steering. Much of it is based on mathematical analysis of the possibility of different steering criteria~\cite{cs16,ucn20}. The other is devoted to determining in which conditions are possible to reveal steering and exploring how useful it is for practical applications. Most work on demonstration of steering deals with optical systems~\cite{zc19,hx20,hx21,hs21,gh15,wh16,he12,dx17,aw15,sj10,be12,sx14,kh15,wn16,sy16,xy17,tg18,cs15,wu20,wp18,cj18,zw18,qw21,qw22,zs20,ws11,wx20,cx20,ds21}. This is easily understood since there are practical sources of entangled light beams available, which can be easily applied to optics experiments. 
 
 {\red A comprehensive analysis of all the issues considered is practically impossible. An extensive literature on various aspects of EPR steering now exists and is reviewed in several  articles~\cite{rd09,cs16,ucn20,xsh21}.} Little work has been done, however, to study the effects of steering in multipartite, high-dimensional and, in particular, systems where atoms or ions are encoded as qubits. There are two reasons for this. On the experimental side, generation of an entangled state or mapping an entangled state of photons on states of high-dimensional systems or massive objects with perfect efficiency is difficult to achieve in practice. On the theoretical side, the effects of steering have not been so well explored mostly due to difficulties to characterize entangled states of a high degree of entanglement in high dimension systems, between atoms and between massive objects. Thus, in these areas there are still many problems and challenges remaining to be solved.

{\red Therefore, the purpose of this perspective is to provide a comprehensive discussion on challenges and possible directions of future developments of the subject of EPR steering.}
We will focus our attention on practical applications and experimentally tested aspects of EPR steering in multipartite, high-dimensional, and atomic and macroscopic systems. We believe that an extensive study of these systems is appropriate for the present experimental situation, in particular, for atomic ensembles~\cite{pk15,fz18,kp18} and atomic lattices~\cite{ib08,dy16} which could be found efficient for steered (directional) information transfer.

This perspective article is organized as follows. In Sec.~\ref{sec2} we begin our analysis with a brief description of what do we mean by EPR steering. Some description of criteria to identify steering and their experimental verifications required for suggestions of further developments is appropriate, so in Sec.~\ref{sec3} we give an overview of steering criteria which are commonly used in the analysis of EPR steering and their applications adapted to different systems. Next, in Sec.~\ref{sec4} we review some experiments to indicate the experimental progress in the verification of the steering criteria and manifestation of quantum steering. The final section~\ref{sec5} presents a discussion of the current developments, challenges and perspectives.

\section{EPR Steering} \label{sec2}
	
Suppose that two space-like separated observers, say Alice and Bob, share a bipartite physical state described by the density operator $\rho_{AB}$.  We are interested in the steering scenario where Alice tries to convince Bob that the shared state is entangled, even though Bob does not trust her. This entanglement verification task can be mathematically formulated as the problem of checking whether the measurement statistics is incompatible with the local hidden state (LHS) model~\cite{wj07}. Particularly, Alice is required to perform measurements on system $A$, labelled $X$, and announces her outcome $x$. Correspondingly, Bob performs some measurements $Y$ on his system $B$ and obtains outcome $y$. Then, Bob will be successfully convinced by Alice that they share entanglement, or equivalently, $A$ can steer $B$, if and only if the joint probability $p(x, y|X, Y)$ cannot be explained by LHS models of the form~\cite{wj07}
\begin{align}
\label{atob}
p(x, y|X, Y)=&\sum_{\lambda}p(\lambda)\,p(x|X, \lambda)p_{\text{Q}}(y|Y, \rho_{\lambda}) \nn \\
=&\sum_{\lambda}p(\lambda)\,p(x|X, \lambda)\tr{E^B_{y|Y}\rho_\lambda},
\end{align}
where the hidden parameter is described by a random variable $\lambda$ which also specifies the response function $p(x|X, \lambda)$ for system $A$, $p_{\text{Q}}(y|Y, \rho_{\lambda})$ refers to a quantum probability distribution generated from Bob by performing a positive-operator-valued measure (POVM) {\red $\{E^{B}_{y|Y}\}$} on a local hidden state $\rho_\lambda$. Hence, EPR steering for the direction $A\rightarrow B$ is demonstrated if there exists a set of measurements without admitting the above explanation. However, the conclusive proof of no steerability usually requires searching over all possible measurements, which generically is a challenging task to deal with, even for simple two-qubit states~\cite{jm15,nt16,na18}.
	
Noting from Eq.~(\ref{atob}) that system $B$ follows from the description of quantum theory, with a set of tomographically complete measurements, Bob can reconstruct all received states after Alice announcing her measurement outcomes. Consequently, Bob would obtain state {\it assemblages} $\{\sigma_{x|X}\}$~\cite{m13} that the sub-normalized states, each associated with probability $p(x|X)=\tr{\sigma_{x|X}}$, satisfy $\sum_x\sigma_{x|X}=\rho_B={\rm Tr}_A[\rho_{AB}]$ for Alice's measurement $X$. Thus, the LHS model as described in Eq.~(\ref{atob}) is equivalent to there exists an ensemble $\{\rho_\lambda\}_\lambda$ and positive distribution $p(x|X, \lambda)$ such that
\beq
\sigma_{x|X}=\sum_{\lambda} p(\lambda)p(x|X,\lambda) \rho_\lambda,~~\forall~x,~X. \label{assemblage}
\eeq
Here, the state ensemble indexed by the hidden variable $\lambda$ satisfies $\sum_{\lambda}p(\lambda)\rho_\lambda\equiv\sum_{\lambda}\tilde{\rho}_\lambda=\rho_B$. If such a model exists, then Bob may be cheated by Alice via only sending unnormalized states $\tilde{\rho}_\lambda$ with a conditional probability $p(x|X, \lambda)$. Otherwise, Alice succeeds in convincing Bob that she must perform local measurements on an entangled state. Fortunately, determining whether the state assemblages are compatible with an LHS model can be tackled via numerical tools, such as the semidefinite programming~\cite{cs16} or conic programming~\cite{nn19}.

{\red We might mention that EPR steering verified by LHS models of the form (\ref{atob}) lies intermediate between Bell nonlocality, which cannot be explained by local hidden variable (LHV) models of the form~\cite{jb64,cj09,qv15,jw07} 
\begin{align}
\label{atobell}
p_{n}(x, y|X, Y) =& \sum_{\lambda}p(\lambda)\,p(x|X, \lambda)p(y|Y,\lambda) ,
\end{align}
where $p(x|X, \lambda)$ and $p(y|Y, \lambda)$ are response functions of systems $A$ and $B$ respectively, and quantum inseparability (entanglement), which cannot be explained by quantum separable models of the form~\cite{wj07,jw07}
\begin{align}
\label{atoben}
p_{e}(x, y|X, Y) =& \sum_{\lambda}p(\lambda)\,p_{\text{Q}}(x|X, \sigma_{\lambda})p_{\text{Q}}(y|Y, \rho_{\lambda}) ,
\end{align}
where $\sigma_{\lambda}$ and $\rho_{\lambda}$ are some quantum states of systems $A$ and $B$, respectively.}

This hierarchical relation can be illustrated via a class of Werner states~\cite{rw89} 
\beq \label{werner}
\rho_W={\mu}\ket{\Psi}\bra{\Psi}+\frac{1-{\mu}}{4} {\bf I}_{AB}, ~~{\mu}\in[0,1]
\eeq
with $\ket{\Psi} = \left(\ket{01}-\ket{10}\right)/\sqrt{2}$. {\red The state (\ref{werner}) is entangled if and only if $\mu>1/3$~\cite{rw89}, EPR steerable under projective measurements if and only if $\mu>1/2$~\cite{wj07}, and cannot violate any Bell inequality if $\mu<0.6595$~\cite{ag06}. The lower bound for steering, $\mu =1/2$, can be achieved exactly in the limit of infinitely many measurement settings~\cite{wj07}. For a finite number of settings, e.g. $n=10$, $\mu_{10}=0.5236$~\cite{sj10}, and $n=30$, $\mu_{30}=0.5058$~\cite{bn17}.} Generally, their inequivalence can also be proven under POVM measurements~\cite{qv15}. {\red It should be pointed out that the exact critical value of $\mu =\mu_{c}$ at which the state (\ref{werner}) ceases to be the Bell nonlocal under projective measurements is unknown~\cite{ag06,bn17,hq17}. However, there are bounds established, the best currently known $0.6964 > \mu_{c} > 0.6829$ that when $\mu>0.6964$ the state (\ref{werner}) is Bell nonlocal.}  
	
Similar to the steering from $A$ to $B$ ($A\rightarrow B$), steering in the opposite direction $B\rightarrow A$ is verified if the outcome statistics $p(x, y|X,Y)$ does not admit an LHS model described by
\begin{equation}
\label{btoa}
p(a, b|A, B) =\sum_{\lambda}p(\lambda)\tr{E^A_{x|X}\rho_\lambda} p(y|Y, \lambda).
\end{equation}
Here the task is for Bob to convince Alice they share entanglement. It is clear that steering is a directional form of correlation, describing the EPR idea of one party apparently adjusting the state of another by way of local measurements, which is fundamentally defined differently to entanglement. Consequently, the steerability for two directions is not always the same, sometimes even appears only in one direction, i.e., the so-called one-way steering. For example, consider the class of {\it loss-depleted} states~\cite{bh16}
\begin{equation}
\rho_{L} = p\ket{\Psi(\theta)}\bra{\Psi(\theta)} + (1-p)\frac{{\bf I}_{A}}{2}\otimes \rho_{B}^{\theta} ,\label{lossdeplete}
\end{equation}
where $\ket{\Psi(\theta)}=\cos\theta\ket{00}+\sin\theta\ket{11}$ and $\rho_{B}^{\theta} = {\rm Tr}_{A}[\ket{\Psi(\theta)}\bra{\Psi(\theta)}]$. It is one-way steerable for $\theta\in [0,\pi/4]$ and $\cos^{2}(2\theta) \geq (2p-1)/[(2-p)p^{3}]$ under projective measurements. 

{\red The above considerations of EPR steering refer to bipartite systems, but it is not difficult to extend the considerations to multipartite systems. There are a number of theoretical approaches determining criteria for multipartite steering~\cite{cs15,js21,cu18a,ch11,hd11,hr13a}. 
For example, Cavalcanti {\it et al.}~\cite{ch11} have proposed a criterion for multipartite steering of $N$ spatially separated parties among which $T$ are trusted and $N -T$ are untrusted. Each party can perform a measurement $X_j\in\mathcal{M}_j$ and achieve outcome $x_j\in\mathcal{O}_{X_j}$. Following the LHS model (\ref{atob}), if the observed correlations cannot be reproduced by LHS$(T,N)$ models of the form
\begin{align}
\label{msta1}
p(x_1,\ldots,x_{N}|X_1,\ldots,X_{N})=&\nonumber\\ 
 \sum_{\lambda}p(\lambda)\,\prod_{j=1}^{T}p_{\text{Q}}(x_j|X_j, \rho_{\lambda,j})&\prod_{j=T+1}^{N} p(x_j|X_j, \lambda) ,
\end{align}
then the entanglement between $N$ parties in the presence of $N-T$ untrusted parties is confirmed. The violation of an LHS$(1,2)$ model demonstrates EPR steering in a bipartite scenario, and the violation of LHS$(1,N)$ is a demonstration of EPR steering in a multipartite scenario. In the case of a tripartite system, a violation of LHS($2,3$) model is referred to as {\it one-to-two steering} where two trusted parties are steered by the remain third one. Similarly, if the joint probability distribution cannot be explained by a LHS($1,3$) model, the system is called {\it two-to-one steering}. Of a particular interest is a ``collective steering” that the steering of one party ($T=1$) by a group of remaining $N-1$ parties cannot  be demonstrated by measurements involving fewer than $N-1$ parties~\cite{hr13a}. 
The collective steering has been identified as a necessary resource for 1SDI quantum secret sharing~\cite{hr13,aw15,xk17,kx17}.

The above discussed approach to multipartite systems has been further developed by He {\it et al.}~\cite{hd11} to a set of spin observables defined on a discrete Hilbert space, i.e., multisite qudits. Moreover, diverse steering structures in multipartite systems have also been studied~\cite{mdr13}.}

\section{Criteria for EPR steering}\label{sec3}

The definition of EPR steering introduced in the previous section is not efficient for testing steering conditions in physical systems. Therefore, various criteria for determining EPR steering have been proposed, depending on the size of a given system and the detection method. However, it would be inappropriate here to attempt a survey of the large literature on this topic. Readers interested are referred to review papers~\cite{rd09,cs16,ucn20}. Sufficient is to focus attention on only these criteria which could be applied to a large class of systems, particularly to practical systems. 

The first criterion to identify EPR steering was introduced by Reid~\cite{mr89}, which is determined in terms of continuous variables, in-phase $X$ and out-of-phase $P$ quadrature components satisfying the commutation relation $[X,P] =i$ ($\hbar=1$).  Although not explicitly concerned with EPR steering, the criterion was originally formulated to confirm the EPR paradox. Further investigations carried by Wiseman {\it et al.}~\cite{wj07} showed that the EPR paradox is merely a particular case of steering. For bipartite systems the criterion corresponds to a violation of the Heisenberg uncertainty relation on one of the systems, say system $B$, held by Bob, when measurements are performed on the other system, say system $A$, manipulated by Alice. According to the criterion the system $B$ is steered by $A$ if the correlations between them are so strong that the product of the inferred variances falls below the Heisenberg uncertainty bound for system $B$, i.e., there is a violation of the relation
\begin{equation}
S_{B|A} = \Delta_{\rm inf,A}X_{B}\Delta_{\rm inf,A}P_{B}\geq \frac{1}{2}.\label{bis}
\end{equation}
Here $[\Delta_{\rm inf, A}X_{B}]^2=\sum_ap_A(a)[\Delta (X_{B}|a)]^2$, where $[\Delta (X_{B}|a)]^2$ is the variance of the conditional distribution for Bob's $X_B$ conditioned on the Alice's outcome $a$. The measurement at $A$ is chosen to minimize the quantity $[\Delta_{\rm inf, A}X_{B}]^2$. The $[\Delta_{\rm inf, A}P_{B}]^2$ is defined similarly. The condition (\ref{bis}) also holds if we define 
\begin{equation}
\Delta_{\rm inf,i}Q_{j} =\Delta(Q_{j}+u_{i}U_{i}),\label{r1}
\end{equation}
where $Q_{j}$ and $U_{i}$ are quadrature components, either $X_{i(j)}$ or $P_{i(j)}$, and $u_{i}$ is real constant, selected to minimize the variances~\cite{mr89,rd09}. The variances in Eq.~(\ref{bis}) quantify the uncertainty with which Alice at $A$ can predict (infer) the outcome of a measurement at $B$ through her choice of measurement basis and are therefore called inferred variances. 

The violation of the inequality (\ref{bis}) is a sufficient criterion for steering and the largest violation of the inequality (\ref{bis}) means the strongest steerability of Alice. The limit $S_{B|A}=0$ corresponds to perfect EPR steering. Since the inferred variances $\Delta_{\rm inf,B}X_{A}$ and $\Delta_{\rm inf,A}X_{B}$ may not be equal, which may result from the presence of an asymmetry between subsystems $A$ and $B$, we see from Eq.~(\ref{bis}) that the measure of steering $S_{A|B}$ may not be equal to $S_{B|A}$. This implies that steering can be asymmetric to the extend that even if $A$ can steer $B$, the system $B$ may not necessarily be able to steer $A$.

The significance of the Reid criterion is that it is directly accessible experimentally since it involves quadrature components which can be measured accurately by optical homodyning~\cite{af09,rd09}.
The criterion requires one of the two-mode variances (\ref{r1}) to be reduced below threshold for quantum squeezing which is also requirement for an entangled two-mode squeezed vacuum state. Therefore, a direct way to experimentally realize steering is to use output beams of a source of squeezed light such as an optical nondegenerate parametric down-converter~\cite{op92}.

From the theoretical side, the two-mode variances (\ref{r1}) are easy to calculate. This has encouraged theoretical research into applications of the inferred variances criterion in searching for steering conditions in a variety of systems~\cite{rd09}. In this way, a significant body of work has accumulated, in particular, on searching for asymmetric steering conditions in two-mode Gaussian systems ranging from optical fields~\cite{mf10,mo13,hg15,zt15,o17}, Bose-Einstein condensates (BEC)~\cite{hd12,dg20} to hybrid optomechanical systems~\cite{hr13,kh14,hf14,tz15,xs15,kt17,zs19,gm19}. 

{\red The Reid criterion can be extended to multipartite systems.
The most direct approach taken to generalize the Reid criterion to situations of multipartite systems is to divide the set of systems (modes) into two groups~\cite{ch11,hr13a}. 

To outline this approach, we concentrate on a system composed of three modes and divide the modes into two groups, mode $i$ constituting group $A$ and modes $j$ and $k$ 
constituting group $B$. The extension of the Reid criterion for steering to the case of multipartite steering refers to the Heisenberg uncertainty that involves variances of linear combinations of the quadrature components of the group of modes 
\begin{equation}
S_{A|B}=\Delta_{\rm inf,B}(X_{i}|O_{jk})\Delta_{\rm inf,B}(P_{i}|O^{\prime}_{jk}) \geq \frac{1}{2} ,\label{ms2}
\end{equation}
where
\begin{equation}
\Delta_{\rm inf,B}(U_{i}|O_{jk})=\Delta\left[U_{i}+\left(g_{j}O_{j}+g_{k}Q_{k}\right)\right] ,\label{ms1}
\end{equation}
are inferred variances of a linear combination of the quadrature components of the modes. The weight factors $g_{j},g_{k}$ are estimated to minimize the variance. 
A violation of the inequality (\ref{ms2}) implies the presence of multipartite steering between the modes.

The measure of multipartite steering $S_{i|jk}$ is very similar to the measure $S_{i|j}$ of bipartite steering. However, it involves a superposition of the modes $j$ and $k$ 
of the group $B$. This suggests that the superposition can be treated as a single ``collective'' mode so that we can express the variance (\ref{ms1}) in the form
\begin{equation}
\Delta_{\rm inf,B}(U_{i}|O_{jk})=\Delta\left(U_{i} +g_{c}C_{jk}\right) ,\label{ms3}
\end{equation}
where $C_{jk} = O_{j} +g_{jk}Q_{k}$. Thus, the collective mode $C_{jk}$ can be treated as a single mode that the condition for multipartite steering can be converted into a bipartite type steering condition. While it appears to be formally similar to bipartite variance, the mode $C_{jk}$ has now a nature of a collective mode which may not necessarily reflect the nature of the individual modes $j$ and $k$. 
 
The violation of the inequality (\ref{ms2}) is the sufficient condition for tripartite steering without any requirements about possible bipartite steerings between modes $i$ and $j$, and between $i$ and~$k$. This means that in general for a tripartite steering we can have two distinct possibilities. Namely, we could have that the mode $i$ is steered by either mode $j$ or $k$. In this case, we would have that the tripartite steering is accompanied by a bipartite steering. This situation is referred to as ordinary tripartite steering.
The other case corresponds to the inequality $S_{i|jk}<1/2$ with both $S_{i|j}$ and $S_{i|k}$ greater than $1/2$. In this case, the tripartite steering is not accompanied by the bipartite steering and is referred to as collective steering or genuine tripartite steering~\cite{hr13a}. The collective steering is thus a generalization of the ordinary tripartite steering to the case when a given mode is steered only by the collective mode of a linear superposition of the remaining modes. In this sense, the collective steering is a stronger form of the tripartite steering. 

It should be pointed out that in the multimode case there are some limitations, constraints for distribution of steering between different systems (modes) imposed 
by the monogamy relations~\cite{mdr13,jk15,as16,lc16,cm16,xk17}.  For example, two individual modes cannot simultaneously steer a third mode using the same two-setting steering witness~\cite{mdr13}.}

It is convenient to determine properties of bipartite CV systems in terms of covariance matrix, which is composed of the second statistical moments of the quadrature operators, $V=\left(
\begin{array}{cc}
V_{A} & V_{AB}\\
V_{AB}^\mathrm{T} & V_{B}
\end{array}
\right)$, where $V_{A}$ and $V_{B}$ are the covariance matrices corresponding to the reduced states of each subsystem, respectively, and $V_{AB}$ contains correlations between them. 
In this way, the state determined by $V$ is steerable by Gaussian measurements on $A$ if and only if the inequality $V_{B|A}+i\Omega_B\geq 0 $ is violated~\cite{wj07}. Here, $V_{B|A} = V_{B}-V_{AB}V_{A}^{-1}V_{AB}^{\rm T}$ is the Schur complement of $V$, which describes the conditional state of the subsystem $B$ after a measurement was made on subsystem $A$, and $\Omega_{B}=\oplus_{i=1}^{n_{B}}w$ is symplectic structure of the two-dimensional matrix $w= \left(
\begin{array}{cc}
0 & 1 \\
-1 & 0%
\end{array}
\right)$.

{\red This method yielded Adesso \textit{et. al.}~\cite{kl15,ka15} to introduce a necessary and sufficient criterion for steering of arbitrary Gaussian states in the form of two subsets of finite numbers of modes under Gaussian measurements expressed in terms of the symplectic spectra $\{\bar{\nu}_j^{B|A}\}$ of the Schur complement $V_{B|A}$}, which reads as 
\begin{equation}  \label{GSAtoB1}
\mathcal{G}^{A\rightarrow B}:=\max \bigg\{0, {-{\sum}_{j:\bar{\nu}_{j}^{B|A}<1}} \ln \left(\bar{\nu}_{j}^{B|A}\right)\bigg\}.
\end{equation}
The quantity $\mathcal{G}^{A\rightarrow B}$ is a monotone under Gaussian local operations and classical communication and the larger value implies the stronger Gaussian steerability. We will refer to this quantifier as the Adesso criterion. {\red The criterion (\ref{GSAtoB1}) refers to multimode systems  so it allows to study multiple forms of EPR steering. 
If we restrict the subsets to be composed of only a single mode $(n_{A}=m_{B}=1)$, the criterion then refers to bipartite steering. It is found that the bipartite steering criterion is especially useful in classifying the steerability of two-mode Gaussian states based on their purity.}

It should be noted that there are many similarities between the Reid and Adesso criteria. For instance, both criteria are sufficient and necessary for all-Gaussian scenarios, i.e., they are proven to be equivalent to some extent. However, from a practical point of view, the description of Reid criterion in terms of the inferred variances is more convenient experimentally since it only involves quadrature components which are measured directly in laboratory by means of homodyne detections. Since the tomographic reconstruction of covariance matrix can be obtained after the accurate measurements of the quadrature components{\blu~\cite{lk05,af09,dx17},} it follows that Adesso criterion is also experimentally accessible. 

The Reid and Adesso criteria refer to Gaussian systems with Gaussian measurements. However, these criteria are not necessary under non-Gaussian measurements. That is, there are Gaussian systems whose states are one-way steerable with Gaussian measurements but are two-way steerable with non-Gaussian measurements~\cite{wn16,xx17}. Regarding this issue the question has been raised, ``Do there exist states which are only one-way steerable regardless of the measurement?"  To resolve this question theoretical analysis has been taken which shows that this situation might be encountered when states are detected by performing arbitrary projective measurements~\cite{bv14} or arbitrary measurement settings (measurement directions)~\cite{sn14}.

In this situation steering conditions can be verified by the linear EPR steering inequalities~\cite{ew14}. The measure is the expectation value of a correlation function between measurement results in the subsystems summed over the number of measurements $n$. For instance, for measurement setting taken by observer $B$ to correspond to the Pauli observable $\sigma_{k}^{B}$ and a random variable $a_{k}\in \{-1,1\}$ corresponding to declared result the observer $A$ submits to $B$, the steering parameter for system $B$ satisfies the inequality~\cite{sj10}
\begin{equation}
S^{B}_{n} =\frac{1}{n}\sum_{k=1}^{n}\langle a_{k}\sigma_{k}^{B}\rangle \leq C_{n}(\epsilon_{A}) .\label{ls1}
\end{equation}
By exchanging the labels $A\leftrightarrow B$, we can invert the roles of the systems and obtain the steering parameter for $A$
\begin{equation}
S^{A}_{n} =\frac{1}{n}\sum_{k=1}^{n}\langle b_{k}\sigma_{k}^{A}\rangle \leq C_{n}(\epsilon_{B}).\label{ls2}
\end{equation}
Here $C_{n}(\epsilon_{i})\, (i=A,B)$ is the bound derived from the optimal cheating strategy for that efficiency and choice of measurements, and $\epsilon_{i}$ is the proportion of rounds where observer $i$ reports an outcome of his/her measurements to the other observer.  
{\red The questions whether according to the above inequalities $A$ is steering $B$ and/or $B$ is steering $A$ depend on the number of measurements $n$ and a measurement setting that the inequalities not violated for some numbers of measurements they may be violated for a different number of measurements or a different measurement setting.  
However, in any case a violation of both of the inequalities (\ref{ls1}) and (\ref{ls2}) signals the existence of two-way steering.}

Instead of determining variances which tell the statistical uncertainty in measuring an observable of a given system one can, in principle, use entropy as a measure of uncertainty (disorder). Since entropy is a measure of disorder in a system, it obviously depends on the correlation properties which are not limited to the second-order correlations~\cite{ws11}. Therefore, the measure of uncertainty in terms of information entropy contains the contribution of the higher-order correlations. This observation led to the study of quantum steering in terms of information entropy as a more general approach. This also indicates that the general entropy criteria for steering represents a natural progression from the work on. There are different types of entropy, e.g. Shannon, R\'{e}nyi or Tsallis entropy, which have been proposed to obtain criteria for steering because they satisfy certain inequalities called entropic uncertainty relations. Of particular relevance to the subject of quantum steering are uncertainty relations for conditional entropies~ \cite{sb13,ws11,kf18,cu18,cb17,wu20,xx20}.

{\red Let us briefly explain how entropic criteria for steering have been developed. For the Shannon entropy the steering criterion was first developed by Walborn {\it et al.}~\cite{ws11} and for continuous variables by Schneeloch {\it et al.}~\cite{sb13} for discrete observables.}
The starting point is to introduce two observables that one observer measures observables $X_{A}$ and $Z_{A}$ on subsystem $A$ and the other observer measures observables $X_{B}$ and $Z_{B}$ on subsystem $B$. A measurement e.g. $X_{j}\, (j=A,B)$ produces a probability distribution $(p_{1},\ldots, p_{ n})$ of $n$ outcomes of observable $X_j$ for which we can consider the Shannon entropy $H(X_{j}) = -\sum_{i=1}^np_{i}\log(p_{i})$. For measurements of the two observables on side $j$ the uncertainty of the measurements can be expressed as an entropic uncertainty relation 
\begin{equation}
H(X_{j}) + H(Z_{j}) \geq q(X_{j},Z_{j})  ,\label{ec1}
\end{equation}
where $q(X_{j},Z_{j})=-\log_{2}(c^{2}_{X_{j},Z_{j}})$ is the lower bound for this uncertainty relation, and $c^{2}_{X_{j},Z_{j}}$ is the maximum absolute overlap of the eigenvectors 
of $X_{j}$ and $Z_{j}$. 
If two observers $A$ and $B$ are sharing a separable state $\rho_{AB}=\rho_{A}\otimes \rho_{B}$ and observer $B$ is testing the entropic uncertainty relation using side information provided by $A$ then the conditional entropic uncertainty relation holds~\cite{ws11,sb13}
\begin{equation}
H(X_{B}|X_{A}) + H(Z_{B}|Z_{A}) \geq q(X_{B},Z_{B})  ,\label{ec2}
\end{equation}
where $H(X_{B}|X_{A}) = H(X_{A},X_{B}) - H(X_{A})$ is the conditional Shanon entropy which quantifies the uncertainty involved in predicting values of the observables, and $H(X_{A},X_{B})$ is the entropy of measurements $X_{A}$ and $X_{B}$ performed on the total system. Since the measurements of $A$ are not specified, the inequality (\ref{ec2}) can be regarded as a steering inequality that a violation of the inequality (\ref{ec2}) constitutes a demonstration of steering from $A$ to~$B$.

{\red Similarly, we may define a steering parameter for R\'{e}nyi entropy of order $\alpha\geq 0$}, which for probability distribution $(p_{1},\ldots, p_{n})$ of $n$ measurements of observable $X_{j}$ on side $j$ is defined by $H_{\alpha}(X_{j}) = \frac{1}{1-\alpha}\ln\left[\sum_{k=1}^{n}p_{k}^{\alpha}\right]$, and satisfies an entropic uncertainty relation 
\begin{equation}
H_{\alpha}(X_{j}) + H_{\beta}(Z_{j}) \geq q(X_{j},Z_{j})  ,\label{ec1r}
\end{equation}
where $\beta$ is such that $\alpha^{-1} +\beta^{-1}=2$, and the limit $\alpha\rightarrow 1$ the R\'{e}nyi entropy becomes the Shannon entropy. Thus, in terms of the R\'{e}nyi entropy separable states tested on side $B$ satisfy the following conditional entropic uncertainty relation~{\red \cite{kf18}}
\begin{equation}
H_{\alpha}(X_{B}|X_{A}) + H_{\beta}(Z_{B}|Z_{A}) \geq q(X_{B},Z_{B})  ,\label{ecr2}
\end{equation}
where $H_{\alpha}(X|Y)$ is the conditional R\'{e}nyi entropy. Hence, if we introduce the notation 
${\cal H}_{R} = q(X_{B},Z_{B}) - H_{\alpha}(X_{B}|X_{A}) - H_{\alpha}(Z_{B}|Z_{A})$, it follows from (\ref{ecr2}) that for separable states ${\cal H}_{R}\leq 0$. Accordingly, an entangled state shared between $A$ and $B$ is steerable if ${\cal H}_{R}>0$. 

Finally, consider uncertainty relations for the Tsallis entropy of order $\alpha\neq 1$, $S_{\alpha} = -\sum_{k=1}^{n}p_{k}^{\alpha}\ln_{\alpha}\left(p_{k}\right)$,
in which the $\alpha$ logarithm is defined as $\ln_{\alpha}(x) = (x^{1-\alpha}-1)/(1-\alpha)$. The above analysis for the Shannon and R\'{e}nyi entropies cannot be adopted for the Tsallis entropy, because the Tsallis entropy for two independent systems is not additive, i.e., for two independent systems $A$ and $B$
\begin{equation}
S_{\alpha}(A,B) =S_{\alpha}(A) +S_{\alpha}(B) + (1-\alpha)S_{\alpha}(A)S_{\alpha}(B).\label{ec4t}
\end{equation}
Therefore, we cannot write the conditional entropic uncertainty relation from the entropic uncertainty relation. This led to formulate a conditional entropic uncertainty relation with added correction term~\cite{cu18}. In other words, for a given set of $m$ measurements $A_{m}, B_{m}$ on two separate systems $A$ and $B$, one can define an inequality~{\red \cite{cu18}}
\begin{equation}
\sum_{m} S_{\alpha}( B_{m}|A_{m}) +(1-\alpha)C(A_{m},B_{m}) \geq {\cal C}^{\alpha}_{B}(m),\label{ec4t}
\end{equation}
where ${\cal C}^{\alpha}_{B}(m)= m\ln_{\alpha}[md/(m+d-1)]$ for a set of $m$ mutually unbiased basis of dimension $d$, $\alpha\in (0,2]$, and $C(A_{m},B_{m})$ is the correction term. As before, it is convenient to introduce a parameter ${\cal H}_{T}= {\cal C}^{\alpha}_{B}(m) -\sum_{m} S_{\alpha}( B_{m}|A_{m}) +(1-\alpha)C(A_{m},B_{m})$, which for separable states ${\cal H}_{T}\leq 0$, and ${\cal H}_{T}>0$ whenever a shared entangled state is steerable. 

It should be noted that each of the entropic criteria has advantages and disadvantages connected with its use. For example, if the measurements settings in subsystems $A$ and $B$ are bounded by a local hidden state then the R\'{e}nyi entropic steering parameter only holds for the two-measurement settings, whereas the Tsallis entropic steering criteria do not have such restriction~\cite{kf18}. Nevertheless, the analysis of the conditional entropic uncertainty relations is becoming an important technique for the determination of steerable states. 

Interesting studies have been done to determine criteria for high-dimensional systems, i.e. $d$-dimensional systems beyond qubits $(d>2)$, such as qutrits and qudits~\cite{lc15,gc19}. For example, if two measurements are allowed for each side of a bipartite system, EPR steering can be certified through the criterion~\cite{gc19}
\begin{equation}
S_d=\sum_{a=b}p(a,b|j=1)+\sum_{a+b=0}p(a,b|j=2)\leq C_{2}\equiv 1+\frac{1}{\sqrt{d}} ,\label{highl}
\end{equation}
where $b$ denotes the outcomes of two mutually unbiased measurements on the $d$-dimensional system $B$, and $a+b$ denotes sum modulo $d$. 

Another class of high-dimensional steering criteria are the so-called dimension bounded criteria which allow for the detection of steering from correlations with minimal assumptions about measurement devices on one of the sides, either $A$ or $B$. The name of dimension-bounded steering came from the fact that the requirement of trusted measurements can be removed if all measurements are made on quantities belonging to the same Hilbert space of a finite dimension. In other words, the dimension-bounded steering criterion requires that measurements made on side $B$ act on a fixed finite-dimensional Hilbert space. 
The dimension-bounded steering criterion can be evaluated through the data matrix $D$. We are not going into the rather lengthy analysis, which can be found in Refs.~\cite{wu20} and~\cite{mg16}, but merely note that in the case when the data matrix $D$ is build up of the observed data which are not steerable, the determinant of $D$ satisfies an inequality, called the dimension-bounded steering inequality
\begin{equation}
{\cal DB}_{m} = |{\rm det}D| -\frac{1}{\sqrt{d_{A}}}\left(\frac{\sqrt{2d_{A}}-1}{m\sqrt{d_{A}}}\right)^{m} \leq 0.\label{db1}
\end{equation}
Here $m$ is the number of measurements performed in the subsystem $B$ and $d_{A}$ is the dimension of the chosen operators. If, in particular, the system is composed of  two qubits $(d_{A}=d_{B}=2)$, (\ref{db1}) then gives the following inequality
\begin{equation}
{\cal DB}_{m} = |{\rm det}D| -\frac{1}{\sqrt{2}}\left(\frac{\sqrt{1}}{\sqrt{2}m}\right)^{m} \leq 0.\label{db2}
\end{equation}
This criterion can be experimentally tested since for qubits one can use as a tool the Bloch sphere representation. 

Finally, we remark that there are other methods which have been found useful for detection of EPR steering, such as the argument without inequality~\cite{cy13}, the criteria analog of the Bell inequalities~\cite{cf15,zd15}, criteria based on local uncertainty relations~\cite{jl15}, and criteria based on moment matrix \cite{ks15}, and conditional quantum Fisher information~\cite{yf21}. The later three are suitable for arbitrary dimensional systems~\cite{jl15,ks15,yf21}. {\red Steering weight~\cite{sn14} and steering robustness~\cite{pw15} criteria have been proposed for bipartite multidimensional systems. Both criteria could be efficiently computed via the semidefinite programming and are further proven to be steering monotones under certain conditions~\cite{ga15}.} Recently, the steering robustness has been generalized to test genuine high-dimensional steering by connecting to the Schmidt number~\cite{ds21}. In addition, it has been observed that the steerability of state assemblages has an one-to-one mapping to the measurement incompatibility of the corresponding measurement assamblages~\cite{qv14,ub15,cs16a,kb17}, and thus in general the joint measurability criteria map to steering criteria. One can find more interesting examples in Refs.~\cite{cs16,ucn20}. Note that most of them have limitations in their applications that can be adapted to systems with specific structures. How to generalize a widely applied as well as experiment-friendly measure is still one of the burning issues of this topic.

\section{Experimental demonstrations of EPR steering} \label{sec4}

A number of experiments for demonstrating EPR steering have been proposed and in some cases implemented. One class of experiments is based on the use of squeezed light and passive optical elements, i.e., beamsplitters~\cite{aw15} or large-scale integrated quantum photonic circuits~\cite{wp18} to generate multiple beams of entangled photons. The basis of this class of experiments is to use photons to encode information in their discrete variables such as polarization~\cite{sj10,be12,sx14,kh15,sy16,wn16,xy17,tg18,wu20}, paths~\cite{wp18}, or hybrid of them~\cite{cs15,gc19,zc19,hx20,hx21,hs21}, or orbital angular momentum~\cite{zw18,qw21,zs20,qw22}. By detecting photons, the linear steering inequalities or generalized entropic criteria are used to test EPR steering. Another method is to measure noise of amplitude or phase quadratures of the field of entangled beams of photons~\cite{aw15,he12,dx17,wx20,cx20}. 
In this way variances of the quadrature components can be determined and EPR steering quantified by using the Reid or Adesso criteria. 

A different class of experiments is based on the use of massive (macroscopic size) objects, such as ultra cold atomic ensembles or BEC. Massive objects are of fundamental interest because the original discussion of the EPR paradox was of a phenomenon between two distant massive objects and since massive objects may be more tightly bound to the concept of local realism~\cite{mr89,hf36}. Entanglement~\cite{gz10,rb10}, EPR steering~\cite{pk15,fz18,kp18}, and Bell nonlocality~\cite{sb16} in massive objects have been demonstrated experimentally.

In this section, we review some experiments in which EPR steering has been demonstrated.

\subsection{Demonstration of asymmetric steering in bipartite systems} 

In the preceding section we have discussed criteria which indicate conditions for asymmetry and directionality of EPR steering. We will now illustrate how asymmetry and directionality have been achieved in experiments on EPR steering in bipartite systems.

One of the first experiments demonstrating asymmetric feature of EPR steering in bipartite systems was that by Wagner {\it et al.}~\cite{wj08}. This experiment involved single-mode squeezed beams and beamsplitters to generate spatially separated entangled beams. The balanced homodyne method was used to measure the variances of the quadrature components of the entangled beams. Quantifying steering by the Reid criterion (\ref{bis}), they observed an appreciable asymmetry of the values of the two products of the inferred variances $V(X_{B}|X_{A})V(\theta_{B}|\theta_{A})=S_{B|A}$ and $V(X_{A}|X_{B})V(\theta_{A}|\theta_{B})=S_{A|B}$ of the beam positions $X_{A}, X_{B}$ and the beam directions $\theta_{A},\theta_{B}$, respectively, $S_{B|A}= 0.31$ and $S_{A|B}=0.47$. The observed asymmetry with both $S_{B|A}$ and $S_{A|B}$ reduced below $1/2$ is a manifestation of an asymmetric two-way steering.
\begin{figure}[h]
	\center{\includegraphics[width=\linewidth]{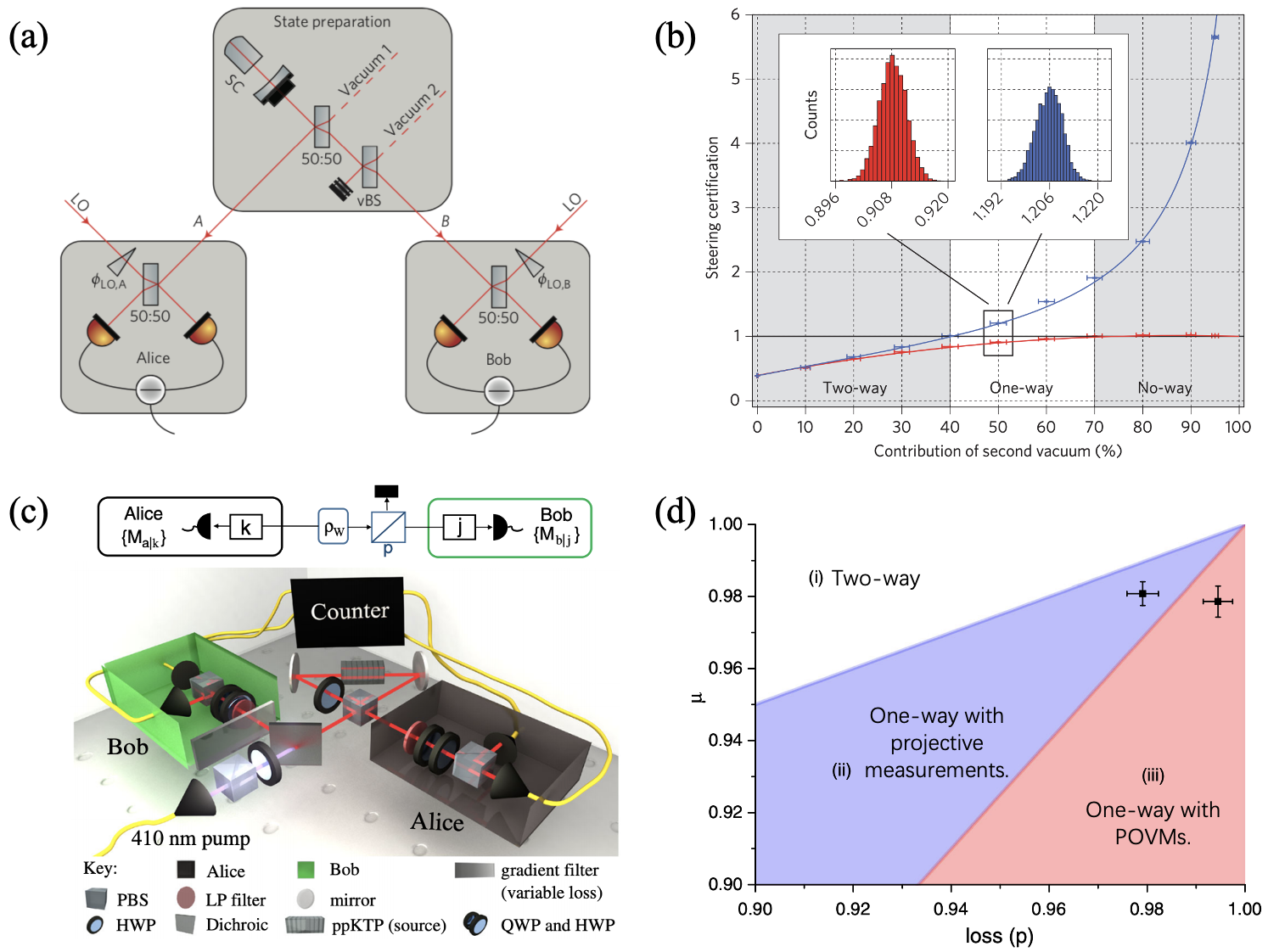}}
	\caption{\label{f1}Experimental demonstrations of one-way steering via linear optics and photonic systems. (a) Outline of the experimental arrangement of H\"{a}ndchen {\it et al.}~\cite{he12} for realization of Gaussian one-way steering. (b) Certification of one-way steering in Ref.~\cite{he12}. Measurement results of the inferred variance products according to Reid criterion versus an increasing contribution of the added vacuum on mode $B$. (c) The experimental scheme of Wollmann {\it et al.}~\cite{wn16} to observe one-way steering in a qubit-qutrit system. (d) Three different steering regimes are parametrized by the purity of Werner state $\mu$ and the loss channel $p$. Reproduced from (a) and (b) Ref.~\cite{he12}, (c) and (d) Ref.~\cite{wn16}.}
\end{figure}

In a slightly different experiment, H\"{a}ndchen {\it et al.}~\cite{he12} succeeded in realization of Gaussian one-way steering between two single-mode squeezed beams. As explained in the previous section, one-way steering between two modes is manifested when simultaneously either $S_{B|A}<1/2$ and  $S_{A|B}\geq1/2$ or $S_{B|A}\geq1/2$ and $S_{A|B}<1/2$. A schematic diagram of the H\"{a}ndchen {\it et al.} experimental configuration is illustrated in Fig.~\ref{f1}(a). The continuous wave squeezed beam generated by type-I parametric down-conversion was superimposed with vacuum on a balanced beamsplitter and divided into two beams $A$ and $B$. Then the beam $B$ was sent through a half-wave plate and a polarizing beamsplitter to prepare mode $B$ with adjustable phase-insensitive loss by mixing with the second vacuum mode. It is apparent that the measured steering parameters fall below the level for steering over appropriate ranges of the contribution of the added vacuum field to the mode $B$. Their measurement results of the conditional variance products versus an increasing contribution of the added vacuum are shown in Fig.~\ref{f1}(b). For loss smaller than $39\%$, steering in both directions was observed. With an increasing added vacuum contribution one-way steering was achieved, i.e., only $S_{B|A}<1/2$. Further increase of the vacuum contribution (larger than $70\%$) resulted in disappearance of steering. 

In the experiments we have discussed so far, the steerability was achieved under the restriction of Gaussian measurements. However, the same may not be true for non-Gaussian measurements. Several theoretical analysis have shown that there exist bipartite Gaussian states which are not steerable by Gaussian measurements, yet whose EPR steering can be revealed by suitable non-Gaussian measurements in certain parameter regimes~\cite{wn16,jl16,xx17}. That is, there are Gaussian systems whose states are one-way steerable with Gaussian measurements which become two-way steerable with non-Gaussian measurements. 

In discrete-variable (DV) systems, following the theoretical predictions, a number of experiments have been performed to demonstrate one-way steering with different measurement settings~\cite{wn16,sy16,xy17}. For example, Fig.~\ref{f1}(c) shows an experiment arrangement of Wollmann {\it et al.}~\cite{wn16} to demonstrate one-way steering by performing arbitrary projective measurements on two-qubit Werner states distributed over a lossy channel which replaces a mode (qubit) with the vacuum state $\ket v$ with probability~$p$:
\begin{equation}
	\rho_{L} = (1-p) \rho_W+ p\frac{{\bf I}_{A}}{2}\otimes \ket{v}\bra{v}.
\end{equation}
Here $\rho_W$ is the Werner state defined in Eq.~(\ref{werner}) with purity parameter $\mu\in[1/2,1]$, ${\bf I}_{A}$ is the identity on the subspace of the subsystem $A$, and $\ket v$ is a vacuum state orthogonal to states of the subsystem $B$. To make the extension to POVMs, it has been proven that when the final state is prepared in the form
\begin{equation}
	\rho_{AB} =\frac{1-p}{3} \rho_W+\frac{p+2}{3}\frac{{\bf I}_{A}}{2}\otimes \ket{v}\bra{v},
\end{equation}
then Bob cannot steer Alice by arbitrary POVMs as long as $p>(2\mu+1)/3$~\cite{qv15}. For the opposite direction, steering from Alice to Bob can be detected by the linear EPR steering measure (\ref{ls1}). Thus, three different steering regimes can be parametrized by the purity of Werner state $\mu$ and the loss channel $p$, as shown in Fig.~\ref{f1} (d). Their data points demonstrated the phenomena of one-way steering with projective measurements and POVMs, respectively.

The above discussed experiments show that control of the added noise to one of the subsystems provides us with a method for varying the directionality of steering. It might therefore be thought that asymmetry in steering can be achieved only if an asymmetry is created in the levels of noise of the subsystems. An entirely different approach to this problem has been adopted which is based on an all-versus-nothing proof of steering without inequalities. The approach offers an elegant argument of the nonexistence of local-hidden-variable models by performing a series of projective measurements on one of the subsystems.

To confirm these predictions, Sun {\it et al.}~\cite{sy16} have performed an experiment which involved two projective measurement settings for detection of steering properties of a family of states
\begin{equation}
	\rho_{AB} = \eta\ket{\Psi(\theta)}\bra{\Psi(\theta)} + (1-\eta)\ket{\Phi(\theta)}\bra{\Phi(\theta)} ,\label{e2}
\end{equation}
where $0\leq \eta\leq 1$, $\ket{\Phi(\theta)} = \cos\theta\ket{10} +\sin\theta\ket{01}$,
and $\ket{\Psi(\theta)}$ is defined in Eq.~(\ref{lossdeplete}). With two settings of projective measurements they have demonstrated that the state (\ref{e2}) with $\theta$ and $\eta$ satisfying the inequality $|\cos 2\theta|\geq |2\eta -1|$ is one-way $(A\rightarrow B)$ steerable which is characterized by \textit{steering radius}. Another experiment of this kind has been reported by Xiao {\it et al.}~\cite{xy17}, who demonstrated the three-setting one-way steering in a two-qubit system for the first time. 

Evidently, the steerability of states depends on the number of experimental settings. Regarding to this, one can ask a question whether there exist states which are only one-way steerable regardless of the measurement? Theoretical analysis have shown that this asymmetric phenomenon can be encountered by performing arbitrary measurement settings for a two-qubit system~\cite{sn14}. Recently, two conclusive experimental certification free of limiting assumptions about the experimental quantum state and measurement assumptions for one-way steering were reported successively in two-qubit systems~\cite{tg18,zs20}.

\subsection{Demonstration of EPR steering in multipartite systems} 

Motivated by the considerable interest in development of quantum networks such as quantum internet, a great efforts have been devoted to demonstrate EPR steering in multipartite and high-dimensional systems. Different kinds of steerable states, such as genuine multipartite steering~\cite{cs15} and collective steering~\cite{aw15} have been achieved experimentally with optical networks. 
\begin{figure}[h]
	\center{\includegraphics[width=\linewidth]{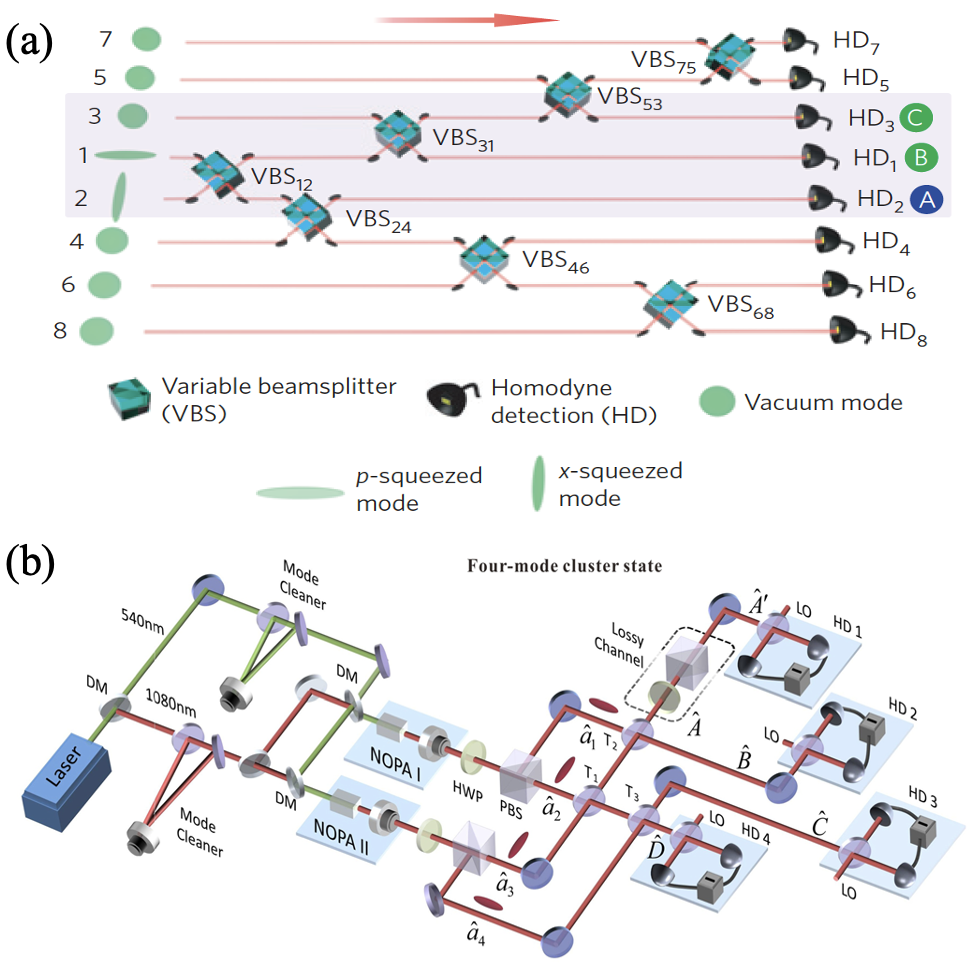}}
	\caption{(a) A schematic diagram of the experimental configuration of Armstrong {\it et al.}~\cite{aw15} to demonstrate multipartite steering in a programmable linear optics circuit. Independent qumodes are shaped to be multiplexed on the same beam. By programmatically changing the measurement basis, the scheme allowed to emulate linear optical networks in real time. (b) Outline of the experiment of Deng {\it et al.}~\cite{dx17} to demonstrate the presence of constrains in the distribution of steering between four modes of a square cluster state imposed by the monogamy relations. Reproduced from (a) Ref.~\cite{aw15}, (b) Ref.~\cite{dx17}. }
\label{f2}
\end{figure}

Evidence of multi-mode steering shared over three or more distinct optical systems has been observed in experiments by Armstrong {\it et al.}~\cite{aw15}. Their apparatus, shown schematically in Fig.~\ref{f2}(a), involved linear optics elements with two quadrature-squeezed qumodes and six vacuum modes as inputs. The multipartite steering was quantified by the Reid criterion (\ref{bis}). In this $k$-mode setting, the inferred variances involve optimized linear combinations of the quadratures of all $k$ modes, i.e., $\Delta_{{\rm inf}}X_{j} =\Delta(X_{j}+\sum_{i=1}^{k-1} u_{i}X_{i})$ $(j\neq i)$, where $u_{i}$ are optimized real constants. Except for multipartite EPR steering, several entanglement and steering related features, such as one-sided device-independent quantum secret sharing, one-sided device-independent quantum key distribution, and genuine tripartite entanglement were also confirmed in this experiment. 

In a multimode systems, directionality of steering, in particular one-way steering can result from constrains imposed on the distribution of steering by the {\it monogamy relations}~\cite{mdr13,jk15,as16,lc16,xk17,cm16}. This comes to another major concern for multipartite systems, that is, how steering can be shared among many parties if there are constrains in its distribution. For example, in a tripartite system, there is a monogamy constrain that two distinct parties cannot simultaneously steer the third one by performing two-setting measurements, but this may be lifted with increasing the number of measurement settings~\cite{mdr13}. For Gaussian systems with Gaussian measurements four types of monogamy relations have been developed~\cite{mdr13,jk15,as16,lc16,xk17} and experimentally tested in a linear optics network~\cite{dx17} and a quantum frequency comb~\cite{cx20}.

The monogamy relations for Gaussian steerability were studied experimentally by Deng {\it et al.}~\cite{dx17}, as shown in Fig.~\ref{f2}(b). The apparatus involved linear optics elements which were used to create a four-mode square cluster state by coupling two phase-squeezed and two amplitude-squeezed states of light generated from two nondegenerate optical parametric amplifiers. The steering properties of different configurations were determined by the Adesso criterion (\ref{GSAtoB1}). In order to introduce an asymmetry in the system, one of the four modes was transmitted through a lossy channel. This experiment validates for the first time general monogamy inequalities for Gaussian steerability with an arbitrary number of modes per party, which establish quantitative constraints on the security of information shared among different parties. 

In a three-qubit system, steering is also proven to be monogamously quantified by the volume of \textit{steering ellipsoids}~\cite{cm16}, which has been verified within an entangled photonic state~\cite{zc19}. Very recently, different configurations of EPR steering in such system were experimentally investigated~\cite{hs21}, in which a proof-of-principle violation of the above strong-monogamy was also demonstrated, i.e., one party can be steered by two other parties simultaneously with three measurement settings.

\subsection{Demonstration of high-dimensional EPR steering} 

In addition to the efforts for generating multipartite steering by entangling more parties, another promising direction is to prepare high-dimensional EPR steering. Interesting aspects of high-dimensional entangled quantum states are the increased channel capacity and improved tolerance against noise~\cite{ll02,ek20}. 
\begin{figure}[h]
	\center{\includegraphics[width=\linewidth]{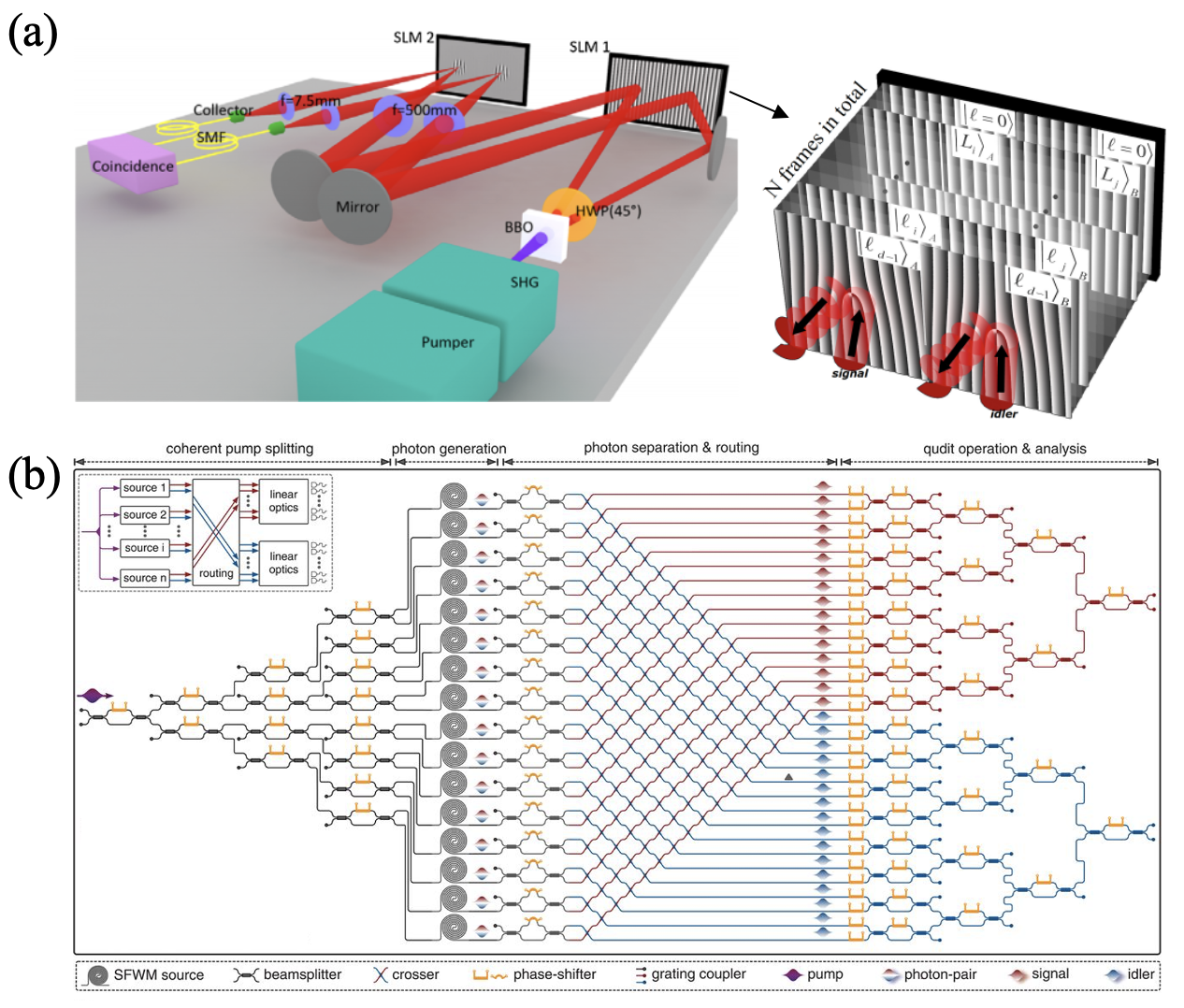}}
	\caption{(a) Experimental arrangement of Zeng {\it et al.}~\cite{zw18} to demonstrate high-dimensional steering with the OAM photons. (b) Circuit diagram of the multidimensional silicon quantum photonic circuit in the experiment of Wang {\it et al.}~\cite{wp18} to demonstrate high-dimensional steering beyond qubits. Reproduced from (a) Ref.~\cite{zw18}, (b) Ref.~\cite{wp18}.}
\label{f3}
\end{figure}

Evidence for high-dimensional steering was certified in experiments by Zeng {\it et al.}~\cite{zw18}. A schematic view of the apparatus used in the experiments is shown in Fig.~ \ref{f3}(a). In this experiment orbital angular momentum photons were generated through a non-collinear type-I spontaneous parametric down-conversion process and two measurements settings were employed. The states were manipulated with spatial light modulator, then their spatial distribution was reversed by placing a mirror in the path of the idler photon. The five-dimensional entangled state $\ket{\Phi}=\sum_{l=-2}^{2}c_l\ket{l}_A\otimes\ket{l}_B$ was obtained, where $l$ is the azimuthal index of the eigenstates. The multi-dimensional EPR steering was determined by violating the linear steering inequality (\ref{highl}). Evidence for the noise-suppression phenomenon caused by the extra dimension was also observed by introducing a tunable isotropic noise into this entanglement system. More recently, Qu {\it et al.}~\cite{qw21} have certified 11-dimensional steering based on photons entangled in their orbital angular momentum and $n$ measurement settings. In addition, it was demonstrated that the extra measurement settings enabled to obtain more information about the underlying quantum state and revealed more strength of steering.

Experiments demonstrating high-dimensional steering beyond qubits were performed by Wang {\it et al.}~\cite{wp18} who used path-entangled photons to encode the desired states beyond the qubit state. In this way a class of photonic qudit states was created to demonstrate $15$-dimensional steering. Figure \ref{f3}(b)  is a view a large-scale silicon chip with $671$ optical components used in the experiment. A total of $16$ spontaneous four-wave mixing sources were coherently pumped, leading to a maximally entangled states in dimensions from $d=2$ to $16$. An integrated reconfigurable interferometric network allowed to perform arbitrary local projective measurements. By measuring the joint probabilities for each measurement, the strong violations of high-dimensional steering inequalities~\cite{sc18} were observed.

Guo {\it et al.}~\cite{gc19} carried out a related experiment in which the trust-free verification of EPR steering was demonstrated beyond qubits by preparing a class of entangled photonic qutrit states. In this experiment entangled two-qutrit states were encoded in the hybrid of the path and polarization degrees of freedom of photons. The noise-suppression phenomenon for high-dimensional EPR steering was also demonstrated.

Distribution of high-dimensional entanglement and steering over a long distance was demonstrated by Hu {\it et al.}~\cite{hx20}. In their experiment a laser beam was separated into two paths and injected into a Sagnac interferometer to pump a type-II nonlinear crystal to generate two-photon polarization entangled state in each path. By encoding polarized photons in different paths they created a 4-dimensional entangled state. The state was then redistributed over a 11 km long fibre and the long-distance path distribution of high-dimensional entanglement and steering demonstrated.  

A different example of high-dimensional steering is work of Designolle~{\it et al.}~\cite{ds21}. The aim of their experiment was to demonstrate a genuine high-dimensional steering, which cannot be achieved by any lower-dimensional entanglement. Steering dimensionality was characterized through the Schmidt number $n$ and used the notion of \textit{steering robustness}~\cite{pw15} to derive a computable quantifier for $n-$dimensional steering under a pair of mutually unbiased bases measurements on each side, which would provide insightful references for the development of multidimensional quantum technologies.

\subsection{Demonstration of EPR steering in massive systems}

In all experiments on the realization of EPR steering that have been discussed so far, photons have been the main carriers of correlations, which are microscopic and massless type objects. This is the most progressed technology so far in demonstrating entanglement and steering. Apart from these experiments based on optical settings, experimental realizations of EPR steering with massive (macroscopic size) objects as carriers of correlations are desirable, because it could be regarded as a test of quantum physics in a regime of macroscopic systems. Despite of a significant experimental progress in the creation of entanglement using massive atomic objects~\cite{zd19,jk01,km11,lp18}, EPR steering as yet had very little experimental investigation with only few experiments carried out on expanding BEC condensates that satisfied Reid criterion (\ref{bis}) in the continuous-variable limit~\cite{pk15,kp18,fz18}. An experiment has also been performed demonstrating bipartite EPR correlations in a hybrid light and atomic ensemble system~\cite{dpw17}.

\begin{figure}[h]
	\includegraphics[width=.9\linewidth]{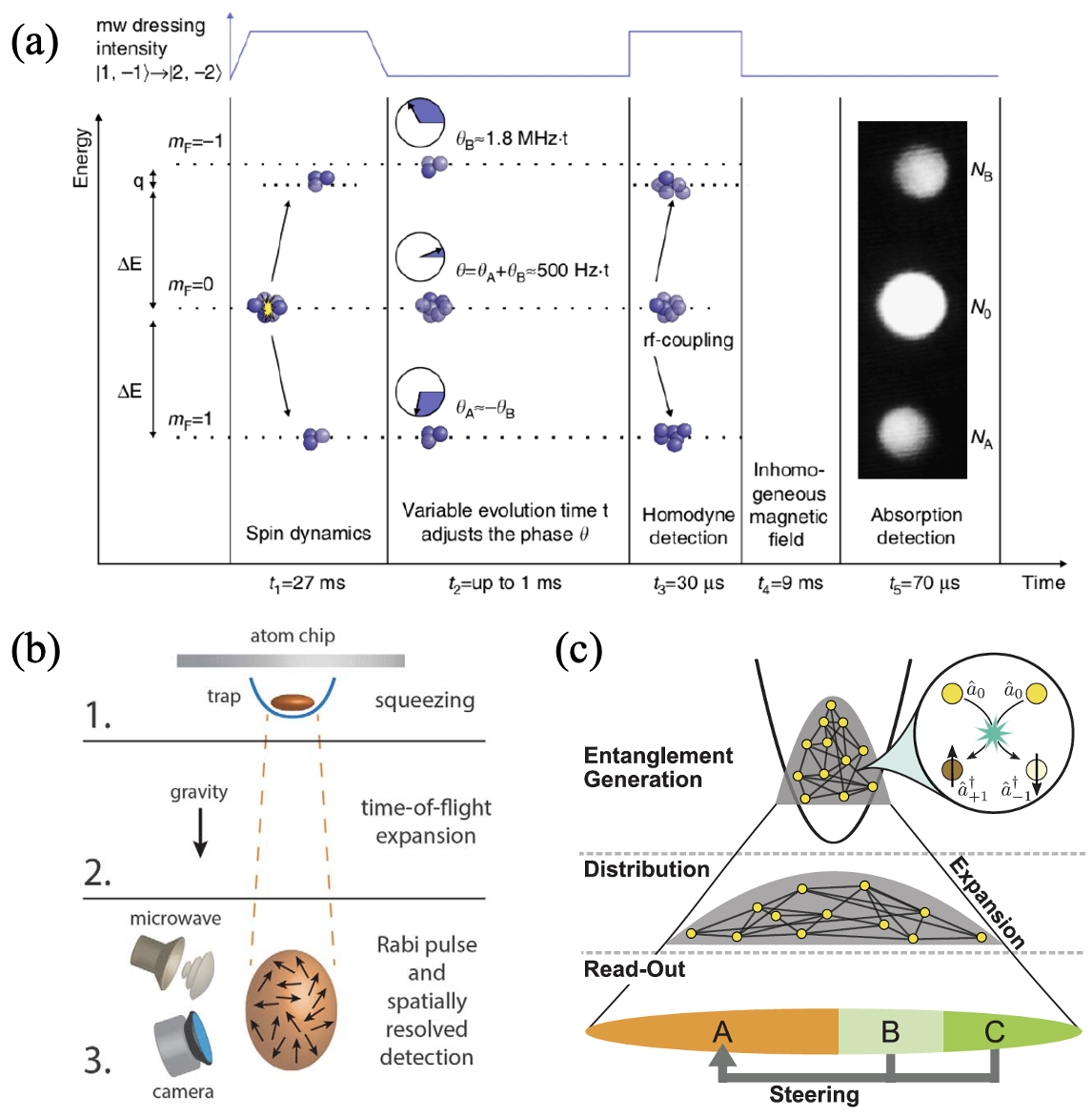}
	\caption{\label{f4} (a) Schematic picture of the experimental sequence used by Peise {\it et al.}~\cite{pk15} to create entangled state of separated clouds of atoms distributed between the three Zeeman states $m_{F}=0$ and $m_{F}=\pm 1$; (b) Outline of the Fadel {\it et al.}~\cite{fz18} experiment to demonstrate entanglement and EPR steering between different parts of an expanded atomic cloud. (c) Experimental setup to detect spatially distributed entanglement in an expanding BEC and a multipartite steering between distinct parts of the expanded cloud. Reproduced from (a) Ref. \cite{pk15}, (b) Ref. \cite{fz18}, (c) Ref. \cite{kp18}.}
\end{figure}

Although not explicitly directed to observe EPR steering, but rather to confirm the EPR paradox, the experiment by Peise {\it et al.}~\cite{pk15} may be classified as the first successful demonstration of EPR steering between massive particles. In the experiment they used a BEC of rubidium atoms magnetically trapped in the Zeeman level $F=1, m_{F}=0$. Spin changing collisions transfer atoms to two spatial modes of the hyperfine Zeeman levels $F = 1, m_{F} = -1$ and $F = 1, m_{F} = 1$ creating a two-component spin-squeezed BEC of spatially correlated (entangled) atoms which resembles to optical parametric down-conversion. The time sequence of spin dynamics performed in the experiment to create and then detect the entangled state is illustrated in Fig.~\ref{f4}(a).

The sum and difference of the quadrature components $X_{A}\pm X_{B}$ and $P_{A}\pm P_{B}$, required to evaluate inferred variances, were experimentally obtained by measuring respectively the sum and the difference of the number of atoms in the $m_{F}\pm 1$ modes. After preparing the state, the BEC was released from the trap and the number of atoms in the modes was detected by absorption imaging on a camera after a spatial separation of the atomic clouds created by applying an inhomogenous magnetic field. The measured values of the product of the inferred variances, needed to quantify steering by the Reid criterion, showed that for intermediate time of the spin dynamics, the correlations created between the components were strong enough to achieve EPR steering. 

A related experiment reporting EPR steering in an expanding two-component BEC, shown in Fig.~\ref{f4}(b), was reported by Fadel {\it et al.}~\cite{fz18}. In their experiment,  BEC of $590\pm30$ $^{87}$Rb atoms was investigated and the two components corresponded to atoms occupying two hyperfine states $\ket{F=1, m_{F}=1}\equiv \ket 1$ and $\ket{F=2, m_{F}=1}\equiv \ket 2$. Each atom effectively behaved as a two-level system with internal states $\ket 1$ and $\ket 2$. As before a spin-squeezed (entangled) state was generated by controlling atomic collisions which coherently populated the states $\ket 1$ and $\ket 2$. By counting the number of atoms $N^{A(B)}_{1}$ and $N_{2}^{A(B)}$ in two chosen regions $A$ and $B$ of the atomic states, they determined local spins and spin variances. EPR steering quantified by the Reid criterion (\ref{bis}) was determined via evaluating the parameters $E_{A|B}$ and $E_{B|A}$ for different positions of the gap between the regions $A$ and $B$, corresponding to different splitting ratios $N^{A}/(N^{A}+N^{B})$, where $N^{A(B)} =N_{1}^{A(B)} +N_{2}^{A(B)}$. Here, EPR steering was observed for intermediate splitting ratios at which regions of both two-way and one-way steerings were achieved.
 
In the above discussed experiments involving the BEC systems the bipartite steering has been demonstrated. Recently, a similar experimental arrangement, illustrated in Fig.~\ref{f4}(c), has been used by Kunkel {\it et al.}~\cite{kp18} to demonstrate multipartite steering between separated regions of an expanded BEC system. Once again steering was investigated on an expanded BEC initially prepared in a spin-squeezed state. But this time the cloud was split into three separate parts by discarding a fraction of atoms. It was observed that each part was collectively steered by the remaining two parts, which demonstrated the presence of thee-way steering. In addition an asymmetry in steering of different parts was observed which could be used in a future experiment to verify if one-way steering between two parts could be generated in this systems.

\section{Outlook, challenges and prospects} \label{sec5}

Now that we have a sufficient knowledge of the direction in which the theoretical studies of EPR steering are heading and on experimental progress in the verification of EPR steering, it would be of interest to extend these considerations to current developments, challenges and future directions. It was gradually become evident from the above review of the theoretical and experimental development that the notion of EPR steering is involved in the whole field of quantum physics and applied mathematics. 

An examination of the literature readily shows that the research work on EPR steering can be divided into two categories, studies which are mostly focused on mathematical investigations of the fundamental and conceptional aspects of EPR steering, and studies which the aim is to determine and investigate possible practical applications. Of these groups, it can be seen that the former has attracted considerable attention and consequently the associated literature is quite extensive, including a number of recent reviews e.g. those by Cavalcanti and Skrzypczyk~\cite{cs16} and Uola {\it et al.}~\cite{ucn20}. The latter group concentrates on practical applications and analysis of the role of steering in practical tasks. Within this group there is to be found two categories, studies which deal with all-optical systems, and studies which deal with atomic and macroscopic particle systems. This is understood because steering requires entangled states of a high degree of entanglement and purity, and there are practical sources of such states, parametric down converters, which can be easily applied to linear optics experiments. Therefore, most theoretical and experimental work on applications of steering deal with photons and linear optics networks.

As we have already noticed in the area of macroscopic objects EPR steering as yet had very little theoretical and experimental investigations, with only few experiments so far carried out on expanding BEC condensates~\cite{pk15,fz18,kp18}, and theoretical interest almost restricted to hybrid optomechanical (mesoscopic) systems~\cite{hr13,hf14,tz15,kh14,xs15,kt17,gm19,zs19} which has recently extended to cavity magnonics~\cite{zsy21,sz21,t19,tl21}. What role does steering play in quantum communication? How one could steer transmission of an entangled state through a quantum network comprised of atoms? These questions are in general not entirely answered yet. Thus, in this area there are still many problems and challenges which remain to be solved. In particular, further investigations and improvements could be done in practical Gaussian systems and in particular in systems involving massive objects. We thus focus our attention on few key systems and phenomena which are potentially promising for future development and applications. Of special interest are challenges and possible future directions in the area of:

{\it Gaussian systems:} This is the area which has been the most explored and an extensive literature on various aspects of EPR steering now exist. However, there are still some aspects into which the study of Gaussian systems could be directed. One of such aspects is the extension of the Reid and Adesso steering parameters into their spectral distributions in analogy to the squeezing criteria extended to the spectrum of squeezing~\cite{cw84,oh87}. The inequalities (\ref{bis}) and (\ref{GSAtoB1}) refer to steering criteria for the total fields or steering in the full sense. For Gaussian systems which are usually represented by broadband fields, it is possible for some specific frequencies to exhibit steering even though $S_{i|j}>1/2$\, $(\mathcal{G}^{j\rightarrow i}<0)$ for the total fields. In addition, when the total field exhibits steering, i.e., $S_{i|j}<1/2$\, $(\mathcal{G}^{j\rightarrow i}>0)$, some modes at selected frequencies may not exhibit steering or may exhibit better steering than the total field. It seems more appropriate to investigate the steering criteria in terms of the frequency components. 

Another aspect is the extension of the investigation of the Gaussian systems to include higher-order moments of the tested observables, which have as yet played a negligible role in the analysis of entanglement and steering, especially in quantifying entanglement and steering still remains to be shown. Studies of a Gaussian system are usually performed in terms of second-order correlation on functions which, as it is well known, contain all the information required for the complete description of the Gaussian characters of the system. The presence of squeezing devices, which are sources of two-photon entangled Gaussian fields, makes properties of Gaussian systems to be easy examined experimentally. However, it has been argued by Walborn {\it et al.}~\cite{ws11} that the entropic steering criteria, which are not limited to the second-order moments of the observables, detect steering in a larger class of states than the criteria relying on variances. In that work, they examined the EPR correlation between the complementary variables-the position and wave vectors-of photon pairs created via the parametric down-conversion process in a nonlinear crystal. Note that these measurements are different to the measurements of the quadrature phase operators tested for Gaussian steering. In a similar spirit, some attempts have also been made to determine steering conditions based on pseudospin measurements~\cite{xx17}, local orthogonal observables~\cite{jl16}, and squared amplitudes~\cite{sa15}. This indicates that a more general approach is required to investigate the contribution of higher-order correlations. In other words, to extract information on entanglement and steering by means of higher-order photon correlation measurements. 

The description of squeezing in terms of the second-order variances of quadrature component is referred as second-order squeezing. However, the variances do not exhaust all possible forms of squeezing. The squeezing concept can be generalized to higher moments of the field. There are possible forms of higher-order squeezing, the higher-order and amplitude-squared squeezing, introduced a long time ago with some analysis of their relations to higher-order correlations~\cite{hma85,hmb85,mh87}. The higher-order squeezing can be detected in a multi-port homodyne experiment. It would seem worthy to investigate these types of squeezing from the prospectives of multi-mode entanglement and steering involving higher-order correlation functions. Of course, the ideas proposed above must be carefully and quantitatively analyzed on examples of Gaussian systems. It should be noted that four-mode squeezing which involves higher-order correlations has been experimentally realized~\cite{ls87}. In an alternative way, multi-photon entangled states could be generated in a cascaded parametric down-conversion~\cite{hh10} or by using multi-port beamsplitters~\cite{sl13}.

{\it Non-Gaussian systems:} Despite the fact that substantial progress has been made in controllable generation of multimode non-Gaussian states via photon subtraction~\cite{wt20,wp20,rd20}, and three-photon spontaneous parametric down-conversion~\cite{cs20}, the field of research on non-Gaussian steering is currently only theoretical~\cite{sa15,yf21,gs21,lg22}, and experiments have yet to be attempted. It is still desire to develop a general method to classify and quantify steering in such systems. Such issues as how to accomplish a characterization of non-Gaussian states as well as their non-Gaussian properties with general and sufficiently ways has not yet been completely understood. Moreover, the classification and applications for different types of quantum correlations in such scenarios are still largely unexplored. 

It has been recently proved that EPR steering proves a necessary requirement to remotely prepare a Wigner-negative state~\cite{wt20,wp20}. Wigner negativity is known to be a necessary resource for reaching a quantum computation advantage~\cite{me12}. Based on the nonlocal effect existing in the steerable system, one can induce Wigner negativity in multi steering modes by performing some appropriate operations on the steered mode~\cite{wp20,xl21}. Very recently, this connection was applied in magnon-photon system to remotely generate a magnon Schr\"{o}dinger cat state~\cite{sz21}. For the obtained non-Gaussian systems, EPR steering can be detected by measuring higher-order moments~\cite{gs21}, which also reveals the steerability in a larger range of parameters than the criteria only involving second-order moments of the observables. Thus, it is still quite an open area for further investigations. 

{\it Atomic ensembles:} As we have already seen EPR steering as yet had very little investigation between massive systems, in fact limited to BEC systems~\cite{pk15,kp18,fz18}. The study of EPR steering as a test of quantum correlations in microscopic systems should be extended to macroscopic systems such as separated atomic ensembles, where entanglement have been realized in experiments~\cite{jk01,km11}. This topic is of interest for two reasons. Firstly, it could demonstrate steering between distant massive objects. Secondly, it could provide a practical scheme for creation of one-way multipartite steerable states. 

In addition, ideas from atomic physics and cavity QED were combined by considering separated atomic ensembles located in a ring cavity, where entanglement was realized by controlled interaction between the ensembles mediated by the field of the cavity~\cite{ps06,sl11}. This idea could be further extended to treat EPR steering. It could be particularly useful for steering because it could be regarded as an example of a two dimensional closed system which resembles a closed loop of qubits in a two dimensional lattice. In a ring cavity field is composed of two degenerate in frequency and overlapped counter-propagating modes, and each arm could be fitted with an atomic ensemble. A question arises: since the modes are not distinguishable, how this could influence on the possibility to create one-way steerable multipartite state between the ensembles?  If there could exist a multipartite state, in which direction it would be steerable, clockwise or anti-clockwise? We would face here a dilemma of a kind of ``steering frustration". Thus, understanding monogamy of steering in closed systems is also of great relevance for practical applications. This is an area where further investigation is obviously very desirable since the exclusion of one of the directions imposed by steering could demonstrate how monogamy relations could put constrains on direction of steering.

{\it Decoherence:} Current problems with experimental realization of EPR steering in macroscopic system is the effect of decoherence due to the coupling of a system with its environment. Studies have been done on the effect of decoherence on separate systems each coupled to own environment~\cite{ye04,ft02}. Interaction with common reservoir (or intermediate mode) has the advantage that the correlations present in the environment can be transferred into the system leading to a creation of symmetric and antisymmetric states between parts of the system~\cite{zs19,zsy21}. The antisymmetric states have this property that their decoherence rates are significantly reduced compared to the rates of individual systems. This fact could serve as a method to reduce decoherence in macroscopic system to create steerable sates. This is at the moment somewhat missing and would seem worthy of investigations. 

{\it One-way steering in atomic lattices:} Recent practical realizations of atomic lattices~\cite{dy16} seem to be quite promising for simulation of one-way steering in atomic chains or atomic lattices. Realization of the directional property of EPR steering would lead to controlled transmission of an information. 

There have not been much interest to determine what role steering could play in transmission of information through a linear chain or two (2D) and three (3D) dimensional lattices of atoms. Steering is a concept that is asymmetric between two systems. Does it mean that the asymmetric nature of steering requires an asymmetry between the systems? We have seen that in linear optics devices, where field modes play the role of qubits, one-way steering was achieved by imposing extra noise on one of the modes, i.e., asymmetric steering between modes was achieved when modes were damped with different rates. Thus, an asymmetry imposed on the systems resulted in the presence of steering. However, dealing with atoms as qubits, one could face a serious problem. If one-way steering between atoms would require that their damping rates must be different, does it mean that it could be achieved only between nonidentical atoms? The most progressed technology so far in creation of atomic chains and atomic lattices is based on identical atoms. Thus it looks that supposably completely new ways of creating one-way steering between identical atoms will have to be devised.  

Alternatively to create an asymmetry between atoms, one could employ cascaded open systems treatment of an entanglement source illuminating the atoms~\cite{cwg93,hjc93}. In this approach two systems are coupled to each other in such a way that one reacts to the photons emitted by the other, while there is no interaction in the reverse direction. Applying this approach to a linear chain of atoms there should not be much difficulty to achieve the directional transmission of an initial state. In 2D and 3D lattices even with the one-way interaction between atoms an initial state would be redistributed among different atoms. Thus, in lattices additional limitations should be imposed. When investigating steering one of such limitations would be imposed by the monogamy relations.

{\it Hybrid systems:} The idea of hybridization has come into focus and an extensive literature already exists on entanglement and steering in such systems. As one typical hybrid system, cavity optomechanics usually consists of a nano-mechanical oscillator interacting with both a cavity mode and an atomic ensemble or a BEC. It offers new possibilities to explore the boundary between classical and quantum physics~\cite{br08}, and has potential applications for quantum sensing. The successful demonstration of cooling optomechanical systems near their ground states provides the possibility for studying quantum effects in mesoscopic massive systems~\cite{oh10,ca11,td11}.  A lot of efforts have been made to investigate the properties of quantum correlations in such systems, e.g., multipartite steering~\cite{kh14,xs15} and genuine steering~\cite{kt17}.  The observation of deterministic entanglement to mechanical systems was only recently realized~\cite{kp21,do21,tp21}. A further improvement to generate a stronger entanglement to achieve a steerable state is challenging as the state of mechanical oscillator is not accessible directly and requires precise control with a vanishingly small error. 

Even though the cavity optomechanics is composed of different physical elements, optical modes and massive mirrors, the whole system remains in the CV Gaussian regime following the standard linearization method~\cite{hw11}. Recently, the hybrid of DV and CV technologies in one task has broad application prospects~\cite{an15}.  For instance, towards the realization of hybrid quantum networks allowing for the information exchange between CV and DV nodes, researchers achieved the demonstration of EPR steering between continuous- and discrete-variable optical qubits~\cite{cj18}. It is interesting to ask that a state-of-the-art quantum experiment would be able to observe steering in a hybrid CV-DV network apart from fully optical implementations.

\textit{High-dimensional and multipartite systems:} One of the greatest challenges is to prepare EPR steering in multipartite systems, especially with high dimensions, which has been clarified to possess the advantages of increasing communication capacity and noise resilience over quantum channels. It is difficult to distinguish a genuine high-dimensional entanglement which cannot be simulated by using multiple copies of low-dimensional systems~\cite{kr18}. Although the computable necessary and sufficient condition for the high-dimensional steering is rare so far, it can be witnessed by the violation of some inequalities.

In addition, the notion of multipartite steering discussed was mostly defined within the bipartition scenario, even though each partition is allowed to comprise multi modes. However, more general scenarios such as arbitrary $K$-partition of $N$-modes ($K=2,\ldots,N$) are needed for classification. A computable witness for $K-$partite entanglement was proposed~\cite{sv13} and experimentally verified via a quantum frequency comb~\cite{gs15}. It could be extended to study $K-$partite steering, which would lead to further development of a powerful tool lifting any configuration across bipartition to a more general scenario with arbitrary partitions. Moreover, other types of {\it monogamy constraints} for distributing steering over many modes, {\it the computable quantifier} for genuine high-dimensional and multipartite steering are important for implementing quantum tasks. 

{\it Recycling entanglement via quantum steering:} The possibility of recycling entanglement resources via sequential weak measurements is a new promising direction. Recent theoretical investigations~\cite{sg15,bc20,cl21} and experiments~\cite{sc17,hz18} have shown naturally, quantum steering offers an alternate way to recycle entanglement and hence to accomplish the corresponding information processing tasks between sequential independent observers. In this sequential measurement scenario where each observer measures his/her part, records the outcome, and then passes the post-measurement state to the next independent observer, the problem is how many observers can steer or be steered by one single observer or multiple independent ones at the other side? It has been partially answered affirmatively in the recent works~\cite{sd18,sd19,ch20,gm21,yr21,zh21,ll21}. Thus, it is an important issue to establish a complete framework to study the sequential sharing of steerability and hence to analyze the possibilities and limitations of sequential protocols based on quantum steering. 

In summary, recent expanding interest in EPR steering highlights the fundamental significance and general interest of this topic, which has opened exciting possibilities for quantum communication and computing and the study of nonclassical properties of quantum states. Many of these have already been observed. Interest in the understanding the general concept of steering, and in applications of steered states to fundamental problems as well as 
measurement techniques, seems certain to continue for some time.

\begin{acknowledgments}
	This work is supported by the National Natural Science Foundation of China (Grants No. 12125402, 11975026, and No. 12004011), the National Key R$\&$D Program of China (Grant No. 2019YFA0308702). Q.H. also thanks partial support from Beijing Natural Science Foundation (Grant No. Z190005) and the Key R$\&$D Program of Guangdong Province (Grant No. 2018B030329001). S.C. acknowledges the Shanghai Municipal Science and Technology Fundamental Project (Grant No. 21JC1405400).

\end{acknowledgments}

\bibliography{reference_1round}

\begin{thebibliography}{189}%
\makeatletter
\providecommand \@ifxundefined [1]{%
 \@ifx{#1\undefined}
}%
\providecommand \@ifnum [1]{%
 \ifnum #1\expandafter \@firstoftwo
 \else \expandafter \@secondoftwo
 \fi
}%
\providecommand \@ifx [1]{%
 \ifx #1\expandafter \@firstoftwo
 \else \expandafter \@secondoftwo
 \fi
}%
\providecommand \natexlab [1]{#1}%
\providecommand \enquote  [1]{``#1''}%
\providecommand \bibnamefont  [1]{#1}%
\providecommand \bibfnamefont [1]{#1}%
\providecommand \citenamefont [1]{#1}%
\providecommand \href@noop [0]{\@secondoftwo}%
\providecommand \href [0]{\begingroup \@sanitize@url \@href}%
\providecommand \@href[1]{\@@startlink{#1}\@@href}%
\providecommand \@@href[1]{\endgroup#1\@@endlink}%
\providecommand \@sanitize@url [0]{\catcode `\\12\catcode `\$12\catcode
  `\&12\catcode `\#12\catcode `\^12\catcode `\_12\catcode `\%12\relax}%
\providecommand \@@startlink[1]{}%
\providecommand \@@endlink[0]{}%
\providecommand \url  [0]{\begingroup\@sanitize@url \@url }%
\providecommand \@url [1]{\endgroup\@href {#1}{\urlprefix }}%
\providecommand \urlprefix  [0]{URL }%
\providecommand \Eprint [0]{\href }%
\providecommand \doibase [0]{https://doi.org/}%
\providecommand \selectlanguage [0]{\@gobble}%
\providecommand \bibinfo  [0]{\@secondoftwo}%
\providecommand \bibfield  [0]{\@secondoftwo}%
\providecommand \translation [1]{[#1]}%
\providecommand \BibitemOpen [0]{}%
\providecommand \bibitemStop [0]{}%
\providecommand \bibitemNoStop [0]{.\EOS\space}%
\providecommand \EOS [0]{\spacefactor3000\relax}%
\providecommand \BibitemShut  [1]{\csname bibitem#1\endcsname}%
\let\auto@bib@innerbib\@empty
\bibitem [{\citenamefont {Schr{\"o}dinger}(1935)}]{s35}%
  \BibitemOpen
  \bibfield  {author} {\bibinfo {author} {\bibfnamefont {E.}~\bibnamefont
  {Schr{\"o}dinger}},\ }\bibfield  {title} {\bibinfo {title} {Discussion of
  probability relations between separated systems},\ }\href
  {https://doi.org/10.1017/S0305004100013554} {\bibfield  {journal} {\bibinfo
  {journal} {Proc. Cambridge Philos. Soc.}\ }\textbf {\bibinfo {volume} {31}},\
  \bibinfo {pages} {555} (\bibinfo {year} {1935})}\BibitemShut {NoStop}%
\bibitem [{\citenamefont {Einstein}\ \emph {et~al.}(1935)\citenamefont
  {Einstein}, \citenamefont {Podolsky},\ and\ \citenamefont {Rosen}}]{epr35}%
  \BibitemOpen
  \bibfield  {author} {\bibinfo {author} {\bibfnamefont {A.}~\bibnamefont
  {Einstein}}, \bibinfo {author} {\bibfnamefont {B.}~\bibnamefont {Podolsky}},\
  and\ \bibinfo {author} {\bibfnamefont {N.}~\bibnamefont {Rosen}},\ }\bibfield
   {title} {\bibinfo {title} {Can quantum-mechanical description of physical
  reality be considered complete?},\ }\href
  {https://doi.org/10.1103/PhysRev.47.777} {\bibfield  {journal} {\bibinfo
  {journal} {Phys. Rev.}\ }\textbf {\bibinfo {volume} {47}},\ \bibinfo {pages}
  {777} (\bibinfo {year} {1935})}\BibitemShut {NoStop}%
\bibitem [{\citenamefont {Reid}(1989)}]{mr89}%
  \BibitemOpen
  \bibfield  {author} {\bibinfo {author} {\bibfnamefont {M.~D.}\ \bibnamefont
  {Reid}},\ }\bibfield  {title} {\bibinfo {title} {Demonstration of the
  einstein-podolsky-rosen paradox using nondegenerate parametric
  amplification},\ }\href@noop {} {\bibfield  {journal} {\bibinfo  {journal}
  {Phys. Rev. A}\ }\textbf {\bibinfo {volume} {40}},\ \bibinfo {pages} {913}
  (\bibinfo {year} {1989})}\BibitemShut {NoStop}%
\bibitem [{\citenamefont {Ou}\ \emph {et~al.}(1992)\citenamefont {Ou},
  \citenamefont {Pereira}, \citenamefont {Kimble},\ and\ \citenamefont
  {Peng}}]{op92}%
  \BibitemOpen
  \bibfield  {author} {\bibinfo {author} {\bibfnamefont {Z.~Y.}\ \bibnamefont
  {Ou}}, \bibinfo {author} {\bibfnamefont {S.~F.}\ \bibnamefont {Pereira}},
  \bibinfo {author} {\bibfnamefont {H.~J.}\ \bibnamefont {Kimble}},\ and\
  \bibinfo {author} {\bibfnamefont {K.~C.}\ \bibnamefont {Peng}},\ }\bibfield
  {title} {\bibinfo {title} {Realization of the einstein-podolsky-rosen paradox
  for continuous variables},\ }\href
  {https://doi.org/10.1103/PhysRevLett.68.3663} {\bibfield  {journal} {\bibinfo
   {journal} {Phys. Rev. Lett.}\ }\textbf {\bibinfo {volume} {68}},\ \bibinfo
  {pages} {3663} (\bibinfo {year} {1992})}\BibitemShut {NoStop}%
\bibitem [{\citenamefont {Reid}\ \emph {et~al.}(2009)\citenamefont {Reid},
  \citenamefont {Drummond}, \citenamefont {Bowen}, \citenamefont {Cavalcanti},
  \citenamefont {Lam}, \citenamefont {Bachor}, \citenamefont {Andersen},\ and\
  \citenamefont {Leuchs}}]{rd09}%
  \BibitemOpen
  \bibfield  {author} {\bibinfo {author} {\bibfnamefont {M.~D.}\ \bibnamefont
  {Reid}}, \bibinfo {author} {\bibfnamefont {P.~D.}\ \bibnamefont {Drummond}},
  \bibinfo {author} {\bibfnamefont {W.~P.}\ \bibnamefont {Bowen}}, \bibinfo
  {author} {\bibfnamefont {E.~G.}\ \bibnamefont {Cavalcanti}}, \bibinfo
  {author} {\bibfnamefont {P.~K.}\ \bibnamefont {Lam}}, \bibinfo {author}
  {\bibfnamefont {H.~A.}\ \bibnamefont {Bachor}}, \bibinfo {author}
  {\bibfnamefont {U.~L.}\ \bibnamefont {Andersen}},\ and\ \bibinfo {author}
  {\bibfnamefont {G.}~\bibnamefont {Leuchs}},\ }\bibfield  {title} {\bibinfo
  {title} {Colloquium: The einstein-podolsky-rosen paradox: From concepts to
  applications},\ }\href@noop {} {\bibfield  {journal} {\bibinfo  {journal}
  {Rev. Mod. Phys.}\ }\textbf {\bibinfo {volume} {81}},\ \bibinfo {pages}
  {1727} (\bibinfo {year} {2009})}\BibitemShut {NoStop}%
\bibitem [{\citenamefont {Cavalcanti}\ and\ \citenamefont
  {Skrzypczyk}(2016{\natexlab{a}})}]{cs16}%
  \BibitemOpen
  \bibfield  {author} {\bibinfo {author} {\bibfnamefont {D.}~\bibnamefont
  {Cavalcanti}}\ and\ \bibinfo {author} {\bibfnamefont {P.}~\bibnamefont
  {Skrzypczyk}},\ }\bibfield  {title} {\bibinfo {title} {Quantum steering: a
  review with focus on semidefinite programming},\ }\href
  {https://doi.org/10.1088/1361-6633/80/2/024001} {\bibfield  {journal}
  {\bibinfo  {journal} {Rep. Prog. Phys.}\ }\textbf {\bibinfo {volume} {80}},\
  \bibinfo {pages} {024001} (\bibinfo {year} {2016}{\natexlab{a}})}\BibitemShut
  {NoStop}%
\bibitem [{\citenamefont {Uola}\ \emph {et~al.}(2020)\citenamefont {Uola},
  \citenamefont {Costa}, \citenamefont {Nguyen},\ and\ \citenamefont
  {G\"{u}hne}}]{ucn20}%
  \BibitemOpen
  \bibfield  {author} {\bibinfo {author} {\bibfnamefont {R.}~\bibnamefont
  {Uola}}, \bibinfo {author} {\bibfnamefont {A.~C.~S.}\ \bibnamefont {Costa}},
  \bibinfo {author} {\bibfnamefont {H.~C.}\ \bibnamefont {Nguyen}},\ and\
  \bibinfo {author} {\bibfnamefont {O.}~\bibnamefont {G\"{u}hne}},\ }\bibfield
  {title} {\bibinfo {title} {Quantum steering},\ }\href@noop {} {\bibfield
  {journal} {\bibinfo  {journal} {Rev. Mod. Phys.}\ }\textbf {\bibinfo {volume}
  {92}},\ \bibinfo {pages} {015001} (\bibinfo {year} {2020})}\BibitemShut
  {NoStop}%
\bibitem [{\citenamefont {Xiang}\ \emph {et~al.}(2021)\citenamefont {Xiang},
  \citenamefont {Sun}, \citenamefont {He},\ and\ \citenamefont {Gong}}]{xsh21}%
  \BibitemOpen
  \bibfield  {author} {\bibinfo {author} {\bibfnamefont {Y.}~\bibnamefont
  {Xiang}}, \bibinfo {author} {\bibfnamefont {F.~X.}\ \bibnamefont {Sun}},
  \bibinfo {author} {\bibfnamefont {Q.~Y.}\ \bibnamefont {He}},\ and\ \bibinfo
  {author} {\bibfnamefont {Q.~H.}\ \bibnamefont {Gong}},\ }\bibfield  {title}
  {\bibinfo {title} {Advances in multipartite and high-dimensional
  einstein-podolsky-rosen steering},\ }\href@noop {} {\bibfield  {journal}
  {\bibinfo  {journal} {Fundamental Research}\ }\textbf {\bibinfo {volume}
  {1}},\ \bibinfo {pages} {99} (\bibinfo {year} {2021})}\BibitemShut {NoStop}%
\bibitem [{\citenamefont {Wiseman}\ \emph {et~al.}(2007)\citenamefont
  {Wiseman}, \citenamefont {Jones},\ and\ \citenamefont {Doherty}}]{wj07}%
  \BibitemOpen
  \bibfield  {author} {\bibinfo {author} {\bibfnamefont {H.~M.}\ \bibnamefont
  {Wiseman}}, \bibinfo {author} {\bibfnamefont {S.~J.}\ \bibnamefont {Jones}},\
  and\ \bibinfo {author} {\bibfnamefont {A.~C.}\ \bibnamefont {Doherty}},\
  }\bibfield  {title} {\bibinfo {title} {Steering, entanglement, nonlocality,
  and the einstein-podolsky-rosen paradox},\ }\href@noop {} {\bibfield
  {journal} {\bibinfo  {journal} {Phys. Rev. Lett.}\ }\textbf {\bibinfo
  {volume} {98}},\ \bibinfo {pages} {140402} (\bibinfo {year}
  {2007})}\BibitemShut {NoStop}%
\bibitem [{\citenamefont {Jones}\ \emph
  {et~al.}(2007{\natexlab{a}})\citenamefont {Jones}, \citenamefont {Wiseman},\
  and\ \citenamefont {Doherty}}]{jwd07}%
  \BibitemOpen
  \bibfield  {author} {\bibinfo {author} {\bibfnamefont {S.~J.}\ \bibnamefont
  {Jones}}, \bibinfo {author} {\bibfnamefont {H.~M.}\ \bibnamefont {Wiseman}},\
  and\ \bibinfo {author} {\bibfnamefont {A.~C.}\ \bibnamefont {Doherty}},\
  }\bibfield  {title} {\bibinfo {title} {Entanglement, einstein-podolsky-rosen
  correlations, bell nonlocality, and steering},\ }\href
  {https://doi.org/10.1103/PhysRevA.76.052116} {\bibfield  {journal} {\bibinfo
  {journal} {Phys. Rev. A}\ }\textbf {\bibinfo {volume} {76}},\ \bibinfo
  {pages} {052116} (\bibinfo {year} {2007}{\natexlab{a}})}\BibitemShut
  {NoStop}%
\bibitem [{\citenamefont {Branciard}\ \emph {et~al.}(2012)\citenamefont
  {Branciard}, \citenamefont {Cavalcanti}, \citenamefont {Walborn},
  \citenamefont {Scarani},\ and\ \citenamefont {Wiseman}}]{bc12}%
  \BibitemOpen
  \bibfield  {author} {\bibinfo {author} {\bibfnamefont {C.}~\bibnamefont
  {Branciard}}, \bibinfo {author} {\bibfnamefont {E.~G.}\ \bibnamefont
  {Cavalcanti}}, \bibinfo {author} {\bibfnamefont {S.~P.}\ \bibnamefont
  {Walborn}}, \bibinfo {author} {\bibfnamefont {V.}~\bibnamefont {Scarani}},\
  and\ \bibinfo {author} {\bibfnamefont {H.~M.}\ \bibnamefont {Wiseman}},\
  }\bibfield  {title} {\bibinfo {title} {One-sided device-independent quantum
  key distribution: Security, feasibility, and the connection with steering},\
  }\href {https://doi.org/10.1103/PhysRevA.85.010301} {\bibfield  {journal}
  {\bibinfo  {journal} {Phys. Rev. A}\ }\textbf {\bibinfo {volume} {85}},\
  \bibinfo {pages} {010301} (\bibinfo {year} {2012})}\BibitemShut {NoStop}%
\bibitem [{\citenamefont {Lo}\ \emph {et~al.}(2014)\citenamefont {Lo},
  \citenamefont {Curty},\ and\ \citenamefont {Tamaki}}]{lct14}%
  \BibitemOpen
  \bibfield  {author} {\bibinfo {author} {\bibfnamefont {H.-K.}\ \bibnamefont
  {Lo}}, \bibinfo {author} {\bibfnamefont {M.}~\bibnamefont {Curty}},\ and\
  \bibinfo {author} {\bibfnamefont {K.}~\bibnamefont {Tamaki}},\ }\bibfield
  {title} {\bibinfo {title} {Secure quantum key distribution},\ }\href@noop {}
  {\bibfield  {journal} {\bibinfo  {journal} {Nat. Photonics}\ }\textbf
  {\bibinfo {volume} {8}},\ \bibinfo {pages} {595} (\bibinfo {year}
  {2014})}\BibitemShut {NoStop}%
\bibitem [{\citenamefont {Gehring}\ \emph {et~al.}(2015)\citenamefont
  {Gehring}, \citenamefont {H\"{a}ndchen}, \citenamefont {Duhme}, \citenamefont
  {Furrer}, \citenamefont {Franz}, \citenamefont {Pacher}, \citenamefont
  {Werner},\ and\ \citenamefont {Schnabel}}]{gh15}%
  \BibitemOpen
  \bibfield  {author} {\bibinfo {author} {\bibfnamefont {T.}~\bibnamefont
  {Gehring}}, \bibinfo {author} {\bibfnamefont {V.}~\bibnamefont
  {H\"{a}ndchen}}, \bibinfo {author} {\bibfnamefont {J.}~\bibnamefont {Duhme}},
  \bibinfo {author} {\bibfnamefont {F.}~\bibnamefont {Furrer}}, \bibinfo
  {author} {\bibfnamefont {T.}~\bibnamefont {Franz}}, \bibinfo {author}
  {\bibfnamefont {C.}~\bibnamefont {Pacher}}, \bibinfo {author} {\bibfnamefont
  {R.~F.}\ \bibnamefont {Werner}},\ and\ \bibinfo {author} {\bibfnamefont
  {R.}~\bibnamefont {Schnabel}},\ }\bibfield  {title} {\bibinfo {title}
  {Implementation of continuous-variable quantum key distribution with
  composable and one-sided-device independent security against coherent
  attacks},\ }\href@noop {} {\bibfield  {journal} {\bibinfo  {journal} {Nat.
  Commun.}\ }\textbf {\bibinfo {volume} {6}},\ \bibinfo {pages} {8795}
  (\bibinfo {year} {2015})}\BibitemShut {NoStop}%
\bibitem [{\citenamefont {Walk}\ \emph {et~al.}(2016)\citenamefont {Walk},
  \citenamefont {Hosseini}, \citenamefont {Geng}, \citenamefont {Thearle},
  \citenamefont {Haw}, \citenamefont {Armstrong}, \citenamefont {Assad},
  \citenamefont {Janou\v{s}ek}, \citenamefont {Ralph}, \citenamefont {Symul},
  \citenamefont {Wiseman},\ and\ \citenamefont {Lam}}]{wh16}%
  \BibitemOpen
  \bibfield  {author} {\bibinfo {author} {\bibfnamefont {N.}~\bibnamefont
  {Walk}}, \bibinfo {author} {\bibfnamefont {S.}~\bibnamefont {Hosseini}},
  \bibinfo {author} {\bibfnamefont {J.}~\bibnamefont {Geng}}, \bibinfo {author}
  {\bibfnamefont {O.}~\bibnamefont {Thearle}}, \bibinfo {author} {\bibfnamefont
  {J.~Y.}\ \bibnamefont {Haw}}, \bibinfo {author} {\bibfnamefont
  {S.}~\bibnamefont {Armstrong}}, \bibinfo {author} {\bibfnamefont {S.~M.}\
  \bibnamefont {Assad}}, \bibinfo {author} {\bibfnamefont {J.}~\bibnamefont
  {Janou\v{s}ek}}, \bibinfo {author} {\bibfnamefont {T.~C.}\ \bibnamefont
  {Ralph}}, \bibinfo {author} {\bibfnamefont {T.}~\bibnamefont {Symul}},
  \bibinfo {author} {\bibfnamefont {H.~M.}\ \bibnamefont {Wiseman}},\ and\
  \bibinfo {author} {\bibfnamefont {P.~K.}\ \bibnamefont {Lam}},\ }\bibfield
  {title} {\bibinfo {title} {Experimental demonstration of gaussian protocols
  for one-sided device-independent quantum key distribution},\ }\href@noop {}
  {\bibfield  {journal} {\bibinfo  {journal} {Optica}\ }\textbf {\bibinfo
  {volume} {3}},\ \bibinfo {pages} {634} (\bibinfo {year} {2016})}\BibitemShut
  {NoStop}%
\bibitem [{\citenamefont {Armstrong}\ \emph {et~al.}(2015)\citenamefont
  {Armstrong}, \citenamefont {Wang}, \citenamefont {Teh}, \citenamefont {Gong},
  \citenamefont {He}, \citenamefont {Janousek}, \citenamefont {Bachor},
  \citenamefont {Reid},\ and\ \citenamefont {Lam}}]{aw15}%
  \BibitemOpen
  \bibfield  {author} {\bibinfo {author} {\bibfnamefont {S.}~\bibnamefont
  {Armstrong}}, \bibinfo {author} {\bibfnamefont {M.}~\bibnamefont {Wang}},
  \bibinfo {author} {\bibfnamefont {R.~Y.}\ \bibnamefont {Teh}}, \bibinfo
  {author} {\bibfnamefont {Q.}~\bibnamefont {Gong}}, \bibinfo {author}
  {\bibfnamefont {Q.~Y.}\ \bibnamefont {He}}, \bibinfo {author} {\bibfnamefont
  {J.}~\bibnamefont {Janousek}}, \bibinfo {author} {\bibfnamefont {H.-A.}\
  \bibnamefont {Bachor}}, \bibinfo {author} {\bibfnamefont {M.~D.}\
  \bibnamefont {Reid}},\ and\ \bibinfo {author} {\bibfnamefont {P.~K.}\
  \bibnamefont {Lam}},\ }\bibfield  {title} {\bibinfo {title} {Multipartite
  einstein-podolsky-rosen steering and genuine tripartite entanglement with
  optical networks},\ }\href@noop {} {\bibfield  {journal} {\bibinfo  {journal}
  {Nat. Phys.}\ }\textbf {\bibinfo {volume} {11}},\ \bibinfo {pages} {167}
  (\bibinfo {year} {2015})}\BibitemShut {NoStop}%
\bibitem [{\citenamefont {Kogias}\ \emph {et~al.}(2017)\citenamefont {Kogias},
  \citenamefont {Xiang}, \citenamefont {He},\ and\ \citenamefont
  {Adesso}}]{kx17}%
  \BibitemOpen
  \bibfield  {author} {\bibinfo {author} {\bibfnamefont {I.}~\bibnamefont
  {Kogias}}, \bibinfo {author} {\bibfnamefont {Y.}~\bibnamefont {Xiang}},
  \bibinfo {author} {\bibfnamefont {Q.~Y.}\ \bibnamefont {He}},\ and\ \bibinfo
  {author} {\bibfnamefont {G.}~\bibnamefont {Adesso}},\ }\bibfield  {title}
  {\bibinfo {title} {Unconditional security of entanglement-based
  continuous-variable quantum secret sharing},\ }\href
  {https://doi.org/10.1103/PhysRevA.95.012315} {\bibfield  {journal} {\bibinfo
  {journal} {Phys. Rev. A}\ }\textbf {\bibinfo {volume} {95}},\ \bibinfo
  {pages} {012315} (\bibinfo {year} {2017})}\BibitemShut {NoStop}%
\bibitem [{\citenamefont {Wilkinson}\ \emph {et~al.}(2021)\citenamefont
  {Wilkinson}, \citenamefont {Thornton},\ and\ \citenamefont
  {Korolkova}}]{wt21}%
  \BibitemOpen
  \bibfield  {author} {\bibinfo {author} {\bibfnamefont {C.}~\bibnamefont
  {Wilkinson}}, \bibinfo {author} {\bibfnamefont {M.}~\bibnamefont
  {Thornton}},\ and\ \bibinfo {author} {\bibfnamefont {N.}~\bibnamefont
  {Korolkova}},\ }\bibfield  {title} {\bibinfo {title} {Quantum steering is the
  resource for secure tripartite quantum state sharing},\ }\href@noop {}
  {\bibfield  {journal} {\bibinfo  {journal} {arXiv preprint arXiv:2106.06337}\
  } (\bibinfo {year} {2021})}\BibitemShut {NoStop}%
\bibitem [{\citenamefont {Reid}(2013{\natexlab{a}})}]{mr13}%
  \BibitemOpen
  \bibfield  {author} {\bibinfo {author} {\bibfnamefont {M.~D.}\ \bibnamefont
  {Reid}},\ }\bibfield  {title} {\bibinfo {title} {Signifying quantum
  benchmarks for qubit teleportation and secure quantum communication using
  einstein-podolsky-rosen steering inequalities},\ }\href
  {https://doi.org/10.1103/PhysRevA.88.062338} {\bibfield  {journal} {\bibinfo
  {journal} {Phys. Rev. A}\ }\textbf {\bibinfo {volume} {88}},\ \bibinfo
  {pages} {062338} (\bibinfo {year} {2013}{\natexlab{a}})}\BibitemShut
  {NoStop}%
\bibitem [{\citenamefont {He}\ \emph {et~al.}(2015{\natexlab{a}})\citenamefont
  {He}, \citenamefont {Rosales-Z\'arate}, \citenamefont {Adesso},\ and\
  \citenamefont {Reid}}]{hz15}%
  \BibitemOpen
  \bibfield  {author} {\bibinfo {author} {\bibfnamefont {Q.~Y.}\ \bibnamefont
  {He}}, \bibinfo {author} {\bibfnamefont {L.}~\bibnamefont
  {Rosales-Z\'arate}}, \bibinfo {author} {\bibfnamefont {G.}~\bibnamefont
  {Adesso}},\ and\ \bibinfo {author} {\bibfnamefont {M.~D.}\ \bibnamefont
  {Reid}},\ }\bibfield  {title} {\bibinfo {title} {Secure continuous variable
  teleportation and einstein-podolsky-rosen steering},\ }\href
  {https://doi.org/10.1103/PhysRevLett.115.180502} {\bibfield  {journal}
  {\bibinfo  {journal} {Phys. Rev. Lett.}\ }\textbf {\bibinfo {volume} {115}},\
  \bibinfo {pages} {180502} (\bibinfo {year} {2015}{\natexlab{a}})}\BibitemShut
  {NoStop}%
\bibitem [{\citenamefont {Law}\ \emph {et~al.}(2014)\citenamefont {Law},
  \citenamefont {Thinh}, \citenamefont {Bancal},\ and\ \citenamefont
  {Scarani}}]{lt14}%
  \BibitemOpen
  \bibfield  {author} {\bibinfo {author} {\bibfnamefont {Y.~Z.}\ \bibnamefont
  {Law}}, \bibinfo {author} {\bibfnamefont {L.~P.}\ \bibnamefont {Thinh}},
  \bibinfo {author} {\bibfnamefont {J.~D.}\ \bibnamefont {Bancal}},\ and\
  \bibinfo {author} {\bibfnamefont {V.}~\bibnamefont {Scarani}},\ }\bibfield
  {title} {\bibinfo {title} {Quantum randomness extraction for various levels
  of characterization of the devices},\ }\href
  {https://iopscience.iop.org/article/10.1088/1751-8113/47/42/424028}
  {\bibfield  {journal} {\bibinfo  {journal} {J. Phys. A}\ }\textbf {\bibinfo
  {volume} {47}},\ \bibinfo {pages} {424028} (\bibinfo {year}
  {2014})}\BibitemShut {NoStop}%
\bibitem [{\citenamefont {Passaro}\ \emph {et~al.}(2015)\citenamefont
  {Passaro}, \citenamefont {Cavalcanti}, \citenamefont {Skrzypczyk},\ and\
  \citenamefont {Ac\'{\i}n}}]{pc15}%
  \BibitemOpen
  \bibfield  {author} {\bibinfo {author} {\bibfnamefont {E.}~\bibnamefont
  {Passaro}}, \bibinfo {author} {\bibfnamefont {D.}~\bibnamefont {Cavalcanti}},
  \bibinfo {author} {\bibfnamefont {P.}~\bibnamefont {Skrzypczyk}},\ and\
  \bibinfo {author} {\bibfnamefont {A.}~\bibnamefont {Ac\'{\i}n}},\ }\bibfield
  {title} {\bibinfo {title} {Optimal randomness certification in the quantum
  steering and prepare-and-measure scenarios},\ }\href
  {https://iopscience.iop.org/article/10.1088/1367-2630/17/11/113010}
  {\bibfield  {journal} {\bibinfo  {journal} {New J. Phys.}\ }\textbf {\bibinfo
  {volume} {17}},\ \bibinfo {pages} {113010} (\bibinfo {year}
  {2015})}\BibitemShut {NoStop}%
\bibitem [{\citenamefont {Skrzypczyk}\ and\ \citenamefont
  {Cavalcanti}(2018)}]{sc18}%
  \BibitemOpen
  \bibfield  {author} {\bibinfo {author} {\bibfnamefont {P.}~\bibnamefont
  {Skrzypczyk}}\ and\ \bibinfo {author} {\bibfnamefont {D.}~\bibnamefont
  {Cavalcanti}},\ }\bibfield  {title} {\bibinfo {title} {Maximal randomness
  generation from steering inequality violations using qudits},\ }\href
  {https://journals.aps.org/prl/abstract/10.1103/PhysRevLett.120.260401}
  {\bibfield  {journal} {\bibinfo  {journal} {Phys. Rev. Lett.}\ }\textbf
  {\bibinfo {volume} {120}},\ \bibinfo {pages} {260401} (\bibinfo {year}
  {2018})}\BibitemShut {NoStop}%
\bibitem [{\citenamefont {Guo}\ \emph {et~al.}(2019)\citenamefont {Guo},
  \citenamefont {Cheng}, \citenamefont {Hu}, \citenamefont {Liu}, \citenamefont
  {Huang}, \citenamefont {Huang}, \citenamefont {Li}, \citenamefont {Guo},\
  and\ \citenamefont {Cavalcanti}}]{gc19}%
  \BibitemOpen
  \bibfield  {author} {\bibinfo {author} {\bibfnamefont {Y.}~\bibnamefont
  {Guo}}, \bibinfo {author} {\bibfnamefont {S.}~\bibnamefont {Cheng}}, \bibinfo
  {author} {\bibfnamefont {X.}~\bibnamefont {Hu}}, \bibinfo {author}
  {\bibfnamefont {B.-H.}\ \bibnamefont {Liu}}, \bibinfo {author} {\bibfnamefont
  {E.-M.}\ \bibnamefont {Huang}}, \bibinfo {author} {\bibfnamefont {Y.-F.}\
  \bibnamefont {Huang}}, \bibinfo {author} {\bibfnamefont {C.-F.}\ \bibnamefont
  {Li}}, \bibinfo {author} {\bibfnamefont {G.-C.}\ \bibnamefont {Guo}},\ and\
  \bibinfo {author} {\bibfnamefont {E.~G.}\ \bibnamefont {Cavalcanti}},\
  }\bibfield  {title} {\bibinfo {title} {Experimental
  measurement-device-independent quantum steering and randomness generation
  beyond qubits},\ }\href@noop {} {\bibfield  {journal} {\bibinfo  {journal}
  {Phys. Rev. Lett.}\ }\textbf {\bibinfo {volume} {123}},\ \bibinfo {pages}
  {170402} (\bibinfo {year} {2019})}\BibitemShut {NoStop}%
\bibitem [{\citenamefont {Piani}\ and\ \citenamefont {Watrous}(2015)}]{pw15}%
  \BibitemOpen
  \bibfield  {author} {\bibinfo {author} {\bibfnamefont {M.}~\bibnamefont
  {Piani}}\ and\ \bibinfo {author} {\bibfnamefont {J.}~\bibnamefont
  {Watrous}},\ }\bibfield  {title} {\bibinfo {title} {Necessary and sufficient
  quantum information characterization of einstein-podolsky-rosen steering},\
  }\href {https://doi.org/10.1103/PhysRevLett.114.060404} {\bibfield  {journal}
  {\bibinfo  {journal} {Phys. Rev. Lett.}\ }\textbf {\bibinfo {volume} {114}},\
  \bibinfo {pages} {060404} (\bibinfo {year} {2015})}\BibitemShut {NoStop}%
\bibitem [{\citenamefont {Yadin}\ \emph {et~al.}(2021)\citenamefont {Yadin},
  \citenamefont {Fadel},\ and\ \citenamefont {Gessner}}]{yf21}%
  \BibitemOpen
  \bibfield  {author} {\bibinfo {author} {\bibfnamefont {B.}~\bibnamefont
  {Yadin}}, \bibinfo {author} {\bibfnamefont {M.}~\bibnamefont {Fadel}},\ and\
  \bibinfo {author} {\bibfnamefont {M.}~\bibnamefont {Gessner}},\ }\bibfield
  {title} {\bibinfo {title} {Metrological complementarity reveals the
  einstein-podolsky-rosen paradox},\ }\href@noop {} {\bibfield  {journal}
  {\bibinfo  {journal} {Nat. Commun.}\ }\textbf {\bibinfo {volume} {12}},\
  \bibinfo {pages} {2410} (\bibinfo {year} {2021})}\BibitemShut {NoStop}%
\bibitem [{\citenamefont {Brunner}\ \emph {et~al.}(2014)\citenamefont
  {Brunner}, \citenamefont {Cavalcanti}, \citenamefont {Pironio}, \citenamefont
  {Scarani},\ and\ \citenamefont {Wehner}}]{bcp14}%
  \BibitemOpen
  \bibfield  {author} {\bibinfo {author} {\bibfnamefont {N.}~\bibnamefont
  {Brunner}}, \bibinfo {author} {\bibfnamefont {D.}~\bibnamefont {Cavalcanti}},
  \bibinfo {author} {\bibfnamefont {S.}~\bibnamefont {Pironio}}, \bibinfo
  {author} {\bibfnamefont {V.}~\bibnamefont {Scarani}},\ and\ \bibinfo {author}
  {\bibfnamefont {S.}~\bibnamefont {Wehner}},\ }\bibfield  {title} {\bibinfo
  {title} {Bell nonlocality},\ }\href
  {https://doi.org/10.1103/RevModPhys.86.419} {\bibfield  {journal} {\bibinfo
  {journal} {Rev. Mod. Phys.}\ }\textbf {\bibinfo {volume} {86}},\ \bibinfo
  {pages} {419} (\bibinfo {year} {2014})}\BibitemShut {NoStop}%
\bibitem [{\citenamefont {Horodecki}\ \emph {et~al.}(2009)\citenamefont
  {Horodecki}, \citenamefont {Horodecki}, \citenamefont {Horodecki},\ and\
  \citenamefont {Horodecki}}]{hhhh09}%
  \BibitemOpen
  \bibfield  {author} {\bibinfo {author} {\bibfnamefont {R.}~\bibnamefont
  {Horodecki}}, \bibinfo {author} {\bibfnamefont {P.}~\bibnamefont
  {Horodecki}}, \bibinfo {author} {\bibfnamefont {M.}~\bibnamefont
  {Horodecki}},\ and\ \bibinfo {author} {\bibfnamefont {K.}~\bibnamefont
  {Horodecki}},\ }\bibfield  {title} {\bibinfo {title} {Quantum entanglement},\
  }\href {https://doi.org/10.1103/RevModPhys.81.865} {\bibfield  {journal}
  {\bibinfo  {journal} {Rev. Mod. Phys.}\ }\textbf {\bibinfo {volume} {81}},\
  \bibinfo {pages} {865} (\bibinfo {year} {2009})}\BibitemShut {NoStop}%
\bibitem [{\citenamefont {Quintino}\ \emph {et~al.}(2015)\citenamefont
  {Quintino}, \citenamefont {V\'ertesi}, \citenamefont {Cavalcanti},
  \citenamefont {Augusiak}, \citenamefont {Demianowicz}, \citenamefont
  {Ac\'{\i}n},\ and\ \citenamefont {Brunner}}]{qv15}%
  \BibitemOpen
  \bibfield  {author} {\bibinfo {author} {\bibfnamefont {M.~T.}\ \bibnamefont
  {Quintino}}, \bibinfo {author} {\bibfnamefont {T.}~\bibnamefont {V\'ertesi}},
  \bibinfo {author} {\bibfnamefont {D.}~\bibnamefont {Cavalcanti}}, \bibinfo
  {author} {\bibfnamefont {R.}~\bibnamefont {Augusiak}}, \bibinfo {author}
  {\bibfnamefont {M.}~\bibnamefont {Demianowicz}}, \bibinfo {author}
  {\bibfnamefont {A.}~\bibnamefont {Ac\'{\i}n}},\ and\ \bibinfo {author}
  {\bibfnamefont {N.}~\bibnamefont {Brunner}},\ }\bibfield  {title} {\bibinfo
  {title} {Inequivalence of entanglement, steering, and bell nonlocality for
  general measurements},\ }\href {https://doi.org/10.1103/PhysRevA.92.032107}
  {\bibfield  {journal} {\bibinfo  {journal} {Phys. Rev. A}\ }\textbf {\bibinfo
  {volume} {92}},\ \bibinfo {pages} {032107} (\bibinfo {year}
  {2015})}\BibitemShut {NoStop}%
\bibitem [{\citenamefont {Wagner}\ \emph {et~al.}(2008)\citenamefont {Wagner},
  \citenamefont {Janousek}, \citenamefont {Delaubert}, \citenamefont {Zou},
  \citenamefont {Harb}, \citenamefont {Treps}, \citenamefont {Morizur},
  \citenamefont {Lam},\ and\ \citenamefont {Bachor}}]{wj08}%
  \BibitemOpen
  \bibfield  {author} {\bibinfo {author} {\bibfnamefont {K.}~\bibnamefont
  {Wagner}}, \bibinfo {author} {\bibfnamefont {J.}~\bibnamefont {Janousek}},
  \bibinfo {author} {\bibfnamefont {V.}~\bibnamefont {Delaubert}}, \bibinfo
  {author} {\bibfnamefont {H.}~\bibnamefont {Zou}}, \bibinfo {author}
  {\bibfnamefont {C.}~\bibnamefont {Harb}}, \bibinfo {author} {\bibfnamefont
  {N.}~\bibnamefont {Treps}}, \bibinfo {author} {\bibfnamefont {J.~F.}\
  \bibnamefont {Morizur}}, \bibinfo {author} {\bibfnamefont {P.~K.}\
  \bibnamefont {Lam}},\ and\ \bibinfo {author} {\bibfnamefont {H.~A.}\
  \bibnamefont {Bachor}},\ }\bibfield  {title} {\bibinfo {title} {Entangling
  the spatial properties of laser beams},\ }\href@noop {} {\bibfield  {journal}
  {\bibinfo  {journal} {Science}\ }\textbf {\bibinfo {volume} {321}},\ \bibinfo
  {pages} {541} (\bibinfo {year} {2008})}\BibitemShut {NoStop}%
\bibitem [{\citenamefont {He}\ \emph {et~al.}(2015{\natexlab{b}})\citenamefont
  {He}, \citenamefont {Gong},\ and\ \citenamefont {Reid}}]{hg15}%
  \BibitemOpen
  \bibfield  {author} {\bibinfo {author} {\bibfnamefont {Q.~Y.}\ \bibnamefont
  {He}}, \bibinfo {author} {\bibfnamefont {Q.~H.}\ \bibnamefont {Gong}},\ and\
  \bibinfo {author} {\bibfnamefont {M.~D.}\ \bibnamefont {Reid}},\ }\bibfield
  {title} {\bibinfo {title} {Classifying directional gaussian entanglement,
  einstein-podolsky-rosen steering, and discord},\ }\href
  {https://doi.org/10.1103/PhysRevLett.114.060402} {\bibfield  {journal}
  {\bibinfo  {journal} {Phys. Rev. Lett.}\ }\textbf {\bibinfo {volume} {114}},\
  \bibinfo {pages} {060402} (\bibinfo {year} {2015}{\natexlab{b}})}\BibitemShut
  {NoStop}%
\bibitem [{\citenamefont {Rosales-Z\'{a}rate}\ \emph
  {et~al.}(2015)\citenamefont {Rosales-Z\'{a}rate}, \citenamefont {Teh},
  \citenamefont {Kiesewetter}, \citenamefont {Brolis}, \citenamefont {Ng},\
  and\ \citenamefont {Reid}}]{zt15}%
  \BibitemOpen
  \bibfield  {author} {\bibinfo {author} {\bibfnamefont {L.}~\bibnamefont
  {Rosales-Z\'{a}rate}}, \bibinfo {author} {\bibfnamefont {R.~Y.}\ \bibnamefont
  {Teh}}, \bibinfo {author} {\bibfnamefont {S.}~\bibnamefont {Kiesewetter}},
  \bibinfo {author} {\bibfnamefont {A.}~\bibnamefont {Brolis}}, \bibinfo
  {author} {\bibfnamefont {K.}~\bibnamefont {Ng}},\ and\ \bibinfo {author}
  {\bibfnamefont {M.~D.}\ \bibnamefont {Reid}},\ }\bibfield  {title} {\bibinfo
  {title} {Decoherence of einstein-podolsky-rosen steering},\ }\href
  {https://doi.org/10.1364/JOSAB.32.000A82} {\bibfield  {journal} {\bibinfo
  {journal} {J. Opt. Soc. Am. B}\ }\textbf {\bibinfo {volume} {32}},\ \bibinfo
  {pages} {A82} (\bibinfo {year} {2015})}\BibitemShut {NoStop}%
\bibitem [{\citenamefont {Midgley}\ \emph {et~al.}(2010)\citenamefont
  {Midgley}, \citenamefont {Ferris},\ and\ \citenamefont {Olsen}}]{mf10}%
  \BibitemOpen
  \bibfield  {author} {\bibinfo {author} {\bibfnamefont {S.~L.~W.}\
  \bibnamefont {Midgley}}, \bibinfo {author} {\bibfnamefont {A.~J.}\
  \bibnamefont {Ferris}},\ and\ \bibinfo {author} {\bibfnamefont {M.~K.}\
  \bibnamefont {Olsen}},\ }\bibfield  {title} {\bibinfo {title} {Asymmetric
  gaussian steering: When alice and bob disagree},\ }\href
  {https://doi.org/10.1103/PhysRevA.81.022101} {\bibfield  {journal} {\bibinfo
  {journal} {Phys. Rev. A}\ }\textbf {\bibinfo {volume} {81}},\ \bibinfo
  {pages} {022101} (\bibinfo {year} {2010})}\BibitemShut {NoStop}%
\bibitem [{\citenamefont {Olsen}(2017)}]{o17}%
  \BibitemOpen
  \bibfield  {author} {\bibinfo {author} {\bibfnamefont {M.~K.}\ \bibnamefont
  {Olsen}},\ }\bibfield  {title} {\bibinfo {title} {Controlled asymmetry of
  einstein-podolsky-rosen steering with an injected nondegenerate optical
  parametric oscillator},\ }\href@noop {} {\bibfield  {journal} {\bibinfo
  {journal} {Phys. Rev. Lett.}\ }\textbf {\bibinfo {volume} {119}},\ \bibinfo
  {pages} {160501} (\bibinfo {year} {2017})}\BibitemShut {NoStop}%
\bibitem [{\citenamefont {Olsen}(2013)}]{mo13}%
  \BibitemOpen
  \bibfield  {author} {\bibinfo {author} {\bibfnamefont {M.~K.}\ \bibnamefont
  {Olsen}},\ }\bibfield  {title} {\bibinfo {title} {Asymmetric gaussian
  harmonic steering in second-harmonic generation},\ }\href@noop {} {\bibfield
  {journal} {\bibinfo  {journal} {Phys. Rev. A}\ }\textbf {\bibinfo {volume}
  {88}},\ \bibinfo {pages} {051802} (\bibinfo {year} {2013})}\BibitemShut
  {NoStop}%
\bibitem [{\citenamefont {Bowles}\ \emph {et~al.}(2014)\citenamefont {Bowles},
  \citenamefont {Vertesi}, \citenamefont {Quintino},\ and\ \citenamefont
  {Brunner}}]{bv14}%
  \BibitemOpen
  \bibfield  {author} {\bibinfo {author} {\bibfnamefont {J.}~\bibnamefont
  {Bowles}}, \bibinfo {author} {\bibfnamefont {T.}~\bibnamefont {Vertesi}},
  \bibinfo {author} {\bibfnamefont {M.~T.}\ \bibnamefont {Quintino}},\ and\
  \bibinfo {author} {\bibfnamefont {N.}~\bibnamefont {Brunner}},\ }\bibfield
  {title} {\bibinfo {title} {One-way einstein-podolsky-rosen steering},\
  }\href@noop {} {\bibfield  {journal} {\bibinfo  {journal} {Phys. Rev. Lett.}\
  }\textbf {\bibinfo {volume} {112}},\ \bibinfo {pages} {200402} (\bibinfo
  {year} {2014})}\BibitemShut {NoStop}%
\bibitem [{\citenamefont {Bowles}\ \emph {et~al.}(2016)\citenamefont {Bowles},
  \citenamefont {Hirsch}, \citenamefont {Quintino},\ and\ \citenamefont
  {Brunner}}]{bh16}%
  \BibitemOpen
  \bibfield  {author} {\bibinfo {author} {\bibfnamefont {J.}~\bibnamefont
  {Bowles}}, \bibinfo {author} {\bibfnamefont {F.}~\bibnamefont {Hirsch}},
  \bibinfo {author} {\bibfnamefont {M.}~\bibnamefont {Quintino}},\ and\
  \bibinfo {author} {\bibfnamefont {N.}~\bibnamefont {Brunner}},\ }\bibfield
  {title} {\bibinfo {title} {Sufficient criterion for guaranteeing that a
  two-qubit state is unsteerable},\ }\href
  {https://doi.org/10.1103/PhysRevA.93.022121} {\bibfield  {journal} {\bibinfo
  {journal} {Phys. Rev. A}\ }\textbf {\bibinfo {volume} {93}},\ \bibinfo
  {pages} {022121} (\bibinfo {year} {2016})}\BibitemShut {NoStop}%
\bibitem [{\citenamefont {Baker}\ \emph {et~al.}(2018)\citenamefont {Baker},
  \citenamefont {Wollmann}, \citenamefont {Pryde},\ and\ \citenamefont
  {Wiseman}}]{bs18}%
  \BibitemOpen
  \bibfield  {author} {\bibinfo {author} {\bibfnamefont {T.~J.}\ \bibnamefont
  {Baker}}, \bibinfo {author} {\bibfnamefont {S.}~\bibnamefont {Wollmann}},
  \bibinfo {author} {\bibfnamefont {G.~J.}\ \bibnamefont {Pryde}},\ and\
  \bibinfo {author} {\bibfnamefont {H.~M.}\ \bibnamefont {Wiseman}},\
  }\bibfield  {title} {\bibinfo {title} {Necessary condition for steerability
  of arbitrary two-qubit states with loss},\ }\href
  {https://doi.org/10.1088/2040-8986/aaaa3c} {\bibfield  {journal} {\bibinfo
  {journal} {Journal of Optics}\ }\textbf {\bibinfo {volume} {20}},\ \bibinfo
  {pages} {034008} (\bibinfo {year} {2018})}\BibitemShut {NoStop}%
\bibitem [{\citenamefont {He}\ and\ \citenamefont
  {Reid}(2013{\natexlab{a}})}]{hr13}%
  \BibitemOpen
  \bibfield  {author} {\bibinfo {author} {\bibfnamefont {Q.~Y.}\ \bibnamefont
  {He}}\ and\ \bibinfo {author} {\bibfnamefont {M.~D.}\ \bibnamefont {Reid}},\
  }\bibfield  {title} {\bibinfo {title} {Einstein-podolsky-rosen paradox and
  quantum steering in pulsed optomechanics},\ }\href@noop {} {\bibfield
  {journal} {\bibinfo  {journal} {Phys. Rev. A}\ }\textbf {\bibinfo {volume}
  {88}},\ \bibinfo {pages} {052121} (\bibinfo {year}
  {2013}{\natexlab{a}})}\BibitemShut {NoStop}%
\bibitem [{\citenamefont {He}\ and\ \citenamefont {Ficek}(2014)}]{hf14}%
  \BibitemOpen
  \bibfield  {author} {\bibinfo {author} {\bibfnamefont {Q.~Y.}\ \bibnamefont
  {He}}\ and\ \bibinfo {author} {\bibfnamefont {Z.}~\bibnamefont {Ficek}},\
  }\bibfield  {title} {\bibinfo {title} {Einstein-podolsky-rosen paradox and
  quantum steering in a three-mode optomechanical system},\ }\href@noop {}
  {\bibfield  {journal} {\bibinfo  {journal} {Phys. Rev. A}\ }\textbf {\bibinfo
  {volume} {89}},\ \bibinfo {pages} {022332} (\bibinfo {year}
  {2014})}\BibitemShut {NoStop}%
\bibitem [{\citenamefont {Tan}\ \emph {et~al.}(2015)\citenamefont {Tan},
  \citenamefont {Zhang},\ and\ \citenamefont {Li}}]{tz15}%
  \BibitemOpen
  \bibfield  {author} {\bibinfo {author} {\bibfnamefont {H.}~\bibnamefont
  {Tan}}, \bibinfo {author} {\bibfnamefont {X.}~\bibnamefont {Zhang}},\ and\
  \bibinfo {author} {\bibfnamefont {G.}~\bibnamefont {Li}},\ }\bibfield
  {title} {\bibinfo {title} {Steady-state one-way einstein-podolsky-rosen
  steering in optomechanical interfaces},\ }\href@noop {} {\bibfield  {journal}
  {\bibinfo  {journal} {Phys. Rev. A}\ }\textbf {\bibinfo {volume} {91}},\
  \bibinfo {pages} {032121} (\bibinfo {year} {2015})}\BibitemShut {NoStop}%
\bibitem [{\citenamefont {Zheng}\ \emph {et~al.}(2019)\citenamefont {Zheng},
  \citenamefont {Sun}, \citenamefont {Lai}, \citenamefont {Gong},\ and\
  \citenamefont {He}}]{zs19}%
  \BibitemOpen
  \bibfield  {author} {\bibinfo {author} {\bibfnamefont {S.~S.}\ \bibnamefont
  {Zheng}}, \bibinfo {author} {\bibfnamefont {F.~X.}\ \bibnamefont {Sun}},
  \bibinfo {author} {\bibfnamefont {Y.~J.}\ \bibnamefont {Lai}}, \bibinfo
  {author} {\bibfnamefont {Q.~H.}\ \bibnamefont {Gong}},\ and\ \bibinfo
  {author} {\bibfnamefont {Q.~Y.}\ \bibnamefont {He}},\ }\bibfield  {title}
  {\bibinfo {title} {Manipulation and enhancement of asymmetric steering via
  interference effects induced by closed-loop coupling},\ }\href
  {https://doi.org/10.1103/PhysRevA.99.022335} {\bibfield  {journal} {\bibinfo
  {journal} {Phys. Rev. A}\ }\textbf {\bibinfo {volume} {99}},\ \bibinfo
  {pages} {022335} (\bibinfo {year} {2019})}\BibitemShut {NoStop}%
\bibitem [{\citenamefont {Wu}\ \emph {et~al.}(2020)\citenamefont {Wu},
  \citenamefont {Cheng},\ and\ \citenamefont {Chen}}]{wc20}%
  \BibitemOpen
  \bibfield  {author} {\bibinfo {author} {\bibfnamefont {K.}~\bibnamefont
  {Wu}}, \bibinfo {author} {\bibfnamefont {G.}~\bibnamefont {Cheng}},\ and\
  \bibinfo {author} {\bibfnamefont {A.}~\bibnamefont {Chen}},\ }\bibfield
  {title} {\bibinfo {title} {Tunable asymmetric einstein-podolsky-rosen
  steering of microwave photons in superconducting circuits},\ }\href@noop {}
  {\bibfield  {journal} {\bibinfo  {journal} {J. Opt. Soc. Am. B}\ }\textbf
  {\bibinfo {volume} {37}},\ \bibinfo {pages} {337} (\bibinfo {year}
  {2020})}\BibitemShut {NoStop}%
\bibitem [{\citenamefont {H\"{a}ndchen}\ \emph {et~al.}(2012)\citenamefont
  {H\"{a}ndchen}, \citenamefont {Eberle}, \citenamefont {Steinlechner},
  \citenamefont {Samblowski}, \citenamefont {Franz}, \citenamefont {Werner},\
  and\ \citenamefont {Schnabel}}]{he12}%
  \BibitemOpen
  \bibfield  {author} {\bibinfo {author} {\bibfnamefont {V.}~\bibnamefont
  {H\"{a}ndchen}}, \bibinfo {author} {\bibfnamefont {T.}~\bibnamefont
  {Eberle}}, \bibinfo {author} {\bibfnamefont {S.}~\bibnamefont
  {Steinlechner}}, \bibinfo {author} {\bibfnamefont {A.}~\bibnamefont
  {Samblowski}}, \bibinfo {author} {\bibfnamefont {T.}~\bibnamefont {Franz}},
  \bibinfo {author} {\bibfnamefont {R.~F.}\ \bibnamefont {Werner}},\ and\
  \bibinfo {author} {\bibfnamefont {R.}~\bibnamefont {Schnabel}},\ }\bibfield
  {title} {\bibinfo {title} {Observation of one-way einstein-podolsky-rosen
  steering},\ }\href@noop {} {\bibfield  {journal} {\bibinfo  {journal} {Nat.
  Photonics}\ }\textbf {\bibinfo {volume} {6}},\ \bibinfo {pages} {596}
  (\bibinfo {year} {2012})}\BibitemShut {NoStop}%
\bibitem [{\citenamefont {Deng}\ \emph {et~al.}(2017)\citenamefont {Deng},
  \citenamefont {Xiang}, \citenamefont {Tian}, \citenamefont {Adesso},
  \citenamefont {He}, \citenamefont {Gong}, \citenamefont {Su}, \citenamefont
  {Xie},\ and\ \citenamefont {Peng}}]{dx17}%
  \BibitemOpen
  \bibfield  {author} {\bibinfo {author} {\bibfnamefont {X.}~\bibnamefont
  {Deng}}, \bibinfo {author} {\bibfnamefont {Y.}~\bibnamefont {Xiang}},
  \bibinfo {author} {\bibfnamefont {C.}~\bibnamefont {Tian}}, \bibinfo {author}
  {\bibfnamefont {G.}~\bibnamefont {Adesso}}, \bibinfo {author} {\bibfnamefont
  {Q.~Y.}\ \bibnamefont {He}}, \bibinfo {author} {\bibfnamefont
  {Q.}~\bibnamefont {Gong}}, \bibinfo {author} {\bibfnamefont {X.}~\bibnamefont
  {Su}}, \bibinfo {author} {\bibfnamefont {C.}~\bibnamefont {Xie}},\ and\
  \bibinfo {author} {\bibfnamefont {K.}~\bibnamefont {Peng}},\ }\bibfield
  {title} {\bibinfo {title} {Demonstration of monogamy relations for
  einstein-podolsky-rosen steering in gaussian cluster states},\ }\href@noop {}
  {\bibfield  {journal} {\bibinfo  {journal} {Phys. Rev. Lett.}\ }\textbf
  {\bibinfo {volume} {118}},\ \bibinfo {pages} {230501} (\bibinfo {year}
  {2017})}\BibitemShut {NoStop}%
\bibitem [{\citenamefont {Wang}\ \emph {et~al.}(2020)\citenamefont {Wang},
  \citenamefont {Xiang}, \citenamefont {Kang}, \citenamefont {Han},
  \citenamefont {Liu}, \citenamefont {He}, \citenamefont {Gong}, \citenamefont
  {Su},\ and\ \citenamefont {Peng}}]{wx20}%
  \BibitemOpen
  \bibfield  {author} {\bibinfo {author} {\bibfnamefont {M.}~\bibnamefont
  {Wang}}, \bibinfo {author} {\bibfnamefont {Y.}~\bibnamefont {Xiang}},
  \bibinfo {author} {\bibfnamefont {H.}~\bibnamefont {Kang}}, \bibinfo {author}
  {\bibfnamefont {D.}~\bibnamefont {Han}}, \bibinfo {author} {\bibfnamefont
  {Y.}~\bibnamefont {Liu}}, \bibinfo {author} {\bibfnamefont {Q.~Y.}\
  \bibnamefont {He}}, \bibinfo {author} {\bibfnamefont {Q.}~\bibnamefont
  {Gong}}, \bibinfo {author} {\bibfnamefont {X.}~\bibnamefont {Su}},\ and\
  \bibinfo {author} {\bibfnamefont {K.}~\bibnamefont {Peng}},\ }\bibfield
  {title} {\bibinfo {title} {Deterministic distribution of multipartite
  entanglement and steering in a quantum network by separable states},\
  }\href@noop {} {\bibfield  {journal} {\bibinfo  {journal} {Phys. Rev. Lett.}\
  }\textbf {\bibinfo {volume} {125}},\ \bibinfo {pages} {260506} (\bibinfo
  {year} {2020})}\BibitemShut {NoStop}%
\bibitem [{\citenamefont {Wollmann}\ \emph {et~al.}(2016)\citenamefont
  {Wollmann}, \citenamefont {Walk}, \citenamefont {Bennet}, \citenamefont
  {Wiseman},\ and\ \citenamefont {Pryde}}]{wn16}%
  \BibitemOpen
  \bibfield  {author} {\bibinfo {author} {\bibfnamefont {S.}~\bibnamefont
  {Wollmann}}, \bibinfo {author} {\bibfnamefont {N.}~\bibnamefont {Walk}},
  \bibinfo {author} {\bibfnamefont {A.~J.}\ \bibnamefont {Bennet}}, \bibinfo
  {author} {\bibfnamefont {H.~M.}\ \bibnamefont {Wiseman}},\ and\ \bibinfo
  {author} {\bibfnamefont {G.~J.}\ \bibnamefont {Pryde}},\ }\bibfield  {title}
  {\bibinfo {title} {Observation of genuine one-way einstein-podolsky-rosen
  steering},\ }\href@noop {} {\bibfield  {journal} {\bibinfo  {journal} {Phys.
  Rev. Lett.}\ }\textbf {\bibinfo {volume} {116}},\ \bibinfo {pages} {160403}
  (\bibinfo {year} {2016})}\BibitemShut {NoStop}%
\bibitem [{\citenamefont {Sun}\ \emph {et~al.}(2016)\citenamefont {Sun},
  \citenamefont {Ye}, \citenamefont {Xu}, \citenamefont {Xu}, \citenamefont
  {Tang}, \citenamefont {Wu}, \citenamefont {Chen}, \citenamefont {Li},\ and\
  \citenamefont {Guo}}]{sy16}%
  \BibitemOpen
  \bibfield  {author} {\bibinfo {author} {\bibfnamefont {K.}~\bibnamefont
  {Sun}}, \bibinfo {author} {\bibfnamefont {X.-J.}\ \bibnamefont {Ye}},
  \bibinfo {author} {\bibfnamefont {J.-S.}\ \bibnamefont {Xu}}, \bibinfo
  {author} {\bibfnamefont {X.-Y.}\ \bibnamefont {Xu}}, \bibinfo {author}
  {\bibfnamefont {J.-S.}\ \bibnamefont {Tang}}, \bibinfo {author}
  {\bibfnamefont {Y.-C.}\ \bibnamefont {Wu}}, \bibinfo {author} {\bibfnamefont
  {J.-L.}\ \bibnamefont {Chen}}, \bibinfo {author} {\bibfnamefont {C.-F.}\
  \bibnamefont {Li}},\ and\ \bibinfo {author} {\bibfnamefont {G.-C.}\
  \bibnamefont {Guo}},\ }\bibfield  {title} {\bibinfo {title} {Experimental
  quantification of asymmetric einstein-podolsky-rosen steering},\ }\href@noop
  {} {\bibfield  {journal} {\bibinfo  {journal} {Phys. Rev. Lett.}\ }\textbf
  {\bibinfo {volume} {116}},\ \bibinfo {pages} {160404} (\bibinfo {year}
  {2016})}\BibitemShut {NoStop}%
\bibitem [{\citenamefont {Xiao}\ \emph {et~al.}(2017)\citenamefont {Xiao},
  \citenamefont {Ye}, \citenamefont {Sun}, \citenamefont {Xu}, \citenamefont
  {Li},\ and\ \citenamefont {Guo}}]{xy17}%
  \BibitemOpen
  \bibfield  {author} {\bibinfo {author} {\bibfnamefont {Y.}~\bibnamefont
  {Xiao}}, \bibinfo {author} {\bibfnamefont {X.-J.}\ \bibnamefont {Ye}},
  \bibinfo {author} {\bibfnamefont {K.}~\bibnamefont {Sun}}, \bibinfo {author}
  {\bibfnamefont {J.-S.}\ \bibnamefont {Xu}}, \bibinfo {author} {\bibfnamefont
  {C.-F.}\ \bibnamefont {Li}},\ and\ \bibinfo {author} {\bibfnamefont {G.-C.}\
  \bibnamefont {Guo}},\ }\bibfield  {title} {\bibinfo {title} {Demonstration of
  multisetting one-way einstein-podolsky-rosen steering in two-qubit systems},\
  }\href@noop {} {\bibfield  {journal} {\bibinfo  {journal} {Phys. Rev. Lett.}\
  }\textbf {\bibinfo {volume} {118}},\ \bibinfo {pages} {140404} (\bibinfo
  {year} {2017})}\BibitemShut {NoStop}%
\bibitem [{\citenamefont {Tischler}\ \emph {et~al.}(2018)\citenamefont
  {Tischler}, \citenamefont {Ghafari}, \citenamefont {Baker}, \citenamefont
  {Slussarenko}, \citenamefont {Patel}, \citenamefont {Weston}, \citenamefont
  {Wollmann}, \citenamefont {Shalm}, \citenamefont {Verma}, \citenamefont
  {Nam}, \citenamefont {Nguyen}, \citenamefont {Wiseman},\ and\ \citenamefont
  {Pryde}}]{tg18}%
  \BibitemOpen
  \bibfield  {author} {\bibinfo {author} {\bibfnamefont {N.}~\bibnamefont
  {Tischler}}, \bibinfo {author} {\bibfnamefont {F.}~\bibnamefont {Ghafari}},
  \bibinfo {author} {\bibfnamefont {T.~J.}\ \bibnamefont {Baker}}, \bibinfo
  {author} {\bibfnamefont {S.}~\bibnamefont {Slussarenko}}, \bibinfo {author}
  {\bibfnamefont {R.~B.}\ \bibnamefont {Patel}}, \bibinfo {author}
  {\bibfnamefont {M.~M.}\ \bibnamefont {Weston}}, \bibinfo {author}
  {\bibfnamefont {S.}~\bibnamefont {Wollmann}}, \bibinfo {author}
  {\bibfnamefont {L.~K.}\ \bibnamefont {Shalm}}, \bibinfo {author}
  {\bibfnamefont {V.~B.}\ \bibnamefont {Verma}}, \bibinfo {author}
  {\bibfnamefont {S.~W.}\ \bibnamefont {Nam}}, \bibinfo {author} {\bibfnamefont
  {H.~C.}\ \bibnamefont {Nguyen}}, \bibinfo {author} {\bibfnamefont {H.~M.}\
  \bibnamefont {Wiseman}},\ and\ \bibinfo {author} {\bibfnamefont {G.~J.}\
  \bibnamefont {Pryde}},\ }\bibfield  {title} {\bibinfo {title} {Conclusive
  experimental demonstration of one-way einstein-podolsky-rosen steering},\
  }\href {https://doi.org/10.1103/PhysRevLett.121.100401} {\bibfield  {journal}
  {\bibinfo  {journal} {Phys. Rev. Lett.}\ }\textbf {\bibinfo {volume} {121}},\
  \bibinfo {pages} {100401} (\bibinfo {year} {2018})}\BibitemShut {NoStop}%
\bibitem [{\citenamefont {He}\ and\ \citenamefont
  {Reid}(2013{\natexlab{b}})}]{hr13a}%
  \BibitemOpen
  \bibfield  {author} {\bibinfo {author} {\bibfnamefont {Q.~Y.}\ \bibnamefont
  {He}}\ and\ \bibinfo {author} {\bibfnamefont {M.~D.}\ \bibnamefont {Reid}},\
  }\bibfield  {title} {\bibinfo {title} {Genuine multipartite
  einstein-podolsky-rosen steering},\ }\href
  {https://doi.org/10.1103/PhysRevLett.111.250403} {\bibfield  {journal}
  {\bibinfo  {journal} {Phys. Rev. Lett.}\ }\textbf {\bibinfo {volume} {111}},\
  \bibinfo {pages} {250403} (\bibinfo {year} {2013}{\natexlab{b}})}\BibitemShut
  {NoStop}%
\bibitem [{\citenamefont {Jones}\ \emph {et~al.}(2021)\citenamefont {Jones},
  \citenamefont {\v{S}upi\'{c}}, \citenamefont {Uola}, \citenamefont
  {Brunner},\ and\ \citenamefont {Skrzypczyk}}]{js21}%
  \BibitemOpen
  \bibfield  {author} {\bibinfo {author} {\bibfnamefont {B.~D.~M.}\
  \bibnamefont {Jones}}, \bibinfo {author} {\bibfnamefont {I.}~\bibnamefont
  {\v{S}upi\'{c}}}, \bibinfo {author} {\bibfnamefont {R.}~\bibnamefont {Uola}},
  \bibinfo {author} {\bibfnamefont {N.}~\bibnamefont {Brunner}},\ and\ \bibinfo
  {author} {\bibfnamefont {P.}~\bibnamefont {Skrzypczyk}},\ }\bibfield  {title}
  {\bibinfo {title} {Network quantum steering},\ }\href
  {https://doi.org/10.1103/PhysRevLett.127.170405} {\bibfield  {journal}
  {\bibinfo  {journal} {Phys. Rev. Lett.}\ }\textbf {\bibinfo {volume} {127}},\
  \bibinfo {pages} {170405} (\bibinfo {year} {2021})}\BibitemShut {NoStop}%
\bibitem [{\citenamefont {Cavalcanti}\ \emph
  {et~al.}(2015{\natexlab{a}})\citenamefont {Cavalcanti}, \citenamefont
  {Skrzypczyk}, \citenamefont {Aguilar}, \citenamefont {Nery}, \citenamefont
  {Ribeiro},\ and\ \citenamefont {Walborn}}]{cs15}%
  \BibitemOpen
  \bibfield  {author} {\bibinfo {author} {\bibfnamefont {D.}~\bibnamefont
  {Cavalcanti}}, \bibinfo {author} {\bibfnamefont {P.}~\bibnamefont
  {Skrzypczyk}}, \bibinfo {author} {\bibfnamefont {G.~H.}\ \bibnamefont
  {Aguilar}}, \bibinfo {author} {\bibfnamefont {R.~V.}\ \bibnamefont {Nery}},
  \bibinfo {author} {\bibfnamefont {P.~H.~S.}\ \bibnamefont {Ribeiro}},\ and\
  \bibinfo {author} {\bibfnamefont {S.~P.}\ \bibnamefont {Walborn}},\
  }\bibfield  {title} {\bibinfo {title} {Detection of entanglement in
  asymmetric quantum networks and multipartite quantum steering},\ }\href@noop
  {} {\bibfield  {journal} {\bibinfo  {journal} {Nature Comm.}\ }\textbf
  {\bibinfo {volume} {6}},\ \bibinfo {pages} {7941} (\bibinfo {year}
  {2015}{\natexlab{a}})}\BibitemShut {NoStop}%
\bibitem [{\citenamefont {Zhang}\ \emph
  {et~al.}(2019{\natexlab{a}})\citenamefont {Zhang}, \citenamefont {Cheng},
  \citenamefont {Li}, \citenamefont {Liang}, \citenamefont {Liu}, \citenamefont
  {Huang}, \citenamefont {Li}, \citenamefont {Guo}, \citenamefont {Hall},
  \citenamefont {Wiseman},\ and\ \citenamefont {Pryde}}]{zc19}%
  \BibitemOpen
  \bibfield  {author} {\bibinfo {author} {\bibfnamefont {C.}~\bibnamefont
  {Zhang}}, \bibinfo {author} {\bibfnamefont {S.~M.}\ \bibnamefont {Cheng}},
  \bibinfo {author} {\bibfnamefont {L.}~\bibnamefont {Li}}, \bibinfo {author}
  {\bibfnamefont {Q.-Y.}\ \bibnamefont {Liang}}, \bibinfo {author}
  {\bibfnamefont {B.-H.}\ \bibnamefont {Liu}}, \bibinfo {author} {\bibfnamefont
  {Y.-F.}\ \bibnamefont {Huang}}, \bibinfo {author} {\bibfnamefont {C.-F.}\
  \bibnamefont {Li}}, \bibinfo {author} {\bibfnamefont {G.-C.}\ \bibnamefont
  {Guo}}, \bibinfo {author} {\bibfnamefont {M.~J.~W.}\ \bibnamefont {Hall}},
  \bibinfo {author} {\bibfnamefont {H.~M.}\ \bibnamefont {Wiseman}},\ and\
  \bibinfo {author} {\bibfnamefont {G.~J.}\ \bibnamefont {Pryde}},\ }\bibfield
  {title} {\bibinfo {title} {Experimental validation of quantum steering
  ellipsoids and tests of volume monogamy relations},\ }\href
  {https://doi.org/10.1103/PhysRevLett.122.070402} {\bibfield  {journal}
  {\bibinfo  {journal} {Phys. Rev. Lett.}\ }\textbf {\bibinfo {volume} {122}},\
  \bibinfo {pages} {070402} (\bibinfo {year} {2019}{\natexlab{a}})}\BibitemShut
  {NoStop}%
\bibitem [{\citenamefont {Hu}\ \emph {et~al.}(2020)\citenamefont {Hu},
  \citenamefont {Xing}, \citenamefont {Liu}, \citenamefont {He}, \citenamefont
  {Cao}, \citenamefont {Guo}, \citenamefont {Zhang}, \citenamefont {Zhang},
  \citenamefont {Huang}, \citenamefont {Li},\ and\ \citenamefont {Guo}}]{hx20}%
  \BibitemOpen
  \bibfield  {author} {\bibinfo {author} {\bibfnamefont {X.-M.}\ \bibnamefont
  {Hu}}, \bibinfo {author} {\bibfnamefont {W.-B.}\ \bibnamefont {Xing}},
  \bibinfo {author} {\bibfnamefont {B.-H.}\ \bibnamefont {Liu}}, \bibinfo
  {author} {\bibfnamefont {D.-Y.}\ \bibnamefont {He}}, \bibinfo {author}
  {\bibfnamefont {H.}~\bibnamefont {Cao}}, \bibinfo {author} {\bibfnamefont
  {Y.}~\bibnamefont {Guo}}, \bibinfo {author} {\bibfnamefont {C.}~\bibnamefont
  {Zhang}}, \bibinfo {author} {\bibfnamefont {H.}~\bibnamefont {Zhang}},
  \bibinfo {author} {\bibfnamefont {Y.-F.}\ \bibnamefont {Huang}}, \bibinfo
  {author} {\bibfnamefont {C.-F.}\ \bibnamefont {Li}},\ and\ \bibinfo {author}
  {\bibfnamefont {G.-C.}\ \bibnamefont {Guo}},\ }\bibfield  {title} {\bibinfo
  {title} {Efficient distribution of high-dimensional entanglement through 11
  km fiber},\ }\href {https://doi.org/10.1364/OPTICA.388773} {\bibfield
  {journal} {\bibinfo  {journal} {Optica}\ }\textbf {\bibinfo {volume} {7}},\
  \bibinfo {pages} {738} (\bibinfo {year} {2020})}\BibitemShut {NoStop}%
\bibitem [{\citenamefont {Huang}\ \emph {et~al.}(2021)\citenamefont {Huang},
  \citenamefont {Xiang}, \citenamefont {Guo}, \citenamefont {Wu}, \citenamefont
  {Liu}, \citenamefont {Li}, \citenamefont {Guo},\ and\ \citenamefont
  {Tavakoli}}]{hx21}%
  \BibitemOpen
  \bibfield  {author} {\bibinfo {author} {\bibfnamefont {C.-J.}\ \bibnamefont
  {Huang}}, \bibinfo {author} {\bibfnamefont {G.-Y.}\ \bibnamefont {Xiang}},
  \bibinfo {author} {\bibfnamefont {Y.}~\bibnamefont {Guo}}, \bibinfo {author}
  {\bibfnamefont {K.-D.}\ \bibnamefont {Wu}}, \bibinfo {author} {\bibfnamefont
  {B.-H.}\ \bibnamefont {Liu}}, \bibinfo {author} {\bibfnamefont {C.-F.}\
  \bibnamefont {Li}}, \bibinfo {author} {\bibfnamefont {G.-C.}\ \bibnamefont
  {Guo}},\ and\ \bibinfo {author} {\bibfnamefont {A.}~\bibnamefont
  {Tavakoli}},\ }\bibfield  {title} {\bibinfo {title} {Nonlocality, steering,
  and quantum state tomography in a single experiment},\ }\href
  {https://doi.org/10.1103/PhysRevLett.127.020401} {\bibfield  {journal}
  {\bibinfo  {journal} {Phys. Rev. Lett.}\ }\textbf {\bibinfo {volume} {127}},\
  \bibinfo {pages} {020401} (\bibinfo {year} {2021})}\BibitemShut {NoStop}%
\bibitem [{\citenamefont {Hao}\ \emph {et~al.}(2022)\citenamefont {Hao},
  \citenamefont {Sun}, \citenamefont {Wang}, \citenamefont {Liu}, \citenamefont
  {Yang}, \citenamefont {Xu}, \citenamefont {Li},\ and\ \citenamefont
  {Guo}}]{hs21}%
  \BibitemOpen
  \bibfield  {author} {\bibinfo {author} {\bibfnamefont {Z.-Y.}\ \bibnamefont
  {Hao}}, \bibinfo {author} {\bibfnamefont {K.}~\bibnamefont {Sun}}, \bibinfo
  {author} {\bibfnamefont {Y.}~\bibnamefont {Wang}}, \bibinfo {author}
  {\bibfnamefont {Z.-H.}\ \bibnamefont {Liu}}, \bibinfo {author} {\bibfnamefont
  {M.}~\bibnamefont {Yang}}, \bibinfo {author} {\bibfnamefont {J.-S.}\
  \bibnamefont {Xu}}, \bibinfo {author} {\bibfnamefont {C.-F.}\ \bibnamefont
  {Li}},\ and\ \bibinfo {author} {\bibfnamefont {G.-C.}\ \bibnamefont {Guo}},\
  }\bibfield  {title} {\bibinfo {title} {Demonstrating shareability of
  multipartite einstein-podolsky-rosen steering},\ }\href
  {https://doi.org/10.1103/PhysRevLett.128.120402} {\bibfield  {journal}
  {\bibinfo  {journal} {Phys. Rev. Lett.}\ }\textbf {\bibinfo {volume} {128}},\
  \bibinfo {pages} {120402} (\bibinfo {year} {2022})}\BibitemShut {NoStop}%
\bibitem [{\citenamefont {Saunders}\ \emph {et~al.}(2010)\citenamefont
  {Saunders}, \citenamefont {Jones}, \citenamefont {Wiseman},\ and\
  \citenamefont {Pryde}}]{sj10}%
  \BibitemOpen
  \bibfield  {author} {\bibinfo {author} {\bibfnamefont {D.~J.}\ \bibnamefont
  {Saunders}}, \bibinfo {author} {\bibfnamefont {S.~J.}\ \bibnamefont {Jones}},
  \bibinfo {author} {\bibfnamefont {H.~M.}\ \bibnamefont {Wiseman}},\ and\
  \bibinfo {author} {\bibfnamefont {G.~J.}\ \bibnamefont {Pryde}},\ }\bibfield
  {title} {\bibinfo {title} {Experimental epr steering using bell-local
  states},\ }\href@noop {} {\bibfield  {journal} {\bibinfo  {journal} {Nat.
  Phys.}\ }\textbf {\bibinfo {volume} {6}},\ \bibinfo {pages} {845} (\bibinfo
  {year} {2010})}\BibitemShut {NoStop}%
\bibitem [{\citenamefont {Bennet}\ \emph {et~al.}(2012)\citenamefont {Bennet},
  \citenamefont {Evans}, \citenamefont {Saunders}, \citenamefont {Branciard},
  \citenamefont {Cavalcanti}, \citenamefont {Wiseman},\ and\ \citenamefont
  {Pryde}}]{be12}%
  \BibitemOpen
  \bibfield  {author} {\bibinfo {author} {\bibfnamefont {A.~J.}\ \bibnamefont
  {Bennet}}, \bibinfo {author} {\bibfnamefont {D.~A.}\ \bibnamefont {Evans}},
  \bibinfo {author} {\bibfnamefont {D.~J.}\ \bibnamefont {Saunders}}, \bibinfo
  {author} {\bibfnamefont {C.}~\bibnamefont {Branciard}}, \bibinfo {author}
  {\bibfnamefont {E.~G.}\ \bibnamefont {Cavalcanti}}, \bibinfo {author}
  {\bibfnamefont {H.~M.}\ \bibnamefont {Wiseman}},\ and\ \bibinfo {author}
  {\bibfnamefont {G.~J.}\ \bibnamefont {Pryde}},\ }\bibfield  {title} {\bibinfo
  {title} {Arbitrarily loss-tolerant einstein-podolsky-rosen steering allowing
  a demonstration over 1 km of optical fiber with no detection loophole},\
  }\href {https://doi.org/10.1103/PhysRevX.2.031003} {\bibfield  {journal}
  {\bibinfo  {journal} {Phys. Rev. X}\ }\textbf {\bibinfo {volume} {2}},\
  \bibinfo {pages} {031003} (\bibinfo {year} {2012})}\BibitemShut {NoStop}%
\bibitem [{\citenamefont {Sun}\ \emph {et~al.}(2014)\citenamefont {Sun},
  \citenamefont {Xu}, \citenamefont {Ye}, \citenamefont {Wu}, \citenamefont
  {Chen}, \citenamefont {Li},\ and\ \citenamefont {Guo}}]{sx14}%
  \BibitemOpen
  \bibfield  {author} {\bibinfo {author} {\bibfnamefont {K.}~\bibnamefont
  {Sun}}, \bibinfo {author} {\bibfnamefont {J.-S.}\ \bibnamefont {Xu}},
  \bibinfo {author} {\bibfnamefont {X.-J.}\ \bibnamefont {Ye}}, \bibinfo
  {author} {\bibfnamefont {Y.-C.}\ \bibnamefont {Wu}}, \bibinfo {author}
  {\bibfnamefont {J.-L.}\ \bibnamefont {Chen}}, \bibinfo {author}
  {\bibfnamefont {C.-F.}\ \bibnamefont {Li}},\ and\ \bibinfo {author}
  {\bibfnamefont {G.-C.}\ \bibnamefont {Guo}},\ }\bibfield  {title} {\bibinfo
  {title} {Experimental demonstration of the einstein-podolsky-rosen steering
  game based on the all-versus-nothing proof},\ }\href
  {https://doi.org/10.1103/PhysRevLett.113.140402} {\bibfield  {journal}
  {\bibinfo  {journal} {Phys. Rev. Lett.}\ }\textbf {\bibinfo {volume} {113}},\
  \bibinfo {pages} {140402} (\bibinfo {year} {2014})}\BibitemShut {NoStop}%
\bibitem [{\citenamefont {Kocsis}\ \emph {et~al.}(2015)\citenamefont {Kocsis},
  \citenamefont {Hall}, \citenamefont {Bennet}, \citenamefont {Saunders},\ and\
  \citenamefont {Pryde}}]{kh15}%
  \BibitemOpen
  \bibfield  {author} {\bibinfo {author} {\bibfnamefont {S.}~\bibnamefont
  {Kocsis}}, \bibinfo {author} {\bibfnamefont {M.~J.~W.}\ \bibnamefont {Hall}},
  \bibinfo {author} {\bibfnamefont {A.~J.}\ \bibnamefont {Bennet}}, \bibinfo
  {author} {\bibfnamefont {D.~J.}\ \bibnamefont {Saunders}},\ and\ \bibinfo
  {author} {\bibfnamefont {G.~J.}\ \bibnamefont {Pryde}},\ }\bibfield  {title}
  {\bibinfo {title} {Experimental measurement-device-independent verification
  of quantum steering},\ }\href@noop {} {\bibfield  {journal} {\bibinfo
  {journal} {Nat. Commun.}\ }\textbf {\bibinfo {volume} {6}},\ \bibinfo {pages}
  {5886} (\bibinfo {year} {2015})}\BibitemShut {NoStop}%
\bibitem [{\citenamefont {Wollmann}\ \emph {et~al.}(2020)\citenamefont
  {Wollmann}, \citenamefont {Uola},\ and\ \citenamefont {Costa}}]{wu20}%
  \BibitemOpen
  \bibfield  {author} {\bibinfo {author} {\bibfnamefont {S.}~\bibnamefont
  {Wollmann}}, \bibinfo {author} {\bibfnamefont {R.}~\bibnamefont {Uola}},\
  and\ \bibinfo {author} {\bibfnamefont {A.}~\bibnamefont {Costa}},\ }\bibfield
   {title} {\bibinfo {title} {Experimental demonstration of robust quantum
  steering},\ }\href@noop {} {\bibfield  {journal} {\bibinfo  {journal} {Phys.
  Rev. Lett.}\ }\textbf {\bibinfo {volume} {125}},\ \bibinfo {pages} {020404}
  (\bibinfo {year} {2020})}\BibitemShut {NoStop}%
\bibitem [{\citenamefont {Wang}\ \emph {et~al.}(2018)\citenamefont {Wang},
  \citenamefont {Paesani}, \citenamefont {Ding}, \citenamefont {Santagati},
  \citenamefont {Skrzypczyk}, \citenamefont {Salavrakos}, \citenamefont {Tura},
  \citenamefont {Augusiak}, \citenamefont {Mancinska}, \citenamefont {Bacco},
  \citenamefont {Bonneau}, \citenamefont {Silverstone}, \citenamefont {Gong},
  \citenamefont {Ac\'{i}n}, \citenamefont {Rottwitt}, \citenamefont {Oxenlowe},
  \citenamefont {O'Brien}, \citenamefont {Laing},\ and\ \citenamefont
  {Thompson}}]{wp18}%
  \BibitemOpen
  \bibfield  {author} {\bibinfo {author} {\bibfnamefont {J.}~\bibnamefont
  {Wang}}, \bibinfo {author} {\bibfnamefont {S.}~\bibnamefont {Paesani}},
  \bibinfo {author} {\bibfnamefont {Y.}~\bibnamefont {Ding}}, \bibinfo {author}
  {\bibfnamefont {R.}~\bibnamefont {Santagati}}, \bibinfo {author}
  {\bibfnamefont {P.}~\bibnamefont {Skrzypczyk}}, \bibinfo {author}
  {\bibfnamefont {A.}~\bibnamefont {Salavrakos}}, \bibinfo {author}
  {\bibfnamefont {J.}~\bibnamefont {Tura}}, \bibinfo {author} {\bibfnamefont
  {R.}~\bibnamefont {Augusiak}}, \bibinfo {author} {\bibfnamefont
  {L.}~\bibnamefont {Mancinska}}, \bibinfo {author} {\bibfnamefont
  {D.}~\bibnamefont {Bacco}}, \bibinfo {author} {\bibfnamefont
  {D.}~\bibnamefont {Bonneau}}, \bibinfo {author} {\bibfnamefont {J.~W.}\
  \bibnamefont {Silverstone}}, \bibinfo {author} {\bibfnamefont
  {Q.}~\bibnamefont {Gong}}, \bibinfo {author} {\bibfnamefont {A.}~\bibnamefont
  {Ac\'{i}n}}, \bibinfo {author} {\bibfnamefont {K.}~\bibnamefont {Rottwitt}},
  \bibinfo {author} {\bibfnamefont {L.~K.}\ \bibnamefont {Oxenlowe}}, \bibinfo
  {author} {\bibfnamefont {J.~L.}\ \bibnamefont {O'Brien}}, \bibinfo {author}
  {\bibfnamefont {A.}~\bibnamefont {Laing}},\ and\ \bibinfo {author}
  {\bibfnamefont {M.~G.}\ \bibnamefont {Thompson}},\ }\bibfield  {title}
  {\bibinfo {title} {Multidimensional quantum entanglement with large-scale
  integrated optics},\ }\href@noop {} {\bibfield  {journal} {\bibinfo
  {journal} {Science}\ }\textbf {\bibinfo {volume} {360}},\ \bibinfo {pages}
  {285} (\bibinfo {year} {2018})}\BibitemShut {NoStop}%
\bibitem [{\citenamefont {Cavaill\`es}\ \emph {et~al.}(2018)\citenamefont
  {Cavaill\`es}, \citenamefont {Le~Jeannic}, \citenamefont {Raskop},
  \citenamefont {Guccione}, \citenamefont {Markham}, \citenamefont {Diamanti},
  \citenamefont {Shaw}, \citenamefont {Verma}, \citenamefont {Nam},\ and\
  \citenamefont {Laurat}}]{cj18}%
  \BibitemOpen
  \bibfield  {author} {\bibinfo {author} {\bibfnamefont {A.}~\bibnamefont
  {Cavaill\`es}}, \bibinfo {author} {\bibfnamefont {H.}~\bibnamefont
  {Le~Jeannic}}, \bibinfo {author} {\bibfnamefont {J.}~\bibnamefont {Raskop}},
  \bibinfo {author} {\bibfnamefont {G.}~\bibnamefont {Guccione}}, \bibinfo
  {author} {\bibfnamefont {D.}~\bibnamefont {Markham}}, \bibinfo {author}
  {\bibfnamefont {E.}~\bibnamefont {Diamanti}}, \bibinfo {author}
  {\bibfnamefont {M.~D.}\ \bibnamefont {Shaw}}, \bibinfo {author}
  {\bibfnamefont {V.~B.}\ \bibnamefont {Verma}}, \bibinfo {author}
  {\bibfnamefont {S.~W.}\ \bibnamefont {Nam}},\ and\ \bibinfo {author}
  {\bibfnamefont {J.}~\bibnamefont {Laurat}},\ }\bibfield  {title} {\bibinfo
  {title} {Demonstration of einstein-podolsky-rosen steering using hybrid
  continuous- and discrete-variable entanglement of light},\ }\href
  {https://doi.org/10.1103/PhysRevLett.121.170403} {\bibfield  {journal}
  {\bibinfo  {journal} {Phys. Rev. Lett.}\ }\textbf {\bibinfo {volume} {121}},\
  \bibinfo {pages} {170403} (\bibinfo {year} {2018})}\BibitemShut {NoStop}%
\bibitem [{\citenamefont {Zeng}\ \emph {et~al.}(2018)\citenamefont {Zeng},
  \citenamefont {Wang}, \citenamefont {Li},\ and\ \citenamefont
  {Zhang}}]{zw18}%
  \BibitemOpen
  \bibfield  {author} {\bibinfo {author} {\bibfnamefont {Q.}~\bibnamefont
  {Zeng}}, \bibinfo {author} {\bibfnamefont {B.}~\bibnamefont {Wang}}, \bibinfo
  {author} {\bibfnamefont {P.}~\bibnamefont {Li}},\ and\ \bibinfo {author}
  {\bibfnamefont {X.}~\bibnamefont {Zhang}},\ }\bibfield  {title} {\bibinfo
  {title} {Experimental high-dimensional einstein-podolsky-rosen steering},\
  }\href {https://doi.org/10.1103/PhysRevLett.120.030401} {\bibfield  {journal}
  {\bibinfo  {journal} {Phys. Rev. Lett.}\ }\textbf {\bibinfo {volume} {120}},\
  \bibinfo {pages} {030401} (\bibinfo {year} {2018})}\BibitemShut {NoStop}%
\bibitem [{\citenamefont {Qu}\ \emph {et~al.}(2022{\natexlab{a}})\citenamefont
  {Qu}, \citenamefont {Wang}, \citenamefont {An}, \citenamefont {Wang},
  \citenamefont {Quan}, \citenamefont {Li}, \citenamefont {Gao}, \citenamefont
  {Li},\ and\ \citenamefont {Zhang}}]{qw21}%
  \BibitemOpen
  \bibfield  {author} {\bibinfo {author} {\bibfnamefont {R.}~\bibnamefont
  {Qu}}, \bibinfo {author} {\bibfnamefont {Y.}~\bibnamefont {Wang}}, \bibinfo
  {author} {\bibfnamefont {M.}~\bibnamefont {An}}, \bibinfo {author}
  {\bibfnamefont {F.}~\bibnamefont {Wang}}, \bibinfo {author} {\bibfnamefont
  {Q.}~\bibnamefont {Quan}}, \bibinfo {author} {\bibfnamefont {H.}~\bibnamefont
  {Li}}, \bibinfo {author} {\bibfnamefont {H.}~\bibnamefont {Gao}}, \bibinfo
  {author} {\bibfnamefont {F.}~\bibnamefont {Li}},\ and\ \bibinfo {author}
  {\bibfnamefont {P.}~\bibnamefont {Zhang}},\ }\bibfield  {title} {\bibinfo
  {title} {Retrieving high-dimensional quantum steering from a noisy
  environment with $n$ measurement settings},\ }\href
  {https://doi.org/10.1103/PhysRevLett.128.240402} {\bibfield  {journal}
  {\bibinfo  {journal} {Phys. Rev. Lett.}\ }\textbf {\bibinfo {volume} {128}},\
  \bibinfo {pages} {240402} (\bibinfo {year} {2022}{\natexlab{a}})}\BibitemShut
  {NoStop}%
\bibitem [{\citenamefont {Qu}\ \emph {et~al.}(2022{\natexlab{b}})\citenamefont
  {Qu}, \citenamefont {Wang}, \citenamefont {Zhang}, \citenamefont {Ru},
  \citenamefont {Wang}, \citenamefont {Gao}, \citenamefont {Li},\ and\
  \citenamefont {Zhang}}]{qw22}%
  \BibitemOpen
  \bibfield  {author} {\bibinfo {author} {\bibfnamefont {R.}~\bibnamefont
  {Qu}}, \bibinfo {author} {\bibfnamefont {Y.}~\bibnamefont {Wang}}, \bibinfo
  {author} {\bibfnamefont {X.}~\bibnamefont {Zhang}}, \bibinfo {author}
  {\bibfnamefont {S.}~\bibnamefont {Ru}}, \bibinfo {author} {\bibfnamefont
  {F.}~\bibnamefont {Wang}}, \bibinfo {author} {\bibfnamefont {H.}~\bibnamefont
  {Gao}}, \bibinfo {author} {\bibfnamefont {F.}~\bibnamefont {Li}},\ and\
  \bibinfo {author} {\bibfnamefont {P.}~\bibnamefont {Zhang}},\ }\bibfield
  {title} {\bibinfo {title} {Robust method for certifying genuine
  high-dimensional quantum steering with multimeasurement settings},\ }\href
  {https://doi.org/10.1364/OPTICA.454597} {\bibfield  {journal} {\bibinfo
  {journal} {Optica}\ }\textbf {\bibinfo {volume} {9}},\ \bibinfo {pages} {473}
  (\bibinfo {year} {2022}{\natexlab{b}})}\BibitemShut {NoStop}%
\bibitem [{\citenamefont {Zeng}\ \emph {et~al.}(2022)\citenamefont {Zeng},
  \citenamefont {Shang}, \citenamefont {Nguyen},\ and\ \citenamefont
  {Zhang}}]{zs20}%
  \BibitemOpen
  \bibfield  {author} {\bibinfo {author} {\bibfnamefont {Q.}~\bibnamefont
  {Zeng}}, \bibinfo {author} {\bibfnamefont {J.}~\bibnamefont {Shang}},
  \bibinfo {author} {\bibfnamefont {H.~C.}\ \bibnamefont {Nguyen}},\ and\
  \bibinfo {author} {\bibfnamefont {X.}~\bibnamefont {Zhang}},\ }\bibfield
  {title} {\bibinfo {title} {Reliable experimental certification of one-way
  einstein-podolsky-rosen steering},\ }\href
  {https://doi.org/10.1103/PhysRevResearch.4.013151} {\bibfield  {journal}
  {\bibinfo  {journal} {Phys. Rev. Research}\ }\textbf {\bibinfo {volume}
  {4}},\ \bibinfo {pages} {013151} (\bibinfo {year} {2022})}\BibitemShut
  {NoStop}%
\bibitem [{\citenamefont {Walborn}\ \emph {et~al.}(2011)\citenamefont
  {Walborn}, \citenamefont {Salles}, \citenamefont {Gomes}, \citenamefont
  {Toscano},\ and\ \citenamefont {Ribeiro}}]{ws11}%
  \BibitemOpen
  \bibfield  {author} {\bibinfo {author} {\bibfnamefont {S.~P.}\ \bibnamefont
  {Walborn}}, \bibinfo {author} {\bibfnamefont {A.}~\bibnamefont {Salles}},
  \bibinfo {author} {\bibfnamefont {R.~M.}\ \bibnamefont {Gomes}}, \bibinfo
  {author} {\bibfnamefont {F.}~\bibnamefont {Toscano}},\ and\ \bibinfo {author}
  {\bibfnamefont {P.~H.~S.}\ \bibnamefont {Ribeiro}},\ }\bibfield  {title}
  {\bibinfo {title} {Revealing hidden einstein-podolsky-rosen nonlocality},\
  }\href@noop {} {\bibfield  {journal} {\bibinfo  {journal} {Phys. Rev. Lett.}\
  }\textbf {\bibinfo {volume} {106}},\ \bibinfo {pages} {130402} (\bibinfo
  {year} {2011})}\BibitemShut {NoStop}%
\bibitem [{\citenamefont {Cai}\ \emph {et~al.}(2020)\citenamefont {Cai},
  \citenamefont {Xiang}, \citenamefont {Liu}, \citenamefont {He},\ and\
  \citenamefont {Treps}}]{cx20}%
  \BibitemOpen
  \bibfield  {author} {\bibinfo {author} {\bibfnamefont {Y.}~\bibnamefont
  {Cai}}, \bibinfo {author} {\bibfnamefont {Y.}~\bibnamefont {Xiang}}, \bibinfo
  {author} {\bibfnamefont {Y.}~\bibnamefont {Liu}}, \bibinfo {author}
  {\bibfnamefont {Q.~Y.}\ \bibnamefont {He}},\ and\ \bibinfo {author}
  {\bibfnamefont {N.}~\bibnamefont {Treps}},\ }\bibfield  {title} {\bibinfo
  {title} {Versatile multipartite einstein-podolsky-rosen steering via a
  quantum frequency comb},\ }\href@noop {} {\bibfield  {journal} {\bibinfo
  {journal} {Phys. Rev. Research}\ }\textbf {\bibinfo {volume} {2}},\ \bibinfo
  {pages} {032046(R)} (\bibinfo {year} {2020})}\BibitemShut {NoStop}%
\bibitem [{\citenamefont {Designolle}\ \emph {et~al.}(2021)\citenamefont
  {Designolle}, \citenamefont {Srivastav}, \citenamefont {Uola}, \citenamefont
  {Valencia}, \citenamefont {McCutcheon}, \citenamefont {Malik},\ and\
  \citenamefont {Brunner}}]{ds21}%
  \BibitemOpen
  \bibfield  {author} {\bibinfo {author} {\bibfnamefont {S.}~\bibnamefont
  {Designolle}}, \bibinfo {author} {\bibfnamefont {V.}~\bibnamefont
  {Srivastav}}, \bibinfo {author} {\bibfnamefont {R.}~\bibnamefont {Uola}},
  \bibinfo {author} {\bibfnamefont {N.~H.}\ \bibnamefont {Valencia}}, \bibinfo
  {author} {\bibfnamefont {W.}~\bibnamefont {McCutcheon}}, \bibinfo {author}
  {\bibfnamefont {M.}~\bibnamefont {Malik}},\ and\ \bibinfo {author}
  {\bibfnamefont {N.}~\bibnamefont {Brunner}},\ }\bibfield  {title} {\bibinfo
  {title} {Genuine high-dimensional quantum steering},\ }\href
  {https://doi.org/10.1103/PhysRevLett.126.200404} {\bibfield  {journal}
  {\bibinfo  {journal} {Phys. Rev. Lett.}\ }\textbf {\bibinfo {volume} {126}},\
  \bibinfo {pages} {200404} (\bibinfo {year} {2021})}\BibitemShut {NoStop}%
\bibitem [{\citenamefont {Peise}\ \emph {et~al.}(2015)\citenamefont {Peise},
  \citenamefont {Kruse}, \citenamefont {Lange}, \citenamefont {L\"ucke},
  \citenamefont {Pezze}, \citenamefont {Arlt}, \citenamefont {Ertmer},
  \citenamefont {Hammerer}, \citenamefont {Santos}, \citenamefont {Smerzi},\
  and\ \citenamefont {Klempt}}]{pk15}%
  \BibitemOpen
  \bibfield  {author} {\bibinfo {author} {\bibfnamefont {J.}~\bibnamefont
  {Peise}}, \bibinfo {author} {\bibfnamefont {I.}~\bibnamefont {Kruse}},
  \bibinfo {author} {\bibfnamefont {K.}~\bibnamefont {Lange}}, \bibinfo
  {author} {\bibfnamefont {B.}~\bibnamefont {L\"ucke}}, \bibinfo {author}
  {\bibfnamefont {L.}~\bibnamefont {Pezze}}, \bibinfo {author} {\bibfnamefont
  {J.}~\bibnamefont {Arlt}}, \bibinfo {author} {\bibfnamefont {W.}~\bibnamefont
  {Ertmer}}, \bibinfo {author} {\bibfnamefont {K.}~\bibnamefont {Hammerer}},
  \bibinfo {author} {\bibfnamefont {L.}~\bibnamefont {Santos}}, \bibinfo
  {author} {\bibfnamefont {A.}~\bibnamefont {Smerzi}},\ and\ \bibinfo {author}
  {\bibfnamefont {C.}~\bibnamefont {Klempt}},\ }\bibfield  {title} {\bibinfo
  {title} {Satisfying the einstein--podolsky--rosen criterion with massive
  particles},\ }\href@noop {} {\bibfield  {journal} {\bibinfo  {journal} {Nat.
  Commun.}\ }\textbf {\bibinfo {volume} {6}},\ \bibinfo {pages} {8984}
  (\bibinfo {year} {2015})}\BibitemShut {NoStop}%
\bibitem [{\citenamefont {Fadel}\ \emph {et~al.}(2018)\citenamefont {Fadel},
  \citenamefont {Zibold}, \citenamefont {Decamps},\ and\ \citenamefont
  {Treutlein}}]{fz18}%
  \BibitemOpen
  \bibfield  {author} {\bibinfo {author} {\bibfnamefont {M.}~\bibnamefont
  {Fadel}}, \bibinfo {author} {\bibfnamefont {T.}~\bibnamefont {Zibold}},
  \bibinfo {author} {\bibfnamefont {B.}~\bibnamefont {Decamps}},\ and\ \bibinfo
  {author} {\bibfnamefont {P.}~\bibnamefont {Treutlein}},\ }\bibfield  {title}
  {\bibinfo {title} {Spatial entanglement patterns and einstein-podolsky-rosen
  steering in bose-einstein condensates},\ }\href@noop {} {\bibfield  {journal}
  {\bibinfo  {journal} {Science}\ }\textbf {\bibinfo {volume} {360}},\ \bibinfo
  {pages} {409} (\bibinfo {year} {2018})}\BibitemShut {NoStop}%
\bibitem [{\citenamefont {Kunkel}\ \emph {et~al.}(2018)\citenamefont {Kunkel},
  \citenamefont {Pr\"ufer}, \citenamefont {Strobel}, \citenamefont {Linnemann},
  \citenamefont {Fr\"olian}, \citenamefont {Gasenzer}, \citenamefont
  {G\"arttner},\ and\ \citenamefont {Oberthaler}}]{kp18}%
  \BibitemOpen
  \bibfield  {author} {\bibinfo {author} {\bibfnamefont {P.}~\bibnamefont
  {Kunkel}}, \bibinfo {author} {\bibfnamefont {M.}~\bibnamefont {Pr\"ufer}},
  \bibinfo {author} {\bibfnamefont {H.}~\bibnamefont {Strobel}}, \bibinfo
  {author} {\bibfnamefont {D.}~\bibnamefont {Linnemann}}, \bibinfo {author}
  {\bibfnamefont {A.}~\bibnamefont {Fr\"olian}}, \bibinfo {author}
  {\bibfnamefont {T.}~\bibnamefont {Gasenzer}}, \bibinfo {author}
  {\bibfnamefont {M.}~\bibnamefont {G\"arttner}},\ and\ \bibinfo {author}
  {\bibfnamefont {M.~K.}\ \bibnamefont {Oberthaler}},\ }\bibfield  {title}
  {\bibinfo {title} {Spatially distributed multipartite entanglement enables
  epr steering of atomic clouds},\ }\href@noop {} {\bibfield  {journal}
  {\bibinfo  {journal} {Science}\ }\textbf {\bibinfo {volume} {360}},\ \bibinfo
  {pages} {413} (\bibinfo {year} {2018})}\BibitemShut {NoStop}%
\bibitem [{\citenamefont {Bloch}(2008)}]{ib08}%
  \BibitemOpen
  \bibfield  {author} {\bibinfo {author} {\bibfnamefont {I.}~\bibnamefont
  {Bloch}},\ }\bibfield  {title} {\bibinfo {title} {Quantum coherence and
  entanglement with ultracold atoms in optical lattices},\ }\href
  {https://doi.org/https://doi.org/10.1038/nature07126} {\bibfield  {journal}
  {\bibinfo  {journal} {Nature}\ }\textbf {\bibinfo {volume} {453}},\ \bibinfo
  {pages} {1016} (\bibinfo {year} {2008})}\BibitemShut {NoStop}%
\bibitem [{\citenamefont {Dai}\ \emph {et~al.}(2016)\citenamefont {Dai},
  \citenamefont {Yang}, \citenamefont {Reingruber}, \citenamefont {Xu},
  \citenamefont {Jiang}, \citenamefont {Chen}, \citenamefont {Yuan},\ and\
  \citenamefont {Pan}}]{dy16}%
  \BibitemOpen
  \bibfield  {author} {\bibinfo {author} {\bibfnamefont {H.-N.}\ \bibnamefont
  {Dai}}, \bibinfo {author} {\bibfnamefont {B.}~\bibnamefont {Yang}}, \bibinfo
  {author} {\bibfnamefont {A.}~\bibnamefont {Reingruber}}, \bibinfo {author}
  {\bibfnamefont {X.-F.}\ \bibnamefont {Xu}}, \bibinfo {author} {\bibfnamefont
  {X.}~\bibnamefont {Jiang}}, \bibinfo {author} {\bibfnamefont {Y.-A.}\
  \bibnamefont {Chen}}, \bibinfo {author} {\bibfnamefont {Z.-S.}\ \bibnamefont
  {Yuan}},\ and\ \bibinfo {author} {\bibfnamefont {J.-W.}\ \bibnamefont
  {Pan}},\ }\bibfield  {title} {\bibinfo {title} {Generation and detection of
  atomic spin entanglement in optical lattices},\ }\href
  {https://doi.org/https://doi.org/10.1038/nphys3705} {\bibfield  {journal}
  {\bibinfo  {journal} {Nat. Phys.}\ }\textbf {\bibinfo {volume} {12}},\
  \bibinfo {pages} {783} (\bibinfo {year} {2016})}\BibitemShut {NoStop}%
\bibitem [{\citenamefont {Jevtic}\ \emph {et~al.}(2015)\citenamefont {Jevtic},
  \citenamefont {Hall}, \citenamefont {Anderson}, \citenamefont {Zwierz},\ and\
  \citenamefont {Wiseman}}]{jm15}%
  \BibitemOpen
  \bibfield  {author} {\bibinfo {author} {\bibfnamefont {S.}~\bibnamefont
  {Jevtic}}, \bibinfo {author} {\bibfnamefont {M.~J.~W.}\ \bibnamefont {Hall}},
  \bibinfo {author} {\bibfnamefont {M.~R.}\ \bibnamefont {Anderson}}, \bibinfo
  {author} {\bibfnamefont {M.}~\bibnamefont {Zwierz}},\ and\ \bibinfo {author}
  {\bibfnamefont {H.~M.}\ \bibnamefont {Wiseman}},\ }\bibfield  {title}
  {\bibinfo {title} {Einstein--podolsky--rosen steering and the steering
  ellipsoid},\ }\href {https://doi.org/10.1364/JOSAB.32.000A40} {\bibfield
  {journal} {\bibinfo  {journal} {J. Opt. Soc. Am. B}\ }\textbf {\bibinfo
  {volume} {32}},\ \bibinfo {pages} {A40} (\bibinfo {year} {2015})}\BibitemShut
  {NoStop}%
\bibitem [{\citenamefont {Nguyen}\ and\ \citenamefont {Vu}(2016)}]{nt16}%
  \BibitemOpen
  \bibfield  {author} {\bibinfo {author} {\bibfnamefont {H.~C.}\ \bibnamefont
  {Nguyen}}\ and\ \bibinfo {author} {\bibfnamefont {T.}~\bibnamefont {Vu}},\
  }\bibfield  {title} {\bibinfo {title} {Necessary and sufficient condition for
  steerability of two-qubit states by the geometry of steering outcomes},\
  }\href {https://doi.org/10.1209/0295-5075/115/10003} {\bibfield  {journal}
  {\bibinfo  {journal} {{EPL} (Europhysics Letters)}\ }\textbf {\bibinfo
  {volume} {115}},\ \bibinfo {pages} {10003} (\bibinfo {year}
  {2016})}\BibitemShut {NoStop}%
\bibitem [{\citenamefont {Nguyen}\ \emph {et~al.}(2018)\citenamefont {Nguyen},
  \citenamefont {Milne}, \citenamefont {Vu},\ and\ \citenamefont
  {Jevtic}}]{na18}%
  \BibitemOpen
  \bibfield  {author} {\bibinfo {author} {\bibfnamefont {H.~C.}\ \bibnamefont
  {Nguyen}}, \bibinfo {author} {\bibfnamefont {A.}~\bibnamefont {Milne}},
  \bibinfo {author} {\bibfnamefont {T.}~\bibnamefont {Vu}},\ and\ \bibinfo
  {author} {\bibfnamefont {S.}~\bibnamefont {Jevtic}},\ }\bibfield  {title}
  {\bibinfo {title} {Quantum steering with positive operator valued measures},\
  }\href {https://doi.org/10.1088/1751-8121/aad115} {\bibfield  {journal}
  {\bibinfo  {journal} {J. Phys. A: Math. Theor.}\ }\textbf {\bibinfo {volume}
  {51}},\ \bibinfo {pages} {355302} (\bibinfo {year} {2018})}\BibitemShut
  {NoStop}%
\bibitem [{\citenamefont {Pusey}(2013)}]{m13}%
  \BibitemOpen
  \bibfield  {author} {\bibinfo {author} {\bibfnamefont {M.~F.}\ \bibnamefont
  {Pusey}},\ }\bibfield  {title} {\bibinfo {title} {Negativity and steering: A
  stronger peres conjecture},\ }\href
  {https://doi.org/10.1103/PhysRevA.88.032313} {\bibfield  {journal} {\bibinfo
  {journal} {Phys. Rev. A}\ }\textbf {\bibinfo {volume} {88}},\ \bibinfo
  {pages} {032313} (\bibinfo {year} {2013})}\BibitemShut {NoStop}%
\bibitem [{\citenamefont {Nguyen}\ \emph {et~al.}(2019)\citenamefont {Nguyen},
  \citenamefont {Nguyen},\ and\ \citenamefont {G\"uhne}}]{nn19}%
  \BibitemOpen
  \bibfield  {author} {\bibinfo {author} {\bibfnamefont {H.~C.}\ \bibnamefont
  {Nguyen}}, \bibinfo {author} {\bibfnamefont {H.~V.}\ \bibnamefont {Nguyen}},\
  and\ \bibinfo {author} {\bibfnamefont {O.}~\bibnamefont {G\"uhne}},\
  }\bibfield  {title} {\bibinfo {title} {Geometry of einstein-podolsky-rosen
  correlations},\ }\href {https://doi.org/10.1103/PhysRevLett.122.240401}
  {\bibfield  {journal} {\bibinfo  {journal} {Phys. Rev. Lett.}\ }\textbf
  {\bibinfo {volume} {122}},\ \bibinfo {pages} {240401} (\bibinfo {year}
  {2019})}\BibitemShut {NoStop}%
\bibitem [{\citenamefont {Bell}(1964)}]{jb64}%
  \BibitemOpen
  \bibfield  {author} {\bibinfo {author} {\bibfnamefont {J.~S.}\ \bibnamefont
  {Bell}},\ }\bibfield  {title} {\bibinfo {title} {On the einstein podolsky
  rosen paradox},\ }\href {https://doi.org/10.1103/PhysicsPhysiqueFizika.1.195}
  {\bibfield  {journal} {\bibinfo  {journal} {Physics Physique Fizika}\
  }\textbf {\bibinfo {volume} {1}},\ \bibinfo {pages} {195} (\bibinfo {year}
  {1964})}\BibitemShut {NoStop}%
\bibitem [{\citenamefont {Cavalcanti}\ \emph {et~al.}(2009)\citenamefont
  {Cavalcanti}, \citenamefont {Jones}, \citenamefont {Wiseman},\ and\
  \citenamefont {Reid}}]{cj09}%
  \BibitemOpen
  \bibfield  {author} {\bibinfo {author} {\bibfnamefont {E.~G.}\ \bibnamefont
  {Cavalcanti}}, \bibinfo {author} {\bibfnamefont {S.~J.}\ \bibnamefont
  {Jones}}, \bibinfo {author} {\bibfnamefont {H.~M.}\ \bibnamefont {Wiseman}},\
  and\ \bibinfo {author} {\bibfnamefont {M.~D.}\ \bibnamefont {Reid}},\
  }\bibfield  {title} {\bibinfo {title} {Experimental criteria for steering and
  the einstein-podolsky-rosen paradox experimental criteria for steering and
  the einstein-podolsky-rosen paradox},\ }\href@noop {} {\bibfield  {journal}
  {\bibinfo  {journal} {Phys. Rev. A}\ }\textbf {\bibinfo {volume} {80}},\
  \bibinfo {pages} {032112} (\bibinfo {year} {2009})}\BibitemShut {NoStop}%
\bibitem [{\citenamefont {Jones}\ \emph
  {et~al.}(2007{\natexlab{b}})\citenamefont {Jones}, \citenamefont {Wiseman},\
  and\ \citenamefont {Doherty}}]{jw07}%
  \BibitemOpen
  \bibfield  {author} {\bibinfo {author} {\bibfnamefont {S.~J.}\ \bibnamefont
  {Jones}}, \bibinfo {author} {\bibfnamefont {H.~M.}\ \bibnamefont {Wiseman}},\
  and\ \bibinfo {author} {\bibfnamefont {A.~C.}\ \bibnamefont {Doherty}},\
  }\bibfield  {title} {\bibinfo {title} {Entanglement, einstein-podolsky-rosen
  correlations, bell nonlocality, and steering},\ }\href@noop {} {\bibfield
  {journal} {\bibinfo  {journal} {Phys. Rev. A}\ }\textbf {\bibinfo {volume}
  {76}},\ \bibinfo {pages} {052116} (\bibinfo {year}
  {2007}{\natexlab{b}})}\BibitemShut {NoStop}%
\bibitem [{\citenamefont {Werner}(1989)}]{rw89}%
  \BibitemOpen
  \bibfield  {author} {\bibinfo {author} {\bibfnamefont {R.~F.}\ \bibnamefont
  {Werner}},\ }\bibfield  {title} {\bibinfo {title} {Quantum states with
  einstein-podolsky-rosen correlations admitting a hidden-variable model},\
  }\href {https://doi.org/10.1103/PhysRevA.40.4277} {\bibfield  {journal}
  {\bibinfo  {journal} {Phys. Rev. A}\ }\textbf {\bibinfo {volume} {40}},\
  \bibinfo {pages} {4277} (\bibinfo {year} {1989})}\BibitemShut {NoStop}%
\bibitem [{\citenamefont {Ac\'{\i}n}\ \emph {et~al.}(2006)\citenamefont
  {Ac\'{\i}n}, \citenamefont {Gisin},\ and\ \citenamefont {Toner}}]{ag06}%
  \BibitemOpen
  \bibfield  {author} {\bibinfo {author} {\bibfnamefont {A.}~\bibnamefont
  {Ac\'{\i}n}}, \bibinfo {author} {\bibfnamefont {N.}~\bibnamefont {Gisin}},\
  and\ \bibinfo {author} {\bibfnamefont {B.}~\bibnamefont {Toner}},\ }\bibfield
   {title} {\bibinfo {title} {Grothendieck's constant and local models for
  noisy entangled quantum states},\ }\href
  {https://doi.org/10.1103/PhysRevA.73.062105} {\bibfield  {journal} {\bibinfo
  {journal} {Phys. Rev. A}\ }\textbf {\bibinfo {volume} {73}},\ \bibinfo
  {pages} {062105} (\bibinfo {year} {2006})}\BibitemShut {NoStop}%
\bibitem [{\citenamefont {Brierley}\ \emph {et~al.}(2017)\citenamefont
  {Brierley}, \citenamefont {Navascu\'{e}s},\ and\ \citenamefont
  {V\'ertesi}}]{bn17}%
  \BibitemOpen
  \bibfield  {author} {\bibinfo {author} {\bibfnamefont {S.}~\bibnamefont
  {Brierley}}, \bibinfo {author} {\bibfnamefont {M.}~\bibnamefont
  {Navascu\'{e}s}},\ and\ \bibinfo {author} {\bibfnamefont {T.}~\bibnamefont
  {V\'ertesi}},\ }\bibfield  {title} {\bibinfo {title} {Convex separation from
  convex optimization for large-scale problems},\ }\href@noop {} {\bibfield
  {journal} {\bibinfo  {journal} {arXiv:1609.05011v2}\ } (\bibinfo {year}
  {2017})}\BibitemShut {NoStop}%
\bibitem [{\citenamefont {Hirsch}\ \emph {et~al.}(2017)\citenamefont {Hirsch},
  \citenamefont {Quintino}, \citenamefont {V\'ertesi}, \citenamefont
  {Navascu\'es},\ and\ \citenamefont {Brunner}}]{hq17}%
  \BibitemOpen
  \bibfield  {author} {\bibinfo {author} {\bibfnamefont {F.}~\bibnamefont
  {Hirsch}}, \bibinfo {author} {\bibfnamefont {M.~T.}\ \bibnamefont
  {Quintino}}, \bibinfo {author} {\bibfnamefont {T.}~\bibnamefont {V\'ertesi}},
  \bibinfo {author} {\bibfnamefont {M.}~\bibnamefont {Navascu\'es}},\ and\
  \bibinfo {author} {\bibfnamefont {N.}~\bibnamefont {Brunner}},\ }\bibfield
  {title} {\bibinfo {title} {Better local hidden variable models for two-qubit
  werner states and an upper bound on the grothendieck constant kg(3)},\
  }\href@noop {} {\bibfield  {journal} {\bibinfo  {journal} {Quantum}\ }\textbf
  {\bibinfo {volume} {1}},\ \bibinfo {pages} {3} (\bibinfo {year}
  {2017})}\BibitemShut {NoStop}%
\bibitem [{\citenamefont {Costa}\ \emph
  {et~al.}(2018{\natexlab{a}})\citenamefont {Costa}, \citenamefont {Uola},\
  and\ \citenamefont {G\"{u}hne}}]{cu18a}%
  \BibitemOpen
  \bibfield  {author} {\bibinfo {author} {\bibfnamefont {A.~C.~S.}\
  \bibnamefont {Costa}}, \bibinfo {author} {\bibfnamefont {R.}~\bibnamefont
  {Uola}},\ and\ \bibinfo {author} {\bibfnamefont {O.}~\bibnamefont
  {G\"{u}hne}},\ }\bibfield  {title} {\bibinfo {title} {Entropic steering
  criteria: Applications to bipartite and tripartite systems},\ }\href@noop {}
  {\bibfield  {journal} {\bibinfo  {journal} {Entropy}\ }\textbf {\bibinfo
  {volume} {20}},\ \bibinfo {pages} {763} (\bibinfo {year}
  {2018}{\natexlab{a}})}\BibitemShut {NoStop}%
\bibitem [{\citenamefont {Cavalcanti}\ \emph {et~al.}(2011)\citenamefont
  {Cavalcanti}, \citenamefont {He}, \citenamefont {Reid},\ and\ \citenamefont
  {Wiseman}}]{ch11}%
  \BibitemOpen
  \bibfield  {author} {\bibinfo {author} {\bibfnamefont {E.~G.}\ \bibnamefont
  {Cavalcanti}}, \bibinfo {author} {\bibfnamefont {Q.~Y.}\ \bibnamefont {He}},
  \bibinfo {author} {\bibfnamefont {M.~D.}\ \bibnamefont {Reid}},\ and\
  \bibinfo {author} {\bibfnamefont {H.~M.}\ \bibnamefont {Wiseman}},\
  }\bibfield  {title} {\bibinfo {title} {Unified criteria for multipartite
  quantum nonlocality},\ }\href@noop {} {\bibfield  {journal} {\bibinfo
  {journal} {Phys. Rev. A}\ }\textbf {\bibinfo {volume} {84}} (\bibinfo {year}
  {2011})}\BibitemShut {NoStop}%
\bibitem [{\citenamefont {He}\ \emph {et~al.}(2011)\citenamefont {He},
  \citenamefont {Drummond},\ and\ \citenamefont {Reid}}]{hd11}%
  \BibitemOpen
  \bibfield  {author} {\bibinfo {author} {\bibfnamefont {Q.~Y.}\ \bibnamefont
  {He}}, \bibinfo {author} {\bibfnamefont {P.~D.}\ \bibnamefont {Drummond}},\
  and\ \bibinfo {author} {\bibfnamefont {M.~D.}\ \bibnamefont {Reid}},\
  }\bibfield  {title} {\bibinfo {title} {Entanglement, epr steering, and
  bell-nonlocality criteria for multipartite higher-spin systems},\ }\href
  {https://doi.org/10.1103/PhysRevA.83.032120} {\bibfield  {journal} {\bibinfo
  {journal} {Phys. Rev. A}\ }\textbf {\bibinfo {volume} {83}},\ \bibinfo
  {pages} {032120} (\bibinfo {year} {2011})}\BibitemShut {NoStop}%
\bibitem [{\citenamefont {Xiang}\ \emph
  {et~al.}(2017{\natexlab{a}})\citenamefont {Xiang}, \citenamefont {Kogias},
  \citenamefont {Adesso},\ and\ \citenamefont {He}}]{xk17}%
  \BibitemOpen
  \bibfield  {author} {\bibinfo {author} {\bibfnamefont {Y.}~\bibnamefont
  {Xiang}}, \bibinfo {author} {\bibfnamefont {I.}~\bibnamefont {Kogias}},
  \bibinfo {author} {\bibfnamefont {G.}~\bibnamefont {Adesso}},\ and\ \bibinfo
  {author} {\bibfnamefont {Q.~Y.}\ \bibnamefont {He}},\ }\bibfield  {title}
  {\bibinfo {title} {Multipartite gaussian steering: Monogamy constraints and
  quantum cryptography applications},\ }\href@noop {} {\bibfield  {journal}
  {\bibinfo  {journal} {Phys. Rev. A}\ }\textbf {\bibinfo {volume} {95}},\
  \bibinfo {pages} {010101(R)} (\bibinfo {year}
  {2017}{\natexlab{a}})}\BibitemShut {NoStop}%
\bibitem [{\citenamefont {Reid}(2013{\natexlab{b}})}]{mdr13}%
  \BibitemOpen
  \bibfield  {author} {\bibinfo {author} {\bibfnamefont {M.~D.}\ \bibnamefont
  {Reid}},\ }\bibfield  {title} {\bibinfo {title} {Monogamy inequalities for
  the einstein- podolsky-rosen paradox and quantum steering},\ }\href@noop {}
  {\bibfield  {journal} {\bibinfo  {journal} {Phys. Rev. A}\ }\textbf {\bibinfo
  {volume} {88}},\ \bibinfo {pages} {062108} (\bibinfo {year}
  {2013}{\natexlab{b}})}\BibitemShut {NoStop}%
\bibitem [{\citenamefont {D'Auria}\ \emph {et~al.}(2009)\citenamefont
  {D'Auria}, \citenamefont {Fornaro}, \citenamefont {Porzio}, \citenamefont
  {Solimeno}, \citenamefont {Olivares},\ and\ \citenamefont {Paris}}]{af09}%
  \BibitemOpen
  \bibfield  {author} {\bibinfo {author} {\bibfnamefont {V.}~\bibnamefont
  {D'Auria}}, \bibinfo {author} {\bibfnamefont {S.}~\bibnamefont {Fornaro}},
  \bibinfo {author} {\bibfnamefont {A.}~\bibnamefont {Porzio}}, \bibinfo
  {author} {\bibfnamefont {S.}~\bibnamefont {Solimeno}}, \bibinfo {author}
  {\bibfnamefont {S.}~\bibnamefont {Olivares}},\ and\ \bibinfo {author}
  {\bibfnamefont {M.~G.~A.}\ \bibnamefont {Paris}},\ }\bibfield  {title}
  {\bibinfo {title} {Full characterization of gaussian bipartite entangled
  states by a single homodyne detector},\ }\href@noop {} {\bibfield  {journal}
  {\bibinfo  {journal} {Phys. Rev. Lett.}\ }\textbf {\bibinfo {volume} {102}},\
  \bibinfo {pages} {020502} (\bibinfo {year} {2009})}\BibitemShut {NoStop}%
\bibitem [{\citenamefont {He}\ \emph {et~al.}(2012)\citenamefont {He},
  \citenamefont {Drummond}, \citenamefont {Olsen},\ and\ \citenamefont
  {Reid}}]{hd12}%
  \BibitemOpen
  \bibfield  {author} {\bibinfo {author} {\bibfnamefont {Q.~Y.}\ \bibnamefont
  {He}}, \bibinfo {author} {\bibfnamefont {P.~D.}\ \bibnamefont {Drummond}},
  \bibinfo {author} {\bibfnamefont {M.}~\bibnamefont {Olsen}},\ and\ \bibinfo
  {author} {\bibfnamefont {M.~D.}\ \bibnamefont {Reid}},\ }\bibfield  {title}
  {\bibinfo {title} {Einstein-podolsky-rosen entanglement and steering in
  two-well bose-einstein-condensate ground states},\ }\href@noop {} {\bibfield
  {journal} {\bibinfo  {journal} {Phys. Rev. A}\ }\textbf {\bibinfo {volume}
  {86}},\ \bibinfo {pages} {023626} (\bibinfo {year} {2012})}\BibitemShut
  {NoStop}%
\bibitem [{\citenamefont {Dalton}\ \emph {et~al.}(2020)\citenamefont {Dalton},
  \citenamefont {Garraway},\ and\ \citenamefont {Reid}}]{dg20}%
  \BibitemOpen
  \bibfield  {author} {\bibinfo {author} {\bibfnamefont {B.~J.}\ \bibnamefont
  {Dalton}}, \bibinfo {author} {\bibfnamefont {B.~M.}\ \bibnamefont
  {Garraway}},\ and\ \bibinfo {author} {\bibfnamefont {M.~D.}\ \bibnamefont
  {Reid}},\ }\bibfield  {title} {\bibinfo {title} {Tests for
  einstein-podolsky-rosen steering in two-mode systems of identical massive
  bosons},\ }\href@noop {} {\bibfield  {journal} {\bibinfo  {journal} {Phys.
  Rev. A}\ }\textbf {\bibinfo {volume} {101}},\ \bibinfo {pages} {012117}
  (\bibinfo {year} {2020})}\BibitemShut {NoStop}%
\bibitem [{\citenamefont {Kiesewetter}\ \emph {et~al.}(2014)\citenamefont
  {Kiesewetter}, \citenamefont {He}, \citenamefont {Drummond},\ and\
  \citenamefont {Reid}}]{kh14}%
  \BibitemOpen
  \bibfield  {author} {\bibinfo {author} {\bibfnamefont {S.}~\bibnamefont
  {Kiesewetter}}, \bibinfo {author} {\bibfnamefont {Q.~Y.}\ \bibnamefont {He}},
  \bibinfo {author} {\bibfnamefont {P.~D.}\ \bibnamefont {Drummond}},\ and\
  \bibinfo {author} {\bibfnamefont {M.~D.}\ \bibnamefont {Reid}},\ }\bibfield
  {title} {\bibinfo {title} {Scalable quantum simulation of pulsed entanglement
  and einstein-podolsky-rosen steering in optomechanics},\ }\href
  {https://doi.org/10.1103/PhysRevA.90.043805} {\bibfield  {journal} {\bibinfo
  {journal} {Phys. Rev. A}\ }\textbf {\bibinfo {volume} {90}},\ \bibinfo
  {pages} {043805} (\bibinfo {year} {2014})}\BibitemShut {NoStop}%
\bibitem [{\citenamefont {Xiang}\ \emph {et~al.}(2015)\citenamefont {Xiang},
  \citenamefont {Sun}, \citenamefont {Wang}, \citenamefont {Gong},\ and\
  \citenamefont {He}}]{xs15}%
  \BibitemOpen
  \bibfield  {author} {\bibinfo {author} {\bibfnamefont {Y.}~\bibnamefont
  {Xiang}}, \bibinfo {author} {\bibfnamefont {F.~X.}\ \bibnamefont {Sun}},
  \bibinfo {author} {\bibfnamefont {M.}~\bibnamefont {Wang}}, \bibinfo {author}
  {\bibfnamefont {Q.~H.}\ \bibnamefont {Gong}},\ and\ \bibinfo {author}
  {\bibfnamefont {Q.~Y.}\ \bibnamefont {He}},\ }\bibfield  {title} {\bibinfo
  {title} {Detection of genuine tripartite entanglement and steering in hybrid
  optomechanics},\ }\href@noop {} {\bibfield  {journal} {\bibinfo  {journal}
  {Optics Express}\ }\textbf {\bibinfo {volume} {23}},\ \bibinfo {pages}
  {30104} (\bibinfo {year} {2015})}\BibitemShut {NoStop}%
\bibitem [{\citenamefont {Kiesewetter}\ \emph {et~al.}(2017)\citenamefont
  {Kiesewetter}, \citenamefont {Teh}, \citenamefont {Drummond},\ and\
  \citenamefont {Reid}}]{kt17}%
  \BibitemOpen
  \bibfield  {author} {\bibinfo {author} {\bibfnamefont {S.}~\bibnamefont
  {Kiesewetter}}, \bibinfo {author} {\bibfnamefont {R.~Y.}\ \bibnamefont
  {Teh}}, \bibinfo {author} {\bibfnamefont {P.~D.}\ \bibnamefont {Drummond}},\
  and\ \bibinfo {author} {\bibfnamefont {M.~D.}\ \bibnamefont {Reid}},\
  }\bibfield  {title} {\bibinfo {title} {Pulsed entanglement of two
  optomechanical oscillators and furry's hypothesis},\ }\href
  {https://doi.org/10.1103/PhysRevLett.119.023601} {\bibfield  {journal}
  {\bibinfo  {journal} {Phys. Rev. Lett.}\ }\textbf {\bibinfo {volume} {119}},\
  \bibinfo {pages} {023601} (\bibinfo {year} {2017})}\BibitemShut {NoStop}%
\bibitem [{\citenamefont {Gebremariam}\ \emph {et~al.}(2019)\citenamefont
  {Gebremariam}, \citenamefont {Mazaheri}, \citenamefont {Zeng},\ and\
  \citenamefont {Li}}]{gm19}%
  \BibitemOpen
  \bibfield  {author} {\bibinfo {author} {\bibfnamefont {T.}~\bibnamefont
  {Gebremariam}}, \bibinfo {author} {\bibfnamefont {M.}~\bibnamefont
  {Mazaheri}}, \bibinfo {author} {\bibfnamefont {Y.}~\bibnamefont {Zeng}},\
  and\ \bibinfo {author} {\bibfnamefont {C.}~\bibnamefont {Li}},\ }\bibfield
  {title} {\bibinfo {title} {Dynamical quantum steering in a pulsed hybrid
  opto-electro-mechanical system},\ }\href@noop {} {\bibfield  {journal}
  {\bibinfo  {journal} {J. Opt. Soc. Am. B}\ }\textbf {\bibinfo {volume}
  {36}},\ \bibinfo {pages} {168} (\bibinfo {year} {2019})}\BibitemShut
  {NoStop}%
\bibitem [{\citenamefont {Ji}\ \emph {et~al.}(2015{\natexlab{a}})\citenamefont
  {Ji}, \citenamefont {Kim},\ and\ \citenamefont {Nha}}]{jk15}%
  \BibitemOpen
  \bibfield  {author} {\bibinfo {author} {\bibfnamefont {S.~W.}\ \bibnamefont
  {Ji}}, \bibinfo {author} {\bibfnamefont {M.~S.}\ \bibnamefont {Kim}},\ and\
  \bibinfo {author} {\bibfnamefont {H.}~\bibnamefont {Nha}},\ }\bibfield
  {title} {\bibinfo {title} {Quantum steering of multimode gaussian states by
  gaussian measurements: Monogamy relations and the peres conjecture},\
  }\href@noop {} {\bibfield  {journal} {\bibinfo  {journal} {J. Phys. A}\
  }\textbf {\bibinfo {volume} {48}},\ \bibinfo {pages} {135301} (\bibinfo
  {year} {2015}{\natexlab{a}})}\BibitemShut {NoStop}%
\bibitem [{\citenamefont {Adesso}\ and\ \citenamefont {Simon}(2016)}]{as16}%
  \BibitemOpen
  \bibfield  {author} {\bibinfo {author} {\bibfnamefont {G.}~\bibnamefont
  {Adesso}}\ and\ \bibinfo {author} {\bibfnamefont {R.}~\bibnamefont {Simon}},\
  }\bibfield  {title} {\bibinfo {title} {Strong subadditivity for
  log-determinant of covariance matrices and its applications},\ }\href
  {https://doi.org/10.1088/1751-8113/49/34/34lt02} {\bibfield  {journal}
  {\bibinfo  {journal} {J. Phys. A: Math. Theor.}\ }\textbf {\bibinfo {volume}
  {49}},\ \bibinfo {pages} {34LT02} (\bibinfo {year} {2016})}\BibitemShut
  {NoStop}%
\bibitem [{\citenamefont {Lami}\ \emph {et~al.}(2016)\citenamefont {Lami},
  \citenamefont {Hirche}, \citenamefont {Adesso},\ and\ \citenamefont
  {Winter}}]{lc16}%
  \BibitemOpen
  \bibfield  {author} {\bibinfo {author} {\bibfnamefont {L.}~\bibnamefont
  {Lami}}, \bibinfo {author} {\bibfnamefont {C.}~\bibnamefont {Hirche}},
  \bibinfo {author} {\bibfnamefont {G.}~\bibnamefont {Adesso}},\ and\ \bibinfo
  {author} {\bibfnamefont {A.}~\bibnamefont {Winter}},\ }\bibfield  {title}
  {\bibinfo {title} {Schur complement inequalities for covariance matrices and
  monogamy of quantum correlations},\ }\href
  {https://doi.org/10.1103/PhysRevLett.117.220502} {\bibfield  {journal}
  {\bibinfo  {journal} {Phys. Rev. Lett.}\ }\textbf {\bibinfo {volume} {117}},\
  \bibinfo {pages} {220502} (\bibinfo {year} {2016})}\BibitemShut {NoStop}%
\bibitem [{\citenamefont {Cheng}\ \emph {et~al.}(2016)\citenamefont {Cheng},
  \citenamefont {Milne}, \citenamefont {Hall},\ and\ \citenamefont
  {Wiseman}}]{cm16}%
  \BibitemOpen
  \bibfield  {author} {\bibinfo {author} {\bibfnamefont {S.}~\bibnamefont
  {Cheng}}, \bibinfo {author} {\bibfnamefont {A.}~\bibnamefont {Milne}},
  \bibinfo {author} {\bibfnamefont {M.~J.~W.}\ \bibnamefont {Hall}},\ and\
  \bibinfo {author} {\bibfnamefont {H.~M.}\ \bibnamefont {Wiseman}},\
  }\bibfield  {title} {\bibinfo {title} {Volume monogamy of quantum steering
  ellipsoids for multiqubit systems},\ }\href@noop {} {\bibfield  {journal}
  {\bibinfo  {journal} {Phys. Rev. A}\ }\textbf {\bibinfo {volume} {94}},\
  \bibinfo {pages} {042105} (\bibinfo {year} {2016})}\BibitemShut {NoStop}%
\bibitem [{\citenamefont {Kogias}\ \emph
  {et~al.}(2015{\natexlab{a}})\citenamefont {Kogias}, \citenamefont {Lee},
  \citenamefont {Ragy},\ and\ \citenamefont {Adesso}}]{kl15}%
  \BibitemOpen
  \bibfield  {author} {\bibinfo {author} {\bibfnamefont {I.}~\bibnamefont
  {Kogias}}, \bibinfo {author} {\bibfnamefont {A.~R.}\ \bibnamefont {Lee}},
  \bibinfo {author} {\bibfnamefont {S.}~\bibnamefont {Ragy}},\ and\ \bibinfo
  {author} {\bibfnamefont {G.}~\bibnamefont {Adesso}},\ }\bibfield  {title}
  {\bibinfo {title} {Quantification of gaussian quantum steering},\ }\href@noop
  {} {\bibfield  {journal} {\bibinfo  {journal} {Phys. Rev. Lett.}\ }\textbf
  {\bibinfo {volume} {114}},\ \bibinfo {pages} {060403} (\bibinfo {year}
  {2015}{\natexlab{a}})}\BibitemShut {NoStop}%
\bibitem [{\citenamefont {Kogias}\ and\ \citenamefont {Adesso}(2015)}]{ka15}%
  \BibitemOpen
  \bibfield  {author} {\bibinfo {author} {\bibfnamefont {I.}~\bibnamefont
  {Kogias}}\ and\ \bibinfo {author} {\bibfnamefont {G.}~\bibnamefont
  {Adesso}},\ }\bibfield  {title} {\bibinfo {title} {Einstein-podolsky-rosen
  steering measure for two-mode continuous variable states},\ }\href@noop {}
  {\bibfield  {journal} {\bibinfo  {journal} {J. Opt. Soc. Am. B}\ }\textbf
  {\bibinfo {volume} {32}},\ \bibinfo {pages} {A27} (\bibinfo {year}
  {2015})}\BibitemShut {NoStop}%
\bibitem [{\citenamefont {Laurat}\ \emph {et~al.}(2005)\citenamefont {Laurat},
  \citenamefont {Keller}, \citenamefont {Oliveira-Huguenin}, \citenamefont
  {Fabre}, \citenamefont {Coudreau}, \citenamefont {Serafini}, \citenamefont
  {Adesso},\ and\ \citenamefont {Illuminati}}]{lk05}%
  \BibitemOpen
  \bibfield  {author} {\bibinfo {author} {\bibfnamefont {J.}~\bibnamefont
  {Laurat}}, \bibinfo {author} {\bibfnamefont {G.}~\bibnamefont {Keller}},
  \bibinfo {author} {\bibfnamefont {J.~A.}\ \bibnamefont {Oliveira-Huguenin}},
  \bibinfo {author} {\bibfnamefont {C.}~\bibnamefont {Fabre}}, \bibinfo
  {author} {\bibfnamefont {T.}~\bibnamefont {Coudreau}}, \bibinfo {author}
  {\bibfnamefont {A.}~\bibnamefont {Serafini}}, \bibinfo {author}
  {\bibfnamefont {G.}~\bibnamefont {Adesso}},\ and\ \bibinfo {author}
  {\bibfnamefont {F.}~\bibnamefont {Illuminati}},\ }\bibfield  {title}
  {\bibinfo {title} {Entanglement of two-mode gaussian states: characterization
  and experimental production and manipulation},\ }\href
  {https://doi.org/10.1088/1464-4266/7/12/021} {\bibfield  {journal} {\bibinfo
  {journal} {J. Opt. B: Quantum Semiclass. Opt}\ }\textbf {\bibinfo {volume}
  {7}},\ \bibinfo {pages} {S577} (\bibinfo {year} {2005})}\BibitemShut
  {NoStop}%
\bibitem [{\citenamefont {Xiang}\ \emph
  {et~al.}(2017{\natexlab{b}})\citenamefont {Xiang}, \citenamefont {Xu},
  \citenamefont {Mi\ifmmode~\check{s}\else \v{s}\fi{}ta}, \citenamefont
  {Tufarelli}, \citenamefont {He},\ and\ \citenamefont {Adesso}}]{xx17}%
  \BibitemOpen
  \bibfield  {author} {\bibinfo {author} {\bibfnamefont {Y.}~\bibnamefont
  {Xiang}}, \bibinfo {author} {\bibfnamefont {B.~Q.}\ \bibnamefont {Xu}},
  \bibinfo {author} {\bibfnamefont {L.}~\bibnamefont {Mi\ifmmode~\check{s}\else
  \v{s}\fi{}ta}}, \bibinfo {author} {\bibfnamefont {T.}~\bibnamefont
  {Tufarelli}}, \bibinfo {author} {\bibfnamefont {Q.~Y.}\ \bibnamefont {He}},\
  and\ \bibinfo {author} {\bibfnamefont {G.}~\bibnamefont {Adesso}},\
  }\bibfield  {title} {\bibinfo {title} {Investigating einstein-podolsky-rosen
  steering of continuous-variable bipartite states by non-gaussian pseudospin
  measurements},\ }\href {https://doi.org/10.1103/PhysRevA.96.042326}
  {\bibfield  {journal} {\bibinfo  {journal} {Phys. Rev. A}\ }\textbf {\bibinfo
  {volume} {96}},\ \bibinfo {pages} {042326} (\bibinfo {year}
  {2017}{\natexlab{b}})}\BibitemShut {NoStop}%
\bibitem [{\citenamefont {Skrzypczyk}\ \emph {et~al.}(2014)\citenamefont
  {Skrzypczyk}, \citenamefont {Navascu\'{e}s},\ and\ \citenamefont
  {Cavalcanti}}]{sn14}%
  \BibitemOpen
  \bibfield  {author} {\bibinfo {author} {\bibfnamefont {P.}~\bibnamefont
  {Skrzypczyk}}, \bibinfo {author} {\bibfnamefont {M.}~\bibnamefont
  {Navascu\'{e}s}},\ and\ \bibinfo {author} {\bibfnamefont {D.}~\bibnamefont
  {Cavalcanti}},\ }\bibfield  {title} {\bibinfo {title} {Quantifying
  einstein-podolsky-rosen steering},\ }\href@noop {} {\bibfield  {journal}
  {\bibinfo  {journal} {Phys. Rev. Lett.}\ }\textbf {\bibinfo {volume} {112}},\
  \bibinfo {pages} {180404} (\bibinfo {year} {2014})}\BibitemShut {NoStop}%
\bibitem [{\citenamefont {Evans}\ and\ \citenamefont {Wiseman}(2014)}]{ew14}%
  \BibitemOpen
  \bibfield  {author} {\bibinfo {author} {\bibfnamefont {D.~A.}\ \bibnamefont
  {Evans}}\ and\ \bibinfo {author} {\bibfnamefont {H.~M.}\ \bibnamefont
  {Wiseman}},\ }\bibfield  {title} {\bibinfo {title} {Optimal measurements for
  tests of einstein-podolsky-rosen steering with no detection loophole using
  two-qubit werner states},\ }\href@noop {} {\bibfield  {journal} {\bibinfo
  {journal} {Phys. Rev. A}\ }\textbf {\bibinfo {volume} {90}},\ \bibinfo
  {pages} {012114} (\bibinfo {year} {2014})}\BibitemShut {NoStop}%
\bibitem [{\citenamefont {Schneeloch}\ \emph {et~al.}(2013)\citenamefont
  {Schneeloch}, \citenamefont {Broadbent}, \citenamefont {Walborn},
  \citenamefont {Cavalcanti},\ and\ \citenamefont {Howell}}]{sb13}%
  \BibitemOpen
  \bibfield  {author} {\bibinfo {author} {\bibfnamefont {J.}~\bibnamefont
  {Schneeloch}}, \bibinfo {author} {\bibfnamefont {C.~J.}\ \bibnamefont
  {Broadbent}}, \bibinfo {author} {\bibfnamefont {S.~P.}\ \bibnamefont
  {Walborn}}, \bibinfo {author} {\bibfnamefont {E.~G.}\ \bibnamefont
  {Cavalcanti}},\ and\ \bibinfo {author} {\bibfnamefont {J.~C.}\ \bibnamefont
  {Howell}},\ }\bibfield  {title} {\bibinfo {title} {Einstein-podolsky-rosen
  steering inequalities from entropic uncertainty relations},\ }\href
  {https://doi.org/10.1103/PhysRevA.87.062103} {\bibfield  {journal} {\bibinfo
  {journal} {Phys. Rev. A}\ }\textbf {\bibinfo {volume} {87}},\ \bibinfo
  {pages} {062103} (\bibinfo {year} {2013})}\BibitemShut {NoStop}%
\bibitem [{\citenamefont {Kriv\'{a}chy}\ \emph {et~al.}(2018)\citenamefont
  {Kriv\'{a}chy}, \citenamefont {Fr\"{o}wis},\ and\ \citenamefont
  {Brunner}}]{kf18}%
  \BibitemOpen
  \bibfield  {author} {\bibinfo {author} {\bibfnamefont {T.}~\bibnamefont
  {Kriv\'{a}chy}}, \bibinfo {author} {\bibfnamefont {F.}~\bibnamefont
  {Fr\"{o}wis}},\ and\ \bibinfo {author} {\bibfnamefont {N.}~\bibnamefont
  {Brunner}},\ }\bibfield  {title} {\bibinfo {title} {Tight steering
  inequalities from generalized entropic uncertainty relations},\ }\href@noop
  {} {\bibfield  {journal} {\bibinfo  {journal} {Phys. Rev. A}\ }\textbf
  {\bibinfo {volume} {98}},\ \bibinfo {pages} {062111} (\bibinfo {year}
  {2018})}\BibitemShut {NoStop}%
\bibitem [{\citenamefont {Costa}\ \emph
  {et~al.}(2018{\natexlab{b}})\citenamefont {Costa}, \citenamefont {Uola},\
  and\ \citenamefont {G\"{u}hne}}]{cu18}%
  \BibitemOpen
  \bibfield  {author} {\bibinfo {author} {\bibfnamefont {A.~C.~S.}\
  \bibnamefont {Costa}}, \bibinfo {author} {\bibfnamefont {R.}~\bibnamefont
  {Uola}},\ and\ \bibinfo {author} {\bibfnamefont {O.}~\bibnamefont
  {G\"{u}hne}},\ }\bibfield  {title} {\bibinfo {title} {Tight steering
  inequalities from generalized entropic uncertainty relations},\ }\href@noop
  {} {\bibfield  {journal} {\bibinfo  {journal} {Phys. Rev. A}\ }\textbf
  {\bibinfo {volume} {98}},\ \bibinfo {pages} {050104(R)} (\bibinfo {year}
  {2018}{\natexlab{b}})}\BibitemShut {NoStop}%
\bibitem [{\citenamefont {Coles}\ \emph {et~al.}(2017)\citenamefont {Coles},
  \citenamefont {Berta}, \citenamefont {Tomamichel},\ and\ \citenamefont
  {Wehner}}]{cb17}%
  \BibitemOpen
  \bibfield  {author} {\bibinfo {author} {\bibfnamefont {P.~J.}\ \bibnamefont
  {Coles}}, \bibinfo {author} {\bibfnamefont {M.}~\bibnamefont {Berta}},
  \bibinfo {author} {\bibfnamefont {M.}~\bibnamefont {Tomamichel}},\ and\
  \bibinfo {author} {\bibfnamefont {S.}~\bibnamefont {Wehner}},\ }\bibfield
  {title} {\bibinfo {title} {Entropic uncertainty relations and their
  applications},\ }\href@noop {} {\bibfield  {journal} {\bibinfo  {journal}
  {Rev. Mod. Phys.}\ }\textbf {\bibinfo {volume} {89}},\ \bibinfo {pages}
  {015002} (\bibinfo {year} {2017})}\BibitemShut {NoStop}%
\bibitem [{\citenamefont {Xiao}\ \emph {et~al.}(2020)\citenamefont {Xiao},
  \citenamefont {Xiang}, \citenamefont {He},\ and\ \citenamefont
  {Sanders}}]{xx20}%
  \BibitemOpen
  \bibfield  {author} {\bibinfo {author} {\bibfnamefont {Y.~L.}\ \bibnamefont
  {Xiao}}, \bibinfo {author} {\bibfnamefont {Y.}~\bibnamefont {Xiang}},
  \bibinfo {author} {\bibfnamefont {Q.~Y.}\ \bibnamefont {He}},\ and\ \bibinfo
  {author} {\bibfnamefont {B.~C.}\ \bibnamefont {Sanders}},\ }\bibfield
  {title} {\bibinfo {title} {Quasi-fine-grained uncertainty relations},\ }\href
  {https://doi.org/10.1088/1367-2630/ab9d57} {\bibfield  {journal} {\bibinfo
  {journal} {New J. Phys}\ }\textbf {\bibinfo {volume} {22}},\ \bibinfo {pages}
  {073063} (\bibinfo {year} {2020})}\BibitemShut {NoStop}%
\bibitem [{\citenamefont {Li}\ \emph {et~al.}(2015)\citenamefont {Li},
  \citenamefont {Chen}, \citenamefont {Lambert}, \citenamefont {Chiu},\ and\
  \citenamefont {Nori}}]{lc15}%
  \BibitemOpen
  \bibfield  {author} {\bibinfo {author} {\bibfnamefont {C.-M.}\ \bibnamefont
  {Li}}, \bibinfo {author} {\bibfnamefont {Y.-N.}\ \bibnamefont {Chen}},
  \bibinfo {author} {\bibfnamefont {N.}~\bibnamefont {Lambert}}, \bibinfo
  {author} {\bibfnamefont {C.-Y.}\ \bibnamefont {Chiu}},\ and\ \bibinfo
  {author} {\bibfnamefont {F.}~\bibnamefont {Nori}},\ }\bibfield  {title}
  {\bibinfo {title} {Certifying single-system steering for quantum-information
  processing},\ }\href {https://doi.org/10.1103/PhysRevA.92.062310} {\bibfield
  {journal} {\bibinfo  {journal} {Phys. Rev. A}\ }\textbf {\bibinfo {volume}
  {92}},\ \bibinfo {pages} {062310} (\bibinfo {year} {2015})}\BibitemShut
  {NoStop}%
\bibitem [{\citenamefont {Moroder}\ \emph {et~al.}(2016)\citenamefont
  {Moroder}, \citenamefont {Gittsovich}, \citenamefont {Huber}, \citenamefont
  {Uola},\ and\ \citenamefont {G\"{u}hne}}]{mg16}%
  \BibitemOpen
  \bibfield  {author} {\bibinfo {author} {\bibfnamefont {T.}~\bibnamefont
  {Moroder}}, \bibinfo {author} {\bibfnamefont {O.}~\bibnamefont {Gittsovich}},
  \bibinfo {author} {\bibfnamefont {M.}~\bibnamefont {Huber}}, \bibinfo
  {author} {\bibfnamefont {R.}~\bibnamefont {Uola}},\ and\ \bibinfo {author}
  {\bibfnamefont {O.}~\bibnamefont {G\"{u}hne}},\ }\bibfield  {title} {\bibinfo
  {title} {Steering maps and their application to dimension-bounded steering},\
  }\href@noop {} {\bibfield  {journal} {\bibinfo  {journal} {Phys. Rev. Lett.}\
  }\textbf {\bibinfo {volume} {116}},\ \bibinfo {pages} {090403} (\bibinfo
  {year} {2016})}\BibitemShut {NoStop}%
\bibitem [{\citenamefont {Chen}\ \emph {et~al.}(2013)\citenamefont {Chen},
  \citenamefont {Ye}, \citenamefont {Wu}, \citenamefont {Su}, \citenamefont
  {Cabello}, \citenamefont {Kwek},\ and\ \citenamefont {Oh}}]{cy13}%
  \BibitemOpen
  \bibfield  {author} {\bibinfo {author} {\bibfnamefont {J.-L.}\ \bibnamefont
  {Chen}}, \bibinfo {author} {\bibfnamefont {X.-J.}\ \bibnamefont {Ye}},
  \bibinfo {author} {\bibfnamefont {C.~F.}\ \bibnamefont {Wu}}, \bibinfo
  {author} {\bibfnamefont {H.-Y.}\ \bibnamefont {Su}}, \bibinfo {author}
  {\bibfnamefont {A.}~\bibnamefont {Cabello}}, \bibinfo {author} {\bibfnamefont
  {L.~C.}\ \bibnamefont {Kwek}},\ and\ \bibinfo {author} {\bibfnamefont
  {C.~H.}\ \bibnamefont {Oh}},\ }\bibfield  {title} {\bibinfo {title}
  {All-versus-nothing proof of einstein- podolsky-rosen steering},\ }\href@noop
  {} {\bibfield  {journal} {\bibinfo  {journal} {Sci. Reports}\ }\textbf
  {\bibinfo {volume} {3}},\ \bibinfo {pages} {02143} (\bibinfo {year}
  {2013})}\BibitemShut {NoStop}%
\bibitem [{\citenamefont {Cavalcanti}\ \emph
  {et~al.}(2015{\natexlab{b}})\citenamefont {Cavalcanti}, \citenamefont
  {Foster}, \citenamefont {Fuwa},\ and\ \citenamefont {Wiseman}}]{cf15}%
  \BibitemOpen
  \bibfield  {author} {\bibinfo {author} {\bibfnamefont {E.~G.}\ \bibnamefont
  {Cavalcanti}}, \bibinfo {author} {\bibfnamefont {C.~J.}\ \bibnamefont
  {Foster}}, \bibinfo {author} {\bibfnamefont {M.}~\bibnamefont {Fuwa}},\ and\
  \bibinfo {author} {\bibfnamefont {H.~M.}\ \bibnamefont {Wiseman}},\
  }\bibfield  {title} {\bibinfo {title} {Analog of the chsh inequality for
  steering},\ }\href {https://doi.org/10.1364/JOSAB.32.000A74} {\bibfield
  {journal} {\bibinfo  {journal} {J. Opt. Soc. Am. B}\ }\textbf {\bibinfo
  {volume} {32}},\ \bibinfo {pages} {A74} (\bibinfo {year}
  {2015}{\natexlab{b}})}\BibitemShut {NoStop}%
\bibitem [{\citenamefont {\ifmmode~\dot{Z}\else \.{Z}\fi{}ukowski}\ \emph
  {et~al.}(2015)\citenamefont {\ifmmode~\dot{Z}\else \.{Z}\fi{}ukowski},
  \citenamefont {Dutta},\ and\ \citenamefont {Yin}}]{zd15}%
  \BibitemOpen
  \bibfield  {author} {\bibinfo {author} {\bibfnamefont {M.}~\bibnamefont
  {\ifmmode~\dot{Z}\else \.{Z}\fi{}ukowski}}, \bibinfo {author} {\bibfnamefont
  {A.}~\bibnamefont {Dutta}},\ and\ \bibinfo {author} {\bibfnamefont
  {Z.}~\bibnamefont {Yin}},\ }\bibfield  {title} {\bibinfo {title} {Geometric
  bell-like inequalities for steering},\ }\href
  {https://doi.org/10.1103/PhysRevA.91.032107} {\bibfield  {journal} {\bibinfo
  {journal} {Phys. Rev. A}\ }\textbf {\bibinfo {volume} {91}},\ \bibinfo
  {pages} {032107} (\bibinfo {year} {2015})}\BibitemShut {NoStop}%
\bibitem [{\citenamefont {Ji}\ \emph {et~al.}(2015{\natexlab{b}})\citenamefont
  {Ji}, \citenamefont {Lee}, \citenamefont {Park},\ and\ \citenamefont
  {Nha}}]{jl15}%
  \BibitemOpen
  \bibfield  {author} {\bibinfo {author} {\bibfnamefont {S.-W.}\ \bibnamefont
  {Ji}}, \bibinfo {author} {\bibfnamefont {J.}~\bibnamefont {Lee}}, \bibinfo
  {author} {\bibfnamefont {J.}~\bibnamefont {Park}},\ and\ \bibinfo {author}
  {\bibfnamefont {H.}~\bibnamefont {Nha}},\ }\bibfield  {title} {\bibinfo
  {title} {Steering criteria via covariance matrices of local observables in
  arbitrary-dimensional quantum systems},\ }\href
  {https://doi.org/10.1103/PhysRevA.92.062130} {\bibfield  {journal} {\bibinfo
  {journal} {Phys. Rev. A}\ }\textbf {\bibinfo {volume} {92}},\ \bibinfo
  {pages} {062130} (\bibinfo {year} {2015}{\natexlab{b}})}\BibitemShut
  {NoStop}%
\bibitem [{\citenamefont {Kogias}\ \emph
  {et~al.}(2015{\natexlab{b}})\citenamefont {Kogias}, \citenamefont
  {Skrzypczyk}, \citenamefont {Cavalcanti}, \citenamefont {Acin},\ and\
  \citenamefont {Adesso}}]{ks15}%
  \BibitemOpen
  \bibfield  {author} {\bibinfo {author} {\bibfnamefont {I.}~\bibnamefont
  {Kogias}}, \bibinfo {author} {\bibfnamefont {P.}~\bibnamefont {Skrzypczyk}},
  \bibinfo {author} {\bibfnamefont {D.}~\bibnamefont {Cavalcanti}}, \bibinfo
  {author} {\bibfnamefont {A.}~\bibnamefont {Acin}},\ and\ \bibinfo {author}
  {\bibfnamefont {G.}~\bibnamefont {Adesso}},\ }\bibfield  {title} {\bibinfo
  {title} {Hierarchy of steering criteria based on moments for all bipartite
  quantum systems},\ }\href@noop {} {\bibfield  {journal} {\bibinfo  {journal}
  {Phys. Rev. Lett.}\ }\textbf {\bibinfo {volume} {115}},\ \bibinfo {pages}
  {210401} (\bibinfo {year} {2015}{\natexlab{b}})}\BibitemShut {NoStop}%
\bibitem [{\citenamefont {Gallego}\ and\ \citenamefont {Aolita}(2015)}]{ga15}%
  \BibitemOpen
  \bibfield  {author} {\bibinfo {author} {\bibfnamefont {R.}~\bibnamefont
  {Gallego}}\ and\ \bibinfo {author} {\bibfnamefont {L.}~\bibnamefont
  {Aolita}},\ }\bibfield  {title} {\bibinfo {title} {Resource theory of
  steering},\ }\href {https://doi.org/10.1103/PhysRevX.5.041008} {\bibfield
  {journal} {\bibinfo  {journal} {Phys. Rev. X}\ }\textbf {\bibinfo {volume}
  {5}},\ \bibinfo {pages} {041008} (\bibinfo {year} {2015})}\BibitemShut
  {NoStop}%
\bibitem [{\citenamefont {Quintino}\ \emph {et~al.}(2014)\citenamefont
  {Quintino}, \citenamefont {V\'ertesi},\ and\ \citenamefont {Brunner}}]{qv14}%
  \BibitemOpen
  \bibfield  {author} {\bibinfo {author} {\bibfnamefont {M.~T.}\ \bibnamefont
  {Quintino}}, \bibinfo {author} {\bibfnamefont {T.}~\bibnamefont
  {V\'ertesi}},\ and\ \bibinfo {author} {\bibfnamefont {N.}~\bibnamefont
  {Brunner}},\ }\bibfield  {title} {\bibinfo {title} {Joint measurability,
  einstein-podolsky-rosen steering, and bell nonlocality},\ }\href
  {https://doi.org/10.1103/PhysRevLett.113.160402} {\bibfield  {journal}
  {\bibinfo  {journal} {Phys. Rev. Lett.}\ }\textbf {\bibinfo {volume} {113}},\
  \bibinfo {pages} {160402} (\bibinfo {year} {2014})}\BibitemShut {NoStop}%
\bibitem [{\citenamefont {Uola}\ \emph {et~al.}(2015)\citenamefont {Uola},
  \citenamefont {Budroni}, \citenamefont {G\"uhne},\ and\ \citenamefont
  {Pellonp\"a\"a}}]{ub15}%
  \BibitemOpen
  \bibfield  {author} {\bibinfo {author} {\bibfnamefont {R.}~\bibnamefont
  {Uola}}, \bibinfo {author} {\bibfnamefont {C.}~\bibnamefont {Budroni}},
  \bibinfo {author} {\bibfnamefont {O.}~\bibnamefont {G\"uhne}},\ and\ \bibinfo
  {author} {\bibfnamefont {J.-P.}\ \bibnamefont {Pellonp\"a\"a}},\ }\bibfield
  {title} {\bibinfo {title} {One-to-one mapping between steering and joint
  measurability problems},\ }\href
  {https://doi.org/10.1103/PhysRevLett.115.230402} {\bibfield  {journal}
  {\bibinfo  {journal} {Phys. Rev. Lett.}\ }\textbf {\bibinfo {volume} {115}},\
  \bibinfo {pages} {230402} (\bibinfo {year} {2015})}\BibitemShut {NoStop}%
\bibitem [{\citenamefont {Cavalcanti}\ and\ \citenamefont
  {Skrzypczyk}(2016{\natexlab{b}})}]{cs16a}%
  \BibitemOpen
  \bibfield  {author} {\bibinfo {author} {\bibfnamefont {D.}~\bibnamefont
  {Cavalcanti}}\ and\ \bibinfo {author} {\bibfnamefont {P.}~\bibnamefont
  {Skrzypczyk}},\ }\bibfield  {title} {\bibinfo {title} {Quantitative relations
  between measurement incompatibility, quantum steering, and nonlocality},\
  }\href {https://doi.org/10.1103/PhysRevA.93.052112} {\bibfield  {journal}
  {\bibinfo  {journal} {Phys. Rev. A}\ }\textbf {\bibinfo {volume} {93}},\
  \bibinfo {pages} {052112} (\bibinfo {year} {2016}{\natexlab{b}})}\BibitemShut
  {NoStop}%
\bibitem [{\citenamefont {Kiukas}\ \emph {et~al.}(2017)\citenamefont {Kiukas},
  \citenamefont {Budroni}, \citenamefont {Uola},\ and\ \citenamefont
  {Pellonp\"a\"a}}]{kb17}%
  \BibitemOpen
  \bibfield  {author} {\bibinfo {author} {\bibfnamefont {J.}~\bibnamefont
  {Kiukas}}, \bibinfo {author} {\bibfnamefont {C.}~\bibnamefont {Budroni}},
  \bibinfo {author} {\bibfnamefont {R.}~\bibnamefont {Uola}},\ and\ \bibinfo
  {author} {\bibfnamefont {J.-P.}\ \bibnamefont {Pellonp\"a\"a}},\ }\bibfield
  {title} {\bibinfo {title} {Continuous-variable steering and incompatibility
  via state-channel duality},\ }\href
  {https://doi.org/10.1103/PhysRevA.96.042331} {\bibfield  {journal} {\bibinfo
  {journal} {Phys. Rev. A}\ }\textbf {\bibinfo {volume} {96}},\ \bibinfo
  {pages} {042331} (\bibinfo {year} {2017})}\BibitemShut {NoStop}%
\bibitem [{\citenamefont {Furry}(1936)}]{hf36}%
  \BibitemOpen
  \bibfield  {author} {\bibinfo {author} {\bibfnamefont {W.~H.}\ \bibnamefont
  {Furry}},\ }\bibfield  {title} {\bibinfo {title} {Note on the
  quantum-mechanical theory of measurement},\ }\href
  {https://doi.org/10.1103/PhysRev.49.393} {\bibfield  {journal} {\bibinfo
  {journal} {Phys. Rev.}\ }\textbf {\bibinfo {volume} {49}},\ \bibinfo {pages}
  {393} (\bibinfo {year} {1936})}\BibitemShut {NoStop}%
\bibitem [{\citenamefont {Gross}\ \emph {et~al.}(2010)\citenamefont {Gross},
  \citenamefont {Zibold}, \citenamefont {Nicklas}, \citenamefont {Est\`{e}ve},\
  and\ \citenamefont {Oberthaler}}]{gz10}%
  \BibitemOpen
  \bibfield  {author} {\bibinfo {author} {\bibfnamefont {C.}~\bibnamefont
  {Gross}}, \bibinfo {author} {\bibfnamefont {T.}~\bibnamefont {Zibold}},
  \bibinfo {author} {\bibfnamefont {E.}~\bibnamefont {Nicklas}}, \bibinfo
  {author} {\bibfnamefont {J.}~\bibnamefont {Est\`{e}ve}},\ and\ \bibinfo
  {author} {\bibfnamefont {M.~K.}\ \bibnamefont {Oberthaler}},\ }\bibfield
  {title} {\bibinfo {title} {Nonlinear atom interferometer surpasses classical
  precision limit},\ }\href {https://doi.org/10.1038/nature08919} {\bibfield
  {journal} {\bibinfo  {journal} {Nature}\ }\textbf {\bibinfo {volume} {464}},\
  \bibinfo {pages} {1165} (\bibinfo {year} {2010})}\BibitemShut {NoStop}%
\bibitem [{\citenamefont {Riedel}\ \emph {et~al.}(2010)\citenamefont {Riedel},
  \citenamefont {B\"{o}hi}, \citenamefont {Li}, \citenamefont {H\"{a}nsch},
  \citenamefont {Sinatra},\ and\ \citenamefont {Treutlein}}]{rb10}%
  \BibitemOpen
  \bibfield  {author} {\bibinfo {author} {\bibfnamefont {M.~F.}\ \bibnamefont
  {Riedel}}, \bibinfo {author} {\bibfnamefont {P.}~\bibnamefont {B\"{o}hi}},
  \bibinfo {author} {\bibfnamefont {Y.}~\bibnamefont {Li}}, \bibinfo {author}
  {\bibfnamefont {T.~W.}\ \bibnamefont {H\"{a}nsch}}, \bibinfo {author}
  {\bibfnamefont {A.}~\bibnamefont {Sinatra}},\ and\ \bibinfo {author}
  {\bibfnamefont {P.}~\bibnamefont {Treutlein}},\ }\bibfield  {title} {\bibinfo
  {title} {Atom-chip-based generation of entanglement for quantum metrology},\
  }\href {https://doi.org/10.1038/nature08988} {\bibfield  {journal} {\bibinfo
  {journal} {Nature}\ }\textbf {\bibinfo {volume} {464}},\ \bibinfo {pages}
  {1170} (\bibinfo {year} {2010})}\BibitemShut {NoStop}%
\bibitem [{\citenamefont {Schmied}\ \emph {et~al.}(2016)\citenamefont
  {Schmied}, \citenamefont {Bancal}, \citenamefont {Allard}, \citenamefont
  {Fadel}, \citenamefont {Scarani}, \citenamefont {Treutlein},\ and\
  \citenamefont {Sangouard}}]{sb16}%
  \BibitemOpen
  \bibfield  {author} {\bibinfo {author} {\bibfnamefont {R.}~\bibnamefont
  {Schmied}}, \bibinfo {author} {\bibfnamefont {J.-D.}\ \bibnamefont {Bancal}},
  \bibinfo {author} {\bibfnamefont {B.}~\bibnamefont {Allard}}, \bibinfo
  {author} {\bibfnamefont {M.}~\bibnamefont {Fadel}}, \bibinfo {author}
  {\bibfnamefont {V.}~\bibnamefont {Scarani}}, \bibinfo {author} {\bibfnamefont
  {P.}~\bibnamefont {Treutlein}},\ and\ \bibinfo {author} {\bibfnamefont
  {N.}~\bibnamefont {Sangouard}},\ }\bibfield  {title} {\bibinfo {title} {Bell
  correlations in a bose-einstein condensate},\ }\href
  {https://doi.org/10.1126/science.aad8665} {\bibfield  {journal} {\bibinfo
  {journal} {Science}\ }\textbf {\bibinfo {volume} {352}},\ \bibinfo {pages}
  {441} (\bibinfo {year} {2016})}\BibitemShut {NoStop}%
\bibitem [{\citenamefont {Ji}\ \emph {et~al.}(2016)\citenamefont {Ji},
  \citenamefont {Lee}, \citenamefont {Park},\ and\ \citenamefont {Nha}}]{jl16}%
  \BibitemOpen
  \bibfield  {author} {\bibinfo {author} {\bibfnamefont {S.-W.}\ \bibnamefont
  {Ji}}, \bibinfo {author} {\bibfnamefont {J.}~\bibnamefont {Lee}}, \bibinfo
  {author} {\bibfnamefont {J.}~\bibnamefont {Park}},\ and\ \bibinfo {author}
  {\bibfnamefont {H.}~\bibnamefont {Nha}},\ }\bibfield  {title} {\bibinfo
  {title} {Quantum steering of gaussian states via non-gaussian measurements},\
  }\href {https://doi.org/10.1038/srep29729} {\bibfield  {journal} {\bibinfo
  {journal} {Sci. Rep.}\ }\textbf {\bibinfo {volume} {6}},\ \bibinfo {pages}
  {29729} (\bibinfo {year} {2016})}\BibitemShut {NoStop}%
\bibitem [{\citenamefont {Liu}\ \emph {et~al.}(2002)\citenamefont {Liu},
  \citenamefont {Long}, \citenamefont {Tong},\ and\ \citenamefont {Li}}]{ll02}%
  \BibitemOpen
  \bibfield  {author} {\bibinfo {author} {\bibfnamefont {X.~S.}\ \bibnamefont
  {Liu}}, \bibinfo {author} {\bibfnamefont {G.~L.}\ \bibnamefont {Long}},
  \bibinfo {author} {\bibfnamefont {D.~M.}\ \bibnamefont {Tong}},\ and\
  \bibinfo {author} {\bibfnamefont {F.}~\bibnamefont {Li}},\ }\bibfield
  {title} {\bibinfo {title} {General scheme for superdense coding between
  multiparties},\ }\href {https://doi.org/10.1103/PhysRevA.65.022304}
  {\bibfield  {journal} {\bibinfo  {journal} {Phys. Rev. A}\ }\textbf {\bibinfo
  {volume} {65}},\ \bibinfo {pages} {022304} (\bibinfo {year}
  {2002})}\BibitemShut {NoStop}%
\bibitem [{\citenamefont {Erhard}\ \emph {et~al.}(2020)\citenamefont {Erhard},
  \citenamefont {Krenn},\ and\ \citenamefont {Zeilinger}}]{ek20}%
  \BibitemOpen
  \bibfield  {author} {\bibinfo {author} {\bibfnamefont {M.}~\bibnamefont
  {Erhard}}, \bibinfo {author} {\bibfnamefont {M.}~\bibnamefont {Krenn}},\ and\
  \bibinfo {author} {\bibfnamefont {A.}~\bibnamefont {Zeilinger}},\ }\bibfield
  {title} {\bibinfo {title} {Advances in high-dimensional quantum
  entanglement},\ }\href {https://doi.org/10.1038/s42254-020-0193-5} {\bibfield
   {journal} {\bibinfo  {journal} {Nat. Rev. Phys.}\ }\textbf {\bibinfo
  {volume} {2}},\ \bibinfo {pages} {365} (\bibinfo {year} {2020})}\BibitemShut
  {NoStop}%
\bibitem [{\citenamefont {Zhang}\ \emph
  {et~al.}(2019{\natexlab{b}})\citenamefont {Zhang}, \citenamefont {Dong},
  \citenamefont {Ding}, \citenamefont {Shi}, \citenamefont {Wang},
  \citenamefont {Liu}, \citenamefont {Zhou}, \citenamefont {Guo},\ and\
  \citenamefont {Shi}}]{zd19}%
  \BibitemOpen
  \bibfield  {author} {\bibinfo {author} {\bibfnamefont {W.}~\bibnamefont
  {Zhang}}, \bibinfo {author} {\bibfnamefont {M.-X.}\ \bibnamefont {Dong}},
  \bibinfo {author} {\bibfnamefont {D.-S.}\ \bibnamefont {Ding}}, \bibinfo
  {author} {\bibfnamefont {S.}~\bibnamefont {Shi}}, \bibinfo {author}
  {\bibfnamefont {K.}~\bibnamefont {Wang}}, \bibinfo {author} {\bibfnamefont
  {S.-L.}\ \bibnamefont {Liu}}, \bibinfo {author} {\bibfnamefont {Z.-Y.}\
  \bibnamefont {Zhou}}, \bibinfo {author} {\bibfnamefont {G.-C.}\ \bibnamefont
  {Guo}},\ and\ \bibinfo {author} {\bibfnamefont {B.-S.}\ \bibnamefont {Shi}},\
  }\bibfield  {title} {\bibinfo {title} {Einstein-podolsky-rosen entanglement
  between separated atomic ensembles},\ }\href@noop {} {\bibfield  {journal}
  {\bibinfo  {journal} {Phys. Rev. A}\ }\textbf {\bibinfo {volume} {100}},\
  \bibinfo {pages} {012347} (\bibinfo {year} {2019}{\natexlab{b}})}\BibitemShut
  {NoStop}%
\bibitem [{\citenamefont {Julsgaard}\ \emph {et~al.}(2001)\citenamefont
  {Julsgaard}, \citenamefont {Kozhekin},\ and\ \citenamefont {Polzik}}]{jk01}%
  \BibitemOpen
  \bibfield  {author} {\bibinfo {author} {\bibfnamefont {B.}~\bibnamefont
  {Julsgaard}}, \bibinfo {author} {\bibfnamefont {A.}~\bibnamefont
  {Kozhekin}},\ and\ \bibinfo {author} {\bibfnamefont {E.~S.}\ \bibnamefont
  {Polzik}},\ }\bibfield  {title} {\bibinfo {title} {Experimental long-lived
  entanglement of two macroscopic objects},\ }\href@noop {} {\bibfield
  {journal} {\bibinfo  {journal} {Nature}\ }\textbf {\bibinfo {volume} {413}},\
  \bibinfo {pages} {400} (\bibinfo {year} {2001})}\BibitemShut {NoStop}%
\bibitem [{\citenamefont {Krauter}\ \emph {et~al.}(2011)\citenamefont
  {Krauter}, \citenamefont {Muschik}, \citenamefont {Jensen}, \citenamefont
  {Wasilewski}, \citenamefont {Petersen}, \citenamefont {Cirac},\ and\
  \citenamefont {Polzik}}]{km11}%
  \BibitemOpen
  \bibfield  {author} {\bibinfo {author} {\bibfnamefont {H.}~\bibnamefont
  {Krauter}}, \bibinfo {author} {\bibfnamefont {C.~A.}\ \bibnamefont
  {Muschik}}, \bibinfo {author} {\bibfnamefont {K.}~\bibnamefont {Jensen}},
  \bibinfo {author} {\bibfnamefont {W.}~\bibnamefont {Wasilewski}}, \bibinfo
  {author} {\bibfnamefont {J.~M.}\ \bibnamefont {Petersen}}, \bibinfo {author}
  {\bibfnamefont {J.~I.}\ \bibnamefont {Cirac}},\ and\ \bibinfo {author}
  {\bibfnamefont {E.~S.}\ \bibnamefont {Polzik}},\ }\bibfield  {title}
  {\bibinfo {title} {Entanglement generated by dissipation and steady state
  entanglement of two macroscopic objects},\ }\href@noop {} {\bibfield
  {journal} {\bibinfo  {journal} {Phys. Rev. Lett.}\ }\textbf {\bibinfo
  {volume} {107}},\ \bibinfo {pages} {080503} (\bibinfo {year}
  {2011})}\BibitemShut {NoStop}%
\bibitem [{\citenamefont {Lange}\ \emph {et~al.}(2018)\citenamefont {Lange},
  \citenamefont {Peise}, \citenamefont {L\"ucke}, \citenamefont {Kruse},
  \citenamefont {Vitagliano}, \citenamefont {Apellaniz}, \citenamefont
  {Kleinmann}, \citenamefont {Toth},\ and\ \citenamefont {Klempt}}]{lp18}%
  \BibitemOpen
  \bibfield  {author} {\bibinfo {author} {\bibfnamefont {K.}~\bibnamefont
  {Lange}}, \bibinfo {author} {\bibfnamefont {J.}~\bibnamefont {Peise}},
  \bibinfo {author} {\bibfnamefont {B.}~\bibnamefont {L\"ucke}}, \bibinfo
  {author} {\bibfnamefont {I.}~\bibnamefont {Kruse}}, \bibinfo {author}
  {\bibfnamefont {G.}~\bibnamefont {Vitagliano}}, \bibinfo {author}
  {\bibfnamefont {I.}~\bibnamefont {Apellaniz}}, \bibinfo {author}
  {\bibfnamefont {M.}~\bibnamefont {Kleinmann}}, \bibinfo {author}
  {\bibfnamefont {G.}~\bibnamefont {Toth}},\ and\ \bibinfo {author}
  {\bibfnamefont {C.}~\bibnamefont {Klempt}},\ }\bibfield  {title} {\bibinfo
  {title} {Entanglement between two spatially separated atomic modes},\
  }\href@noop {} {\bibfield  {journal} {\bibinfo  {journal} {Science}\ }\textbf
  {\bibinfo {volume} {360}},\ \bibinfo {pages} {416} (\bibinfo {year}
  {2018})}\BibitemShut {NoStop}%
\bibitem [{\citenamefont {Dabrowski}\ \emph {et~al.}(2017)\citenamefont
  {Dabrowski}, \citenamefont {Parniak},\ and\ \citenamefont
  {Wasilewski}}]{dpw17}%
  \BibitemOpen
  \bibfield  {author} {\bibinfo {author} {\bibfnamefont {M.}~\bibnamefont
  {Dabrowski}}, \bibinfo {author} {\bibfnamefont {M.}~\bibnamefont {Parniak}},\
  and\ \bibinfo {author} {\bibfnamefont {W.}~\bibnamefont {Wasilewski}},\
  }\bibfield  {title} {\bibinfo {title} {Einstein--podolsky--rosen paradox in a
  hybrid bipartite system},\ }\href@noop {} {\bibfield  {journal} {\bibinfo
  {journal} {Optica}\ }\textbf {\bibinfo {volume} {4}},\ \bibinfo {pages} {272}
  (\bibinfo {year} {2017})}\BibitemShut {NoStop}%
\bibitem [{\citenamefont {Zheng}\ \emph {et~al.}(2021)\citenamefont {Zheng},
  \citenamefont {Sun}, \citenamefont {Yuan}, \citenamefont {Ficek},
  \citenamefont {Gong},\ and\ \citenamefont {He}}]{zsy21}%
  \BibitemOpen
  \bibfield  {author} {\bibinfo {author} {\bibfnamefont {S.~S.}\ \bibnamefont
  {Zheng}}, \bibinfo {author} {\bibfnamefont {F.~X.}\ \bibnamefont {Sun}},
  \bibinfo {author} {\bibfnamefont {H.~Y.}\ \bibnamefont {Yuan}}, \bibinfo
  {author} {\bibfnamefont {Z.}~\bibnamefont {Ficek}}, \bibinfo {author}
  {\bibfnamefont {Q.~H.}\ \bibnamefont {Gong}},\ and\ \bibinfo {author}
  {\bibfnamefont {Q.~Y.}\ \bibnamefont {He}},\ }\bibfield  {title} {\bibinfo
  {title} {Enhanced entanglement and asymmetric epr steering between magnons},\
  }\href@noop {} {\bibfield  {journal} {\bibinfo  {journal} {Sci. China Phys.
  Mech. Astron.}\ }\textbf {\bibinfo {volume} {64}},\ \bibinfo {pages} {210311}
  (\bibinfo {year} {2021})}\BibitemShut {NoStop}%
\bibitem [{\citenamefont {Sun}\ \emph {et~al.}(2021)\citenamefont {Sun},
  \citenamefont {Zheng}, \citenamefont {Xiao}, \citenamefont {Gong},
  \citenamefont {He},\ and\ \citenamefont {Xia}}]{sz21}%
  \BibitemOpen
  \bibfield  {author} {\bibinfo {author} {\bibfnamefont {F.-X.}\ \bibnamefont
  {Sun}}, \bibinfo {author} {\bibfnamefont {S.-S.}\ \bibnamefont {Zheng}},
  \bibinfo {author} {\bibfnamefont {Y.}~\bibnamefont {Xiao}}, \bibinfo {author}
  {\bibfnamefont {Q.}~\bibnamefont {Gong}}, \bibinfo {author} {\bibfnamefont
  {Q.}~\bibnamefont {He}},\ and\ \bibinfo {author} {\bibfnamefont
  {K.}~\bibnamefont {Xia}},\ }\bibfield  {title} {\bibinfo {title} {Remote
  generation of magnon schr\"odinger cat state via magnon-photon
  entanglement},\ }\href {https://doi.org/10.1103/PhysRevLett.127.087203}
  {\bibfield  {journal} {\bibinfo  {journal} {Phys. Rev. Lett.}\ }\textbf
  {\bibinfo {volume} {127}},\ \bibinfo {pages} {087203} (\bibinfo {year}
  {2021})}\BibitemShut {NoStop}%
\bibitem [{\citenamefont {Tan}(2019)}]{t19}%
  \BibitemOpen
  \bibfield  {author} {\bibinfo {author} {\bibfnamefont {H.}~\bibnamefont
  {Tan}},\ }\bibfield  {title} {\bibinfo {title} {Genuine photon-magnon-phonon
  einstein-podolsky-rosen steerable nonlocality in a continuously-monitored
  cavity magnomechanical system},\ }\href
  {https://doi.org/10.1103/PhysRevResearch.1.033161} {\bibfield  {journal}
  {\bibinfo  {journal} {Phys. Rev. Research}\ }\textbf {\bibinfo {volume}
  {1}},\ \bibinfo {pages} {033161} (\bibinfo {year} {2019})}\BibitemShut
  {NoStop}%
\bibitem [{\citenamefont {Tan}\ and\ \citenamefont {Li}(2021)}]{tl21}%
  \BibitemOpen
  \bibfield  {author} {\bibinfo {author} {\bibfnamefont {H.}~\bibnamefont
  {Tan}}\ and\ \bibinfo {author} {\bibfnamefont {J.}~\bibnamefont {Li}},\
  }\bibfield  {title} {\bibinfo {title} {Einstein-podolsky-rosen entanglement
  and asymmetric steering between distant macroscopic mechanical and magnonic
  systems},\ }\href@noop {} {\bibfield  {journal} {\bibinfo  {journal} {Phys.
  Rev. Research}\ }\textbf {\bibinfo {volume} {3}},\ \bibinfo {pages} {013192}
  (\bibinfo {year} {2021})}\BibitemShut {NoStop}%
\bibitem [{\citenamefont {Collett}\ \emph {et~al.}(1984)\citenamefont
  {Collett}, \citenamefont {Walls},\ and\ \citenamefont {Zoller}}]{cw84}%
  \BibitemOpen
  \bibfield  {author} {\bibinfo {author} {\bibfnamefont {M.~J.}\ \bibnamefont
  {Collett}}, \bibinfo {author} {\bibfnamefont {D.~F.}\ \bibnamefont {Walls}},\
  and\ \bibinfo {author} {\bibfnamefont {P.}~\bibnamefont {Zoller}},\
  }\bibfield  {title} {\bibinfo {title} {Spectrum of squeezing in resonance
  fluorescence},\ }\href@noop {} {\bibfield  {journal} {\bibinfo  {journal}
  {Optics Commun.}\ }\textbf {\bibinfo {volume} {52}},\ \bibinfo {pages} {145}
  (\bibinfo {year} {1984})}\BibitemShut {NoStop}%
\bibitem [{\citenamefont {Ou}\ \emph {et~al.}(1987)\citenamefont {Ou},
  \citenamefont {Hong},\ and\ \citenamefont {Mandel}}]{oh87}%
  \BibitemOpen
  \bibfield  {author} {\bibinfo {author} {\bibfnamefont {Z.~Y.}\ \bibnamefont
  {Ou}}, \bibinfo {author} {\bibfnamefont {C.~K.}\ \bibnamefont {Hong}},\ and\
  \bibinfo {author} {\bibfnamefont {L.}~\bibnamefont {Mandel}},\ }\bibfield
  {title} {\bibinfo {title} {Coherence properties of squeezed light and the
  degree of squeezing},\ }\href@noop {} {\bibfield  {journal} {\bibinfo
  {journal} {J. Opt. Soc. Am. B}\ }\textbf {\bibinfo {volume} {4}},\ \bibinfo
  {pages} {1574} (\bibinfo {year} {1987})}\BibitemShut {NoStop}%
\bibitem [{\citenamefont {Shen}\ \emph {et~al.}(2015)\citenamefont {Shen},
  \citenamefont {Assad}, \citenamefont {Grosse}, \citenamefont {Li},
  \citenamefont {Reid},\ and\ \citenamefont {Lam}}]{sa15}%
  \BibitemOpen
  \bibfield  {author} {\bibinfo {author} {\bibfnamefont {Y.}~\bibnamefont
  {Shen}}, \bibinfo {author} {\bibfnamefont {S.~M.}\ \bibnamefont {Assad}},
  \bibinfo {author} {\bibfnamefont {N.~B.}\ \bibnamefont {Grosse}}, \bibinfo
  {author} {\bibfnamefont {X.~Y.}\ \bibnamefont {Li}}, \bibinfo {author}
  {\bibfnamefont {M.~D.}\ \bibnamefont {Reid}},\ and\ \bibinfo {author}
  {\bibfnamefont {P.~K.}\ \bibnamefont {Lam}},\ }\bibfield  {title} {\bibinfo
  {title} {Nonlinear entanglement and its application to generating cat
  states},\ }\href {https://doi.org/10.1103/PhysRevLett.114.100403} {\bibfield
  {journal} {\bibinfo  {journal} {Phys. Rev. Lett.}\ }\textbf {\bibinfo
  {volume} {114}},\ \bibinfo {pages} {100403} (\bibinfo {year}
  {2015})}\BibitemShut {NoStop}%
\bibitem [{\citenamefont {Hong}\ and\ \citenamefont
  {Mandel}(1985{\natexlab{a}})}]{hma85}%
  \BibitemOpen
  \bibfield  {author} {\bibinfo {author} {\bibfnamefont {C.~K.}\ \bibnamefont
  {Hong}}\ and\ \bibinfo {author} {\bibfnamefont {L.}~\bibnamefont {Mandel}},\
  }\bibfield  {title} {\bibinfo {title} {Higher-order squeezing of a quantum
  field},\ }\href@noop {} {\bibfield  {journal} {\bibinfo  {journal} {Phys.
  Rev. Lett.}\ }\textbf {\bibinfo {volume} {54}},\ \bibinfo {pages} {323}
  (\bibinfo {year} {1985}{\natexlab{a}})}\BibitemShut {NoStop}%
\bibitem [{\citenamefont {Hong}\ and\ \citenamefont
  {Mandel}(1985{\natexlab{b}})}]{hmb85}%
  \BibitemOpen
  \bibfield  {author} {\bibinfo {author} {\bibfnamefont {C.~K.}\ \bibnamefont
  {Hong}}\ and\ \bibinfo {author} {\bibfnamefont {L.}~\bibnamefont {Mandel}},\
  }\bibfield  {title} {\bibinfo {title} {Generation of higher-order squeezing
  of quantum electromagnetic fields},\ }\href@noop {} {\bibfield  {journal}
  {\bibinfo  {journal} {Phys. Rev. A}\ }\textbf {\bibinfo {volume} {32}},\
  \bibinfo {pages} {974} (\bibinfo {year} {1985}{\natexlab{b}})}\BibitemShut
  {NoStop}%
\bibitem [{\citenamefont {Hillery}(1987)}]{mh87}%
  \BibitemOpen
  \bibfield  {author} {\bibinfo {author} {\bibfnamefont {M.}~\bibnamefont
  {Hillery}},\ }\bibfield  {title} {\bibinfo {title} {Amplitude-squared
  squeezing of the electromagnetic field},\ }\href@noop {} {\bibfield
  {journal} {\bibinfo  {journal} {Phys. Rev. A}\ }\textbf {\bibinfo {volume}
  {36}},\ \bibinfo {pages} {3796} (\bibinfo {year} {1987})}\BibitemShut
  {NoStop}%
\bibitem [{\citenamefont {Levenson}\ and\ \citenamefont {Shelby}(1987)}]{ls87}%
  \BibitemOpen
  \bibfield  {author} {\bibinfo {author} {\bibfnamefont {M.~D.}\ \bibnamefont
  {Levenson}}\ and\ \bibinfo {author} {\bibfnamefont {R.~M.}\ \bibnamefont
  {Shelby}},\ }\bibfield  {title} {\bibinfo {title} {Four-mode squeezing and
  applications},\ }\href@noop {} {\bibfield  {journal} {\bibinfo  {journal} {J.
  Mod. Opt.}\ }\textbf {\bibinfo {volume} {34}},\ \bibinfo {pages} {775}
  (\bibinfo {year} {1987})}\BibitemShut {NoStop}%
\bibitem [{\citenamefont {H\"{u}bel}\ \emph {et~al.}(2010)\citenamefont
  {H\"{u}bel}, \citenamefont {Hamel}, \citenamefont {Fedrizzi}, \citenamefont
  {Ramelow}, \citenamefont {J.Resch},\ and\ \citenamefont {Jennewein}}]{hh10}%
  \BibitemOpen
  \bibfield  {author} {\bibinfo {author} {\bibfnamefont {H.}~\bibnamefont
  {H\"{u}bel}}, \bibinfo {author} {\bibfnamefont {D.~R.}\ \bibnamefont
  {Hamel}}, \bibinfo {author} {\bibfnamefont {A.}~\bibnamefont {Fedrizzi}},
  \bibinfo {author} {\bibfnamefont {S.}~\bibnamefont {Ramelow}}, \bibinfo
  {author} {\bibfnamefont {K.}~\bibnamefont {J.Resch}},\ and\ \bibinfo {author}
  {\bibfnamefont {T.}~\bibnamefont {Jennewein}},\ }\bibfield  {title} {\bibinfo
  {title} {Direct generation of photon triplets using cascaded photon-pair
  sources},\ }\href@noop {} {\bibfield  {journal} {\bibinfo  {journal}
  {Nature}\ }\textbf {\bibinfo {volume} {466}},\ \bibinfo {pages} {601}
  (\bibinfo {year} {2010})}\BibitemShut {NoStop}%
\bibitem [{\citenamefont {Stobi\'nska}\ \emph {et~al.}(2013)\citenamefont
  {Stobi\'nska}, \citenamefont {Laskowski}, \citenamefont {Wie\'sniak},\ and\
  \citenamefont {Zukowski}}]{sl13}%
  \BibitemOpen
  \bibfield  {author} {\bibinfo {author} {\bibfnamefont {M.}~\bibnamefont
  {Stobi\'nska}}, \bibinfo {author} {\bibfnamefont {W.}~\bibnamefont
  {Laskowski}}, \bibinfo {author} {\bibfnamefont {M.}~\bibnamefont
  {Wie\'sniak}},\ and\ \bibinfo {author} {\bibfnamefont {M.}~\bibnamefont
  {Zukowski}},\ }\bibfield  {title} {\bibinfo {title} {Multiphoton quantum
  interference with high visibility using multiport beam splitters},\
  }\href@noop {} {\bibfield  {journal} {\bibinfo  {journal} {Phys. Rev. A}\
  }\textbf {\bibinfo {volume} {87}},\ \bibinfo {pages} {053828} (\bibinfo
  {year} {2013})}\BibitemShut {NoStop}%
\bibitem [{\citenamefont {Walschaers}\ and\ \citenamefont
  {Treps}(2020)}]{wt20}%
  \BibitemOpen
  \bibfield  {author} {\bibinfo {author} {\bibfnamefont {M.}~\bibnamefont
  {Walschaers}}\ and\ \bibinfo {author} {\bibfnamefont {N.}~\bibnamefont
  {Treps}},\ }\bibfield  {title} {\bibinfo {title} {Remote generation of wigner
  negativity through einstein-podolsky-rosen steering},\ }\href@noop {}
  {\bibfield  {journal} {\bibinfo  {journal} {Phys. Rev. Lett.}\ }\textbf
  {\bibinfo {volume} {124}},\ \bibinfo {pages} {150501} (\bibinfo {year}
  {2020})}\BibitemShut {NoStop}%
\bibitem [{\citenamefont {Walschaers}\ \emph {et~al.}(2020)\citenamefont
  {Walschaers}, \citenamefont {Parigi},\ and\ \citenamefont {Treps}}]{wp20}%
  \BibitemOpen
  \bibfield  {author} {\bibinfo {author} {\bibfnamefont {M.}~\bibnamefont
  {Walschaers}}, \bibinfo {author} {\bibfnamefont {V.}~\bibnamefont {Parigi}},\
  and\ \bibinfo {author} {\bibfnamefont {N.}~\bibnamefont {Treps}},\ }\bibfield
   {title} {\bibinfo {title} {Practical framework for conditional non-gaussian
  quantum state preparation},\ }\href@noop {} {\bibfield  {journal} {\bibinfo
  {journal} {Phys. Rev. X Quantum}\ }\textbf {\bibinfo {volume} {1}},\ \bibinfo
  {pages} {020305} (\bibinfo {year} {2020})}\BibitemShut {NoStop}%
\bibitem [{\citenamefont {Ra}\ \emph {et~al.}(2020)\citenamefont {Ra},
  \citenamefont {Dufour}, \citenamefont {Walschaers}, \citenamefont {Jacquard},
  \citenamefont {Michel}, \citenamefont {Fabre},\ and\ \citenamefont
  {Treps}}]{rd20}%
  \BibitemOpen
  \bibfield  {author} {\bibinfo {author} {\bibfnamefont {Y.~S.}\ \bibnamefont
  {Ra}}, \bibinfo {author} {\bibfnamefont {A.}~\bibnamefont {Dufour}}, \bibinfo
  {author} {\bibfnamefont {M.}~\bibnamefont {Walschaers}}, \bibinfo {author}
  {\bibfnamefont {C.}~\bibnamefont {Jacquard}}, \bibinfo {author}
  {\bibfnamefont {T.}~\bibnamefont {Michel}}, \bibinfo {author} {\bibfnamefont
  {C.}~\bibnamefont {Fabre}},\ and\ \bibinfo {author} {\bibfnamefont
  {N.}~\bibnamefont {Treps}},\ }\bibfield  {title} {\bibinfo {title}
  {Non-gaussian quantum states of a multimode light field},\ }\href@noop {}
  {\bibfield  {journal} {\bibinfo  {journal} {Nat. Phys.}\ }\textbf {\bibinfo
  {volume} {16}},\ \bibinfo {pages} {144} (\bibinfo {year} {2020})}\BibitemShut
  {NoStop}%
\bibitem [{\citenamefont {Chang}\ \emph {et~al.}(2020)\citenamefont {Chang},
  \citenamefont {Sab\'{i}n}, \citenamefont {Forn-D\'{i}az}, \citenamefont
  {Quijandr\'{i}a}, \citenamefont {Vadiraj}, \citenamefont {Nsanzineza},
  \citenamefont {Johansson},\ and\ \citenamefont {Wilson}}]{cs20}%
  \BibitemOpen
  \bibfield  {author} {\bibinfo {author} {\bibfnamefont {C.~W.~S.}\
  \bibnamefont {Chang}}, \bibinfo {author} {\bibfnamefont {C.}~\bibnamefont
  {Sab\'{i}n}}, \bibinfo {author} {\bibfnamefont {P.}~\bibnamefont
  {Forn-D\'{i}az}}, \bibinfo {author} {\bibfnamefont {F.}~\bibnamefont
  {Quijandr\'{i}a}}, \bibinfo {author} {\bibfnamefont {A.~M.}\ \bibnamefont
  {Vadiraj}}, \bibinfo {author} {\bibfnamefont {I.}~\bibnamefont {Nsanzineza}},
  \bibinfo {author} {\bibfnamefont {G.}~\bibnamefont {Johansson}},\ and\
  \bibinfo {author} {\bibfnamefont {C.~M.}\ \bibnamefont {Wilson}},\ }\bibfield
   {title} {\bibinfo {title} {Observation of three-photon spontaneous
  parametric down-conversion in a superconducting parametric cavity},\
  }\href@noop {} {\bibfield  {journal} {\bibinfo  {journal} {Phys. Rev. X}\
  }\textbf {\bibinfo {volume} {10}},\ \bibinfo {pages} {011011} (\bibinfo
  {year} {2020})}\BibitemShut {NoStop}%
\bibitem [{\citenamefont {Guo}\ \emph {et~al.}(2021)\citenamefont {Guo},
  \citenamefont {Sun}, \citenamefont {Zhu}, \citenamefont {Gessner},
  \citenamefont {He},\ and\ \citenamefont {Fadel}}]{gs21}%
  \BibitemOpen
  \bibfield  {author} {\bibinfo {author} {\bibfnamefont {J.}~\bibnamefont
  {Guo}}, \bibinfo {author} {\bibfnamefont {F.-X.}\ \bibnamefont {Sun}},
  \bibinfo {author} {\bibfnamefont {D.}~\bibnamefont {Zhu}}, \bibinfo {author}
  {\bibfnamefont {M.}~\bibnamefont {Gessner}}, \bibinfo {author} {\bibfnamefont
  {Q.}~\bibnamefont {He}},\ and\ \bibinfo {author} {\bibfnamefont
  {M.}~\bibnamefont {Fadel}},\ }\bibfield  {title} {\bibinfo {title} {Detecting
  einstein-podolsky-rosen steering in non-gaussian spin states from conditional
  spin-squeezing parameters},\ }\href {https://arxiv.org/abs/2106.13106}
  {\bibfield  {journal} {\bibinfo  {journal} {arXiv:2106.13106}\ } (\bibinfo
  {year} {2021})}\BibitemShut {NoStop}%
\bibitem [{\citenamefont {Lopetegui}\ \emph {et~al.}(2022)\citenamefont
  {Lopetegui}, \citenamefont {Gessner}, \citenamefont {Fadel}, \citenamefont
  {Treps},\ and\ \citenamefont {Walschaers}}]{lg22}%
  \BibitemOpen
  \bibfield  {author} {\bibinfo {author} {\bibfnamefont {C.~E.}\ \bibnamefont
  {Lopetegui}}, \bibinfo {author} {\bibfnamefont {M.}~\bibnamefont {Gessner}},
  \bibinfo {author} {\bibfnamefont {M.}~\bibnamefont {Fadel}}, \bibinfo
  {author} {\bibfnamefont {N.}~\bibnamefont {Treps}},\ and\ \bibinfo {author}
  {\bibfnamefont {M.}~\bibnamefont {Walschaers}},\ }\bibfield  {title}
  {\bibinfo {title} {Homodyne detection of non-gaussian quantum steering},\
  }\href@noop {} {\bibfield  {journal} {\bibinfo  {journal} {arXiv:2201.11439}\
  } (\bibinfo {year} {2022})}\BibitemShut {NoStop}%
\bibitem [{\citenamefont {Mari}\ and\ \citenamefont {Eisert}(2012)}]{me12}%
  \BibitemOpen
  \bibfield  {author} {\bibinfo {author} {\bibfnamefont {A.}~\bibnamefont
  {Mari}}\ and\ \bibinfo {author} {\bibfnamefont {J.}~\bibnamefont {Eisert}},\
  }\bibfield  {title} {\bibinfo {title} {Positive wigner functions render
  classical simulation of quantum computation efficient},\ }\href
  {https://doi.org/10.1103/PhysRevLett.109.230503} {\bibfield  {journal}
  {\bibinfo  {journal} {Phys. Rev. Lett.}\ }\textbf {\bibinfo {volume} {109}},\
  \bibinfo {pages} {230503} (\bibinfo {year} {2012})}\BibitemShut {NoStop}%
\bibitem [{\citenamefont {Xiang}\ \emph {et~al.}(2022)\citenamefont {Xiang},
  \citenamefont {Liu}, \citenamefont {Guo}, \citenamefont {Gong}, \citenamefont
  {Treps}, \citenamefont {He},\ and\ \citenamefont {Walschaers}}]{xl21}%
  \BibitemOpen
  \bibfield  {author} {\bibinfo {author} {\bibfnamefont {Y.}~\bibnamefont
  {Xiang}}, \bibinfo {author} {\bibfnamefont {S.}~\bibnamefont {Liu}}, \bibinfo
  {author} {\bibfnamefont {J.}~\bibnamefont {Guo}}, \bibinfo {author}
  {\bibfnamefont {Q.}~\bibnamefont {Gong}}, \bibinfo {author} {\bibfnamefont
  {N.}~\bibnamefont {Treps}}, \bibinfo {author} {\bibfnamefont {Q.~Y.}\
  \bibnamefont {He}},\ and\ \bibinfo {author} {\bibfnamefont {M.}~\bibnamefont
  {Walschaers}},\ }\bibfield  {title} {\bibinfo {title} {Distribution and
  quantification of remotely generated wigner negativity},\ }\href@noop {}
  {\bibfield  {journal} {\bibinfo  {journal} {npj Quantum Information}\
  }\textbf {\bibinfo {volume} {8}},\ \bibinfo {pages} {21} (\bibinfo {year}
  {2022})}\BibitemShut {NoStop}%
\bibitem [{\citenamefont {Parkins}\ \emph {et~al.}(2006)\citenamefont
  {Parkins}, \citenamefont {Solano},\ and\ \citenamefont {Cirac}}]{ps06}%
  \BibitemOpen
  \bibfield  {author} {\bibinfo {author} {\bibfnamefont {A.~S.}\ \bibnamefont
  {Parkins}}, \bibinfo {author} {\bibfnamefont {E.}~\bibnamefont {Solano}},\
  and\ \bibinfo {author} {\bibfnamefont {J.~I.}\ \bibnamefont {Cirac}},\
  }\bibfield  {title} {\bibinfo {title} {Unconditional two-mode squeezing of
  separated atomic ensembles},\ }\href@noop {} {\bibfield  {journal} {\bibinfo
  {journal} {Phys. Rev. Lett.}\ }\textbf {\bibinfo {volume} {96}},\ \bibinfo
  {pages} {053602} (\bibinfo {year} {2006})}\BibitemShut {NoStop}%
\bibitem [{\citenamefont {Sun}\ \emph {et~al.}(2011)\citenamefont {Sun},
  \citenamefont {Li}, \citenamefont {Gu},\ and\ \citenamefont {Ficek}}]{sl11}%
  \BibitemOpen
  \bibfield  {author} {\bibinfo {author} {\bibfnamefont {L.~H.}\ \bibnamefont
  {Sun}}, \bibinfo {author} {\bibfnamefont {G.~X.}\ \bibnamefont {Li}},
  \bibinfo {author} {\bibfnamefont {W.~J.}\ \bibnamefont {Gu}},\ and\ \bibinfo
  {author} {\bibfnamefont {Z.}~\bibnamefont {Ficek}},\ }\bibfield  {title}
  {\bibinfo {title} {Generating coherence and entanglement with a finite-size
  atomic ensemble in a ring cavity},\ }\href@noop {} {\bibfield  {journal}
  {\bibinfo  {journal} {New J. Phys.}\ }\textbf {\bibinfo {volume} {13}},\
  \bibinfo {pages} {093019} (\bibinfo {year} {2011})}\BibitemShut {NoStop}%
\bibitem [{\citenamefont {Yu}\ and\ \citenamefont {Eberly}(2004)}]{ye04}%
  \BibitemOpen
  \bibfield  {author} {\bibinfo {author} {\bibfnamefont {T.}~\bibnamefont
  {Yu}}\ and\ \bibinfo {author} {\bibfnamefont {J.~H.}\ \bibnamefont
  {Eberly}},\ }\bibfield  {title} {\bibinfo {title} {Finite-time
  disentanglement via spontaneous emission},\ }\href
  {https://doi.org/10.1103/PhysRevLett.93.140404} {\bibfield  {journal}
  {\bibinfo  {journal} {Phys. Rev. Lett.}\ }\textbf {\bibinfo {volume} {93}},\
  \bibinfo {pages} {140404} (\bibinfo {year} {2004})}\BibitemShut {NoStop}%
\bibitem [{\citenamefont {Ficek}\ and\ \citenamefont {Tana\'{s}}(2002)}]{ft02}%
  \BibitemOpen
  \bibfield  {author} {\bibinfo {author} {\bibfnamefont {Z.}~\bibnamefont
  {Ficek}}\ and\ \bibinfo {author} {\bibfnamefont {R.}~\bibnamefont
  {Tana\'{s}}},\ }\bibfield  {title} {\bibinfo {title} {Entangled states and
  collective nonclassical effects in two-atom systems},\ }\href
  {https://doi.org/https://doi.org/10.1016/S0370-1573(02)00368-X} {\bibfield
  {journal} {\bibinfo  {journal} {Phys. Reports}\ }\textbf {\bibinfo {volume}
  {372}},\ \bibinfo {pages} {369} (\bibinfo {year} {2002})}\BibitemShut
  {NoStop}%
\bibitem [{\citenamefont {Gardiner}(1993)}]{cwg93}%
  \BibitemOpen
  \bibfield  {author} {\bibinfo {author} {\bibfnamefont {C.~W.}\ \bibnamefont
  {Gardiner}},\ }\bibfield  {title} {\bibinfo {title} {Driving a quantum system
  with the output field from another driven quantum system},\ }\href@noop {}
  {\bibfield  {journal} {\bibinfo  {journal} {Phys. Rev. Lett.}\ }\textbf
  {\bibinfo {volume} {70}},\ \bibinfo {pages} {2269} (\bibinfo {year}
  {1993})}\BibitemShut {NoStop}%
\bibitem [{\citenamefont {Carmichael}(1993)}]{hjc93}%
  \BibitemOpen
  \bibfield  {author} {\bibinfo {author} {\bibfnamefont {H.~J.}\ \bibnamefont
  {Carmichael}},\ }\bibfield  {title} {\bibinfo {title} {Quantum trajectory
  theory for cascaded open systems},\ }\href@noop {} {\bibfield  {journal}
  {\bibinfo  {journal} {Phys. Rev. Lett.}\ }\textbf {\bibinfo {volume} {70}},\
  \bibinfo {pages} {2273} (\bibinfo {year} {1993})}\BibitemShut {NoStop}%
\bibitem [{\citenamefont {Brennecke}\ \emph {et~al.}(2008)\citenamefont
  {Brennecke}, \citenamefont {Ritter}, \citenamefont {Donner},\ and\
  \citenamefont {Esslinger}}]{br08}%
  \BibitemOpen
  \bibfield  {author} {\bibinfo {author} {\bibfnamefont {F.}~\bibnamefont
  {Brennecke}}, \bibinfo {author} {\bibfnamefont {S.}~\bibnamefont {Ritter}},
  \bibinfo {author} {\bibfnamefont {T.}~\bibnamefont {Donner}},\ and\ \bibinfo
  {author} {\bibfnamefont {T.}~\bibnamefont {Esslinger}},\ }\bibfield  {title}
  {\bibinfo {title} {Cavity optomechanics with a bose-einstein condensate},\
  }\href@noop {} {\bibfield  {journal} {\bibinfo  {journal} {Science}\ }\textbf
  {\bibinfo {volume} {322}},\ \bibinfo {pages} {541} (\bibinfo {year}
  {2008})}\BibitemShut {NoStop}%
\bibitem [{\citenamefont {O'Connell}\ \emph {et~al.}(2010)\citenamefont
  {O'Connell}, \citenamefont {Hofheinz}, \citenamefont {Ansmann}, \citenamefont
  {Bialczak}, \citenamefont {Lenander}, \citenamefont {Lucero}, \citenamefont
  {Neeley}, \citenamefont {Sank}, \citenamefont {Wang}, \citenamefont {Weides},
  \citenamefont {Wenner}, \citenamefont {Martinis},\ and\ \citenamefont
  {Cleland}}]{oh10}%
  \BibitemOpen
  \bibfield  {author} {\bibinfo {author} {\bibfnamefont {A.~D.}\ \bibnamefont
  {O'Connell}}, \bibinfo {author} {\bibfnamefont {M.}~\bibnamefont {Hofheinz}},
  \bibinfo {author} {\bibfnamefont {M.}~\bibnamefont {Ansmann}}, \bibinfo
  {author} {\bibfnamefont {R.~C.}\ \bibnamefont {Bialczak}}, \bibinfo {author}
  {\bibfnamefont {M.}~\bibnamefont {Lenander}}, \bibinfo {author}
  {\bibfnamefont {E.}~\bibnamefont {Lucero}}, \bibinfo {author} {\bibfnamefont
  {M.}~\bibnamefont {Neeley}}, \bibinfo {author} {\bibfnamefont
  {D.}~\bibnamefont {Sank}}, \bibinfo {author} {\bibfnamefont {H.}~\bibnamefont
  {Wang}}, \bibinfo {author} {\bibfnamefont {M.}~\bibnamefont {Weides}},
  \bibinfo {author} {\bibfnamefont {J.}~\bibnamefont {Wenner}}, \bibinfo
  {author} {\bibfnamefont {J.~M.}\ \bibnamefont {Martinis}},\ and\ \bibinfo
  {author} {\bibfnamefont {A.~N.}\ \bibnamefont {Cleland}},\ }\bibfield
  {title} {\bibinfo {title} {Quantum ground state and single-phonon control of
  a mechanical resonator},\ }\href@noop {} {\bibfield  {journal} {\bibinfo
  {journal} {Nature}\ }\textbf {\bibinfo {volume} {464}},\ \bibinfo {pages}
  {697} (\bibinfo {year} {2010})}\BibitemShut {NoStop}%
\bibitem [{\citenamefont {Chan}\ \emph {et~al.}(2011)\citenamefont {Chan},
  \citenamefont {Alegre}, \citenamefont {Safavi-Naeini}, \citenamefont {Hill},
  \citenamefont {Krause}, \citenamefont {Gr{\"o}blacher}, \citenamefont
  {Aspelmeyer},\ and\ \citenamefont {Painter}}]{ca11}%
  \BibitemOpen
  \bibfield  {author} {\bibinfo {author} {\bibfnamefont {J.}~\bibnamefont
  {Chan}}, \bibinfo {author} {\bibfnamefont {T.~M.}\ \bibnamefont {Alegre}},
  \bibinfo {author} {\bibfnamefont {A.~H.}\ \bibnamefont {Safavi-Naeini}},
  \bibinfo {author} {\bibfnamefont {J.~T.}\ \bibnamefont {Hill}}, \bibinfo
  {author} {\bibfnamefont {A.}~\bibnamefont {Krause}}, \bibinfo {author}
  {\bibfnamefont {S.}~\bibnamefont {Gr{\"o}blacher}}, \bibinfo {author}
  {\bibfnamefont {M.}~\bibnamefont {Aspelmeyer}},\ and\ \bibinfo {author}
  {\bibfnamefont {O.}~\bibnamefont {Painter}},\ }\bibfield  {title} {\bibinfo
  {title} {Laser cooling of a nanomechanical oscillator into its quantum ground
  state},\ }\href@noop {} {\bibfield  {journal} {\bibinfo  {journal} {Nature}\
  }\textbf {\bibinfo {volume} {478}},\ \bibinfo {pages} {89} (\bibinfo {year}
  {2011})}\BibitemShut {NoStop}%
\bibitem [{\citenamefont {Teufel}\ \emph {et~al.}(2011)\citenamefont {Teufel},
  \citenamefont {Donner}, \citenamefont {Li}, \citenamefont {Harlow},
  \citenamefont {Allman}, \citenamefont {Cicak}, \citenamefont {Sirois},
  \citenamefont {Whittaker}, \citenamefont {Lehnert},\ and\ \citenamefont
  {Simmonds}}]{td11}%
  \BibitemOpen
  \bibfield  {author} {\bibinfo {author} {\bibfnamefont {J.~D.}\ \bibnamefont
  {Teufel}}, \bibinfo {author} {\bibfnamefont {T.}~\bibnamefont {Donner}},
  \bibinfo {author} {\bibfnamefont {D.}~\bibnamefont {Li}}, \bibinfo {author}
  {\bibfnamefont {J.~W.}\ \bibnamefont {Harlow}}, \bibinfo {author}
  {\bibfnamefont {M.}~\bibnamefont {Allman}}, \bibinfo {author} {\bibfnamefont
  {K.}~\bibnamefont {Cicak}}, \bibinfo {author} {\bibfnamefont {A.~J.}\
  \bibnamefont {Sirois}}, \bibinfo {author} {\bibfnamefont {J.~D.}\
  \bibnamefont {Whittaker}}, \bibinfo {author} {\bibfnamefont {K.~W.}\
  \bibnamefont {Lehnert}},\ and\ \bibinfo {author} {\bibfnamefont {R.~W.}\
  \bibnamefont {Simmonds}},\ }\bibfield  {title} {\bibinfo {title} {Sideband
  cooling of micromechanical motion to the quantum ground state},\ }\href@noop
  {} {\bibfield  {journal} {\bibinfo  {journal} {Nature}\ }\textbf {\bibinfo
  {volume} {475}},\ \bibinfo {pages} {359} (\bibinfo {year}
  {2011})}\BibitemShut {NoStop}%
\bibitem [{\citenamefont {Kotler}\ \emph {et~al.}(2021)\citenamefont {Kotler},
  \citenamefont {Peterson}, \citenamefont {Shojaee}, \citenamefont {Lecocq},
  \citenamefont {Cicak}, \citenamefont {Kwiatkowski}, \citenamefont {Geller},
  \citenamefont {Glancy}, \citenamefont {Knill}, \citenamefont {Simmonds},
  \citenamefont {Aumentado},\ and\ \citenamefont {Teufel}}]{kp21}%
  \BibitemOpen
  \bibfield  {author} {\bibinfo {author} {\bibfnamefont {S.}~\bibnamefont
  {Kotler}}, \bibinfo {author} {\bibfnamefont {G.~A.}\ \bibnamefont
  {Peterson}}, \bibinfo {author} {\bibfnamefont {E.}~\bibnamefont {Shojaee}},
  \bibinfo {author} {\bibfnamefont {F.}~\bibnamefont {Lecocq}}, \bibinfo
  {author} {\bibfnamefont {K.}~\bibnamefont {Cicak}}, \bibinfo {author}
  {\bibfnamefont {A.}~\bibnamefont {Kwiatkowski}}, \bibinfo {author}
  {\bibfnamefont {S.}~\bibnamefont {Geller}}, \bibinfo {author} {\bibfnamefont
  {S.}~\bibnamefont {Glancy}}, \bibinfo {author} {\bibfnamefont
  {E.}~\bibnamefont {Knill}}, \bibinfo {author} {\bibfnamefont {R.~W.}\
  \bibnamefont {Simmonds}}, \bibinfo {author} {\bibfnamefont {J.}~\bibnamefont
  {Aumentado}},\ and\ \bibinfo {author} {\bibfnamefont {J.~D.}\ \bibnamefont
  {Teufel}},\ }\bibfield  {title} {\bibinfo {title} {Direct observation of
  deterministic macroscopic entanglement},\ }\href@noop {} {\bibfield
  {journal} {\bibinfo  {journal} {Science}\ }\textbf {\bibinfo {volume}
  {372}},\ \bibinfo {pages} {622} (\bibinfo {year} {2021})}\BibitemShut
  {NoStop}%
\bibitem [{\citenamefont {de~L{\'e}pinay}\ \emph {et~al.}(2021)\citenamefont
  {de~L{\'e}pinay}, \citenamefont {Ockeloen-Korppi}, \citenamefont {Woolley},\
  and\ \citenamefont {Sillanp{\"a}{\"a}}}]{do21}%
  \BibitemOpen
  \bibfield  {author} {\bibinfo {author} {\bibfnamefont {L.~M.}\ \bibnamefont
  {de~L{\'e}pinay}}, \bibinfo {author} {\bibfnamefont {C.~F.}\ \bibnamefont
  {Ockeloen-Korppi}}, \bibinfo {author} {\bibfnamefont {M.~J.}\ \bibnamefont
  {Woolley}},\ and\ \bibinfo {author} {\bibfnamefont {M.~A.}\ \bibnamefont
  {Sillanp{\"a}{\"a}}},\ }\bibfield  {title} {\bibinfo {title} {Quantum
  mechanics--free subsystem with mechanical oscillators},\ }\href@noop {}
  {\bibfield  {journal} {\bibinfo  {journal} {Science}\ }\textbf {\bibinfo
  {volume} {372}},\ \bibinfo {pages} {625} (\bibinfo {year}
  {2021})}\BibitemShut {NoStop}%
\bibitem [{\citenamefont {Thomas}\ \emph {et~al.}(2021)\citenamefont {Thomas},
  \citenamefont {Parniak}, \citenamefont {\O{}stfeldt}, \citenamefont
  {M\o{}ller}, \citenamefont {B\ae{}rentsen}, \citenamefont {Tsaturyan},
  \citenamefont {Schliesser}, \citenamefont {J.~Appel},\ and\ \citenamefont
  {Polzik}}]{tp21}%
  \BibitemOpen
  \bibfield  {author} {\bibinfo {author} {\bibfnamefont {R.~A.}\ \bibnamefont
  {Thomas}}, \bibinfo {author} {\bibfnamefont {M.}~\bibnamefont {Parniak}},
  \bibinfo {author} {\bibfnamefont {C.}~\bibnamefont {\O{}stfeldt}}, \bibinfo
  {author} {\bibfnamefont {C.~B.}\ \bibnamefont {M\o{}ller}}, \bibinfo {author}
  {\bibfnamefont {C.}~\bibnamefont {B\ae{}rentsen}}, \bibinfo {author}
  {\bibfnamefont {Y.}~\bibnamefont {Tsaturyan}}, \bibinfo {author}
  {\bibfnamefont {A.}~\bibnamefont {Schliesser}}, \bibinfo {author}
  {\bibfnamefont {E.~Z.}\ \bibnamefont {J.~Appel}},\ and\ \bibinfo {author}
  {\bibfnamefont {E.~S.}\ \bibnamefont {Polzik}},\ }\bibfield  {title}
  {\bibinfo {title} {Entanglement between distant macroscopic mechanical and
  spin systems},\ }\href {https://doi.org/10.1088/1751-8113/49/34/34lt02}
  {\bibfield  {journal} {\bibinfo  {journal} {Nat. Phys.}\ }\textbf {\bibinfo
  {volume} {17}},\ \bibinfo {pages} {228} (\bibinfo {year} {2021})}\BibitemShut
  {NoStop}%
\bibitem [{\citenamefont {Hofer}\ \emph {et~al.}(2011)\citenamefont {Hofer},
  \citenamefont {Wieczorek}, \citenamefont {Aspelmeyer},\ and\ \citenamefont
  {Hammerer}}]{hw11}%
  \BibitemOpen
  \bibfield  {author} {\bibinfo {author} {\bibfnamefont {S.~G.}\ \bibnamefont
  {Hofer}}, \bibinfo {author} {\bibfnamefont {W.}~\bibnamefont {Wieczorek}},
  \bibinfo {author} {\bibfnamefont {M.}~\bibnamefont {Aspelmeyer}},\ and\
  \bibinfo {author} {\bibfnamefont {K.}~\bibnamefont {Hammerer}},\ }\bibfield
  {title} {\bibinfo {title} {Quantum entanglement and teleportation in pulsed
  cavity optomechanics},\ }\href {https://doi.org/10.1103/PhysRevA.84.052327}
  {\bibfield  {journal} {\bibinfo  {journal} {Phys. Rev. A}\ }\textbf {\bibinfo
  {volume} {84}},\ \bibinfo {pages} {052327} (\bibinfo {year}
  {2011})}\BibitemShut {NoStop}%
\bibitem [{\citenamefont {Andersen}\ \emph {et~al.}(2015)\citenamefont
  {Andersen}, \citenamefont {Neergaard-Nielsen}, \citenamefont {Van~Loock},\
  and\ \citenamefont {Furusawa}}]{an15}%
  \BibitemOpen
  \bibfield  {author} {\bibinfo {author} {\bibfnamefont {U.~L.}\ \bibnamefont
  {Andersen}}, \bibinfo {author} {\bibfnamefont {J.~S.}\ \bibnamefont
  {Neergaard-Nielsen}}, \bibinfo {author} {\bibfnamefont {P.}~\bibnamefont
  {Van~Loock}},\ and\ \bibinfo {author} {\bibfnamefont {A.}~\bibnamefont
  {Furusawa}},\ }\bibfield  {title} {\bibinfo {title} {Hybrid discrete-and
  continuous-variable quantum information},\ }\href@noop {} {\bibfield
  {journal} {\bibinfo  {journal} {Nat. Phys.}\ }\textbf {\bibinfo {volume}
  {11}},\ \bibinfo {pages} {713} (\bibinfo {year} {2015})}\BibitemShut
  {NoStop}%
\bibitem [{\citenamefont {Kraft}\ \emph {et~al.}(2018)\citenamefont {Kraft},
  \citenamefont {Ritz}, \citenamefont {Brunner}, \citenamefont {Huber},\ and\
  \citenamefont {G\"uhne}}]{kr18}%
  \BibitemOpen
  \bibfield  {author} {\bibinfo {author} {\bibfnamefont {T.}~\bibnamefont
  {Kraft}}, \bibinfo {author} {\bibfnamefont {C.}~\bibnamefont {Ritz}},
  \bibinfo {author} {\bibfnamefont {N.}~\bibnamefont {Brunner}}, \bibinfo
  {author} {\bibfnamefont {M.}~\bibnamefont {Huber}},\ and\ \bibinfo {author}
  {\bibfnamefont {O.}~\bibnamefont {G\"uhne}},\ }\bibfield  {title} {\bibinfo
  {title} {Characterizing genuine multilevel entanglement},\ }\href
  {https://doi.org/10.1103/PhysRevLett.120.060502} {\bibfield  {journal}
  {\bibinfo  {journal} {Phys. Rev. Lett.}\ }\textbf {\bibinfo {volume} {120}},\
  \bibinfo {pages} {060502} (\bibinfo {year} {2018})}\BibitemShut {NoStop}%
\bibitem [{\citenamefont {Sperling}\ and\ \citenamefont {Vogel}(2013)}]{sv13}%
  \BibitemOpen
  \bibfield  {author} {\bibinfo {author} {\bibfnamefont {J.}~\bibnamefont
  {Sperling}}\ and\ \bibinfo {author} {\bibfnamefont {W.}~\bibnamefont
  {Vogel}},\ }\bibfield  {title} {\bibinfo {title} {Multipartite entanglement
  witnesses},\ }\href {https://doi.org/10.1103/PhysRevLett.111.110503}
  {\bibfield  {journal} {\bibinfo  {journal} {Phys. Rev. Lett.}\ }\textbf
  {\bibinfo {volume} {111}},\ \bibinfo {pages} {110503} (\bibinfo {year}
  {2013})}\BibitemShut {NoStop}%
\bibitem [{\citenamefont {Gerke}\ \emph {et~al.}(2015)\citenamefont {Gerke},
  \citenamefont {Sperling}, \citenamefont {Vogel}, \citenamefont {Cai},
  \citenamefont {Roslund}, \citenamefont {Treps},\ and\ \citenamefont
  {Fabre}}]{gs15}%
  \BibitemOpen
  \bibfield  {author} {\bibinfo {author} {\bibfnamefont {S.}~\bibnamefont
  {Gerke}}, \bibinfo {author} {\bibfnamefont {J.}~\bibnamefont {Sperling}},
  \bibinfo {author} {\bibfnamefont {W.}~\bibnamefont {Vogel}}, \bibinfo
  {author} {\bibfnamefont {Y.}~\bibnamefont {Cai}}, \bibinfo {author}
  {\bibfnamefont {J.}~\bibnamefont {Roslund}}, \bibinfo {author} {\bibfnamefont
  {N.}~\bibnamefont {Treps}},\ and\ \bibinfo {author} {\bibfnamefont
  {C.}~\bibnamefont {Fabre}},\ }\bibfield  {title} {\bibinfo {title} {Full
  multipartite entanglement of frequency-comb gaussian states},\ }\href
  {https://doi.org/10.1103/PhysRevLett.114.050501} {\bibfield  {journal}
  {\bibinfo  {journal} {Phys. Rev. Lett.}\ }\textbf {\bibinfo {volume} {114}},\
  \bibinfo {pages} {050501} (\bibinfo {year} {2015})}\BibitemShut {NoStop}%
\bibitem [{\citenamefont {Silva}\ \emph {et~al.}(2015)\citenamefont {Silva},
  \citenamefont {Gisin}, \citenamefont {Guryanova},\ and\ \citenamefont
  {Popescu}}]{sg15}%
  \BibitemOpen
  \bibfield  {author} {\bibinfo {author} {\bibfnamefont {R.}~\bibnamefont
  {Silva}}, \bibinfo {author} {\bibfnamefont {N.}~\bibnamefont {Gisin}},
  \bibinfo {author} {\bibfnamefont {Y.}~\bibnamefont {Guryanova}},\ and\
  \bibinfo {author} {\bibfnamefont {S.}~\bibnamefont {Popescu}},\ }\bibfield
  {title} {\bibinfo {title} {Multiple observers can share the nonlocality of
  half of an entangled pair by using optimal weak measurements},\ }\href
  {https://doi.org/10.1103/PhysRevLett.114.250401} {\bibfield  {journal}
  {\bibinfo  {journal} {Phys. Rev. Lett.}\ }\textbf {\bibinfo {volume} {114}},\
  \bibinfo {pages} {250401} (\bibinfo {year} {2015})}\BibitemShut {NoStop}%
\bibitem [{\citenamefont {Brown}\ and\ \citenamefont {Colbeck}(2020)}]{bc20}%
  \BibitemOpen
  \bibfield  {author} {\bibinfo {author} {\bibfnamefont {P.~J.}\ \bibnamefont
  {Brown}}\ and\ \bibinfo {author} {\bibfnamefont {R.}~\bibnamefont
  {Colbeck}},\ }\bibfield  {title} {\bibinfo {title} {Arbitrarily many
  independent observers can share the nonlocality of a single maximally
  entangled qubit pair},\ }\href
  {https://doi.org/10.1103/PhysRevLett.125.090401} {\bibfield  {journal}
  {\bibinfo  {journal} {Phys. Rev. Lett.}\ }\textbf {\bibinfo {volume} {125}},\
  \bibinfo {pages} {090401} (\bibinfo {year} {2020})}\BibitemShut {NoStop}%
\bibitem [{\citenamefont {Cheng}\ \emph {et~al.}(2021)\citenamefont {Cheng},
  \citenamefont {Liu}, \citenamefont {Baker},\ and\ \citenamefont
  {Hall}}]{cl21}%
  \BibitemOpen
  \bibfield  {author} {\bibinfo {author} {\bibfnamefont {S.}~\bibnamefont
  {Cheng}}, \bibinfo {author} {\bibfnamefont {L.}~\bibnamefont {Liu}}, \bibinfo
  {author} {\bibfnamefont {T.~J.}\ \bibnamefont {Baker}},\ and\ \bibinfo
  {author} {\bibfnamefont {M.~J.~W.}\ \bibnamefont {Hall}},\ }\bibfield
  {title} {\bibinfo {title} {Limitations on sharing bell nonlocality between
  sequential pairs of observers},\ }\href
  {https://doi.org/10.1103/PhysRevA.104.L060201} {\bibfield  {journal}
  {\bibinfo  {journal} {Phys. Rev. A}\ }\textbf {\bibinfo {volume} {104}},\
  \bibinfo {pages} {L060201} (\bibinfo {year} {2021})}\BibitemShut {NoStop}%
\bibitem [{\citenamefont {Schiavon}\ \emph {et~al.}(2017)\citenamefont
  {Schiavon}, \citenamefont {Calderaro}, \citenamefont {Pittaluga},
  \citenamefont {Vallone},\ and\ \citenamefont {Villoresi}}]{sc17}%
  \BibitemOpen
  \bibfield  {author} {\bibinfo {author} {\bibfnamefont {M.}~\bibnamefont
  {Schiavon}}, \bibinfo {author} {\bibfnamefont {L.}~\bibnamefont {Calderaro}},
  \bibinfo {author} {\bibfnamefont {M.}~\bibnamefont {Pittaluga}}, \bibinfo
  {author} {\bibfnamefont {G.}~\bibnamefont {Vallone}},\ and\ \bibinfo {author}
  {\bibfnamefont {P.}~\bibnamefont {Villoresi}},\ }\bibfield  {title} {\bibinfo
  {title} {Three-observer bell inequality violation on a two-qubit entangled
  state},\ }\href@noop {} {\bibfield  {journal} {\bibinfo  {journal} {Quantum
  Science and Technology}\ }\textbf {\bibinfo {volume} {2}},\ \bibinfo {pages}
  {015010} (\bibinfo {year} {2017})}\BibitemShut {NoStop}%
\bibitem [{\citenamefont {Hu}\ \emph {et~al.}(2018)\citenamefont {Hu},
  \citenamefont {Zhou}, \citenamefont {Hu}, \citenamefont {Li}, \citenamefont
  {Guo},\ and\ \citenamefont {Zhang}}]{hz18}%
  \BibitemOpen
  \bibfield  {author} {\bibinfo {author} {\bibfnamefont {M.-J.}\ \bibnamefont
  {Hu}}, \bibinfo {author} {\bibfnamefont {Z.-Y.}\ \bibnamefont {Zhou}},
  \bibinfo {author} {\bibfnamefont {X.-M.}\ \bibnamefont {Hu}}, \bibinfo
  {author} {\bibfnamefont {C.-F.}\ \bibnamefont {Li}}, \bibinfo {author}
  {\bibfnamefont {G.-C.}\ \bibnamefont {Guo}},\ and\ \bibinfo {author}
  {\bibfnamefont {Y.-S.}\ \bibnamefont {Zhang}},\ }\bibfield  {title} {\bibinfo
  {title} {Observation of non-locality sharing among three observers with one
  entangled pair via optimal weak measurement},\ }\href@noop {} {\bibfield
  {journal} {\bibinfo  {journal} {npj Quantum Information}\ }\textbf {\bibinfo
  {volume} {4}},\ \bibinfo {pages} {1} (\bibinfo {year} {2018})}\BibitemShut
  {NoStop}%
\bibitem [{\citenamefont {Sasmal}\ \emph {et~al.}(2018)\citenamefont {Sasmal},
  \citenamefont {Das}, \citenamefont {Mal},\ and\ \citenamefont
  {Majumdar}}]{sd18}%
  \BibitemOpen
  \bibfield  {author} {\bibinfo {author} {\bibfnamefont {S.}~\bibnamefont
  {Sasmal}}, \bibinfo {author} {\bibfnamefont {D.}~\bibnamefont {Das}},
  \bibinfo {author} {\bibfnamefont {S.}~\bibnamefont {Mal}},\ and\ \bibinfo
  {author} {\bibfnamefont {A.~S.}\ \bibnamefont {Majumdar}},\ }\bibfield
  {title} {\bibinfo {title} {Steering a single system sequentially by multiple
  observers},\ }\href {https://doi.org/10.1103/PhysRevA.98.012305} {\bibfield
  {journal} {\bibinfo  {journal} {Phys. Rev. A}\ }\textbf {\bibinfo {volume}
  {98}},\ \bibinfo {pages} {012305} (\bibinfo {year} {2018})}\BibitemShut
  {NoStop}%
\bibitem [{\citenamefont {Shenoy~H.}\ \emph {et~al.}(2019)\citenamefont
  {Shenoy~H.}, \citenamefont {Designolle}, \citenamefont {Hirsch},
  \citenamefont {Silva}, \citenamefont {Gisin},\ and\ \citenamefont
  {Brunner}}]{sd19}%
  \BibitemOpen
  \bibfield  {author} {\bibinfo {author} {\bibfnamefont {A.}~\bibnamefont
  {Shenoy~H.}}, \bibinfo {author} {\bibfnamefont {S.}~\bibnamefont
  {Designolle}}, \bibinfo {author} {\bibfnamefont {F.}~\bibnamefont {Hirsch}},
  \bibinfo {author} {\bibfnamefont {R.}~\bibnamefont {Silva}}, \bibinfo
  {author} {\bibfnamefont {N.}~\bibnamefont {Gisin}},\ and\ \bibinfo {author}
  {\bibfnamefont {N.}~\bibnamefont {Brunner}},\ }\bibfield  {title} {\bibinfo
  {title} {Unbounded sequence of observers exhibiting einstein-podolsky-rosen
  steering},\ }\href {https://doi.org/10.1103/PhysRevA.99.022317} {\bibfield
  {journal} {\bibinfo  {journal} {Phys. Rev. A}\ }\textbf {\bibinfo {volume}
  {99}},\ \bibinfo {pages} {022317} (\bibinfo {year} {2019})}\BibitemShut
  {NoStop}%
\bibitem [{\citenamefont {Choi}\ \emph {et~al.}(2020)\citenamefont {Choi},
  \citenamefont {Hong}, \citenamefont {Pramanik}, \citenamefont {Lim},
  \citenamefont {Kim}, \citenamefont {Jung}, \citenamefont {Han}, \citenamefont
  {Moon},\ and\ \citenamefont {Cho}}]{ch20}%
  \BibitemOpen
  \bibfield  {author} {\bibinfo {author} {\bibfnamefont {Y.-H.}\ \bibnamefont
  {Choi}}, \bibinfo {author} {\bibfnamefont {S.}~\bibnamefont {Hong}}, \bibinfo
  {author} {\bibfnamefont {T.}~\bibnamefont {Pramanik}}, \bibinfo {author}
  {\bibfnamefont {H.-T.}\ \bibnamefont {Lim}}, \bibinfo {author} {\bibfnamefont
  {Y.-S.}\ \bibnamefont {Kim}}, \bibinfo {author} {\bibfnamefont
  {H.}~\bibnamefont {Jung}}, \bibinfo {author} {\bibfnamefont {S.-W.}\
  \bibnamefont {Han}}, \bibinfo {author} {\bibfnamefont {S.}~\bibnamefont
  {Moon}},\ and\ \bibinfo {author} {\bibfnamefont {Y.-W.}\ \bibnamefont
  {Cho}},\ }\bibfield  {title} {\bibinfo {title} {Demonstration of simultaneous
  quantum steering by multiple observers via sequential weak measurements},\
  }\href@noop {} {\bibfield  {journal} {\bibinfo  {journal} {Optica}\ }\textbf
  {\bibinfo {volume} {7}},\ \bibinfo {pages} {675} (\bibinfo {year}
  {2020})}\BibitemShut {NoStop}%
\bibitem [{\citenamefont {Gupta}\ \emph {et~al.}(2021)\citenamefont {Gupta},
  \citenamefont {Maity}, \citenamefont {Das}, \citenamefont {Roy},\ and\
  \citenamefont {Majumdar}}]{gm21}%
  \BibitemOpen
  \bibfield  {author} {\bibinfo {author} {\bibfnamefont {S.}~\bibnamefont
  {Gupta}}, \bibinfo {author} {\bibfnamefont {A.~G.}\ \bibnamefont {Maity}},
  \bibinfo {author} {\bibfnamefont {D.}~\bibnamefont {Das}}, \bibinfo {author}
  {\bibfnamefont {A.}~\bibnamefont {Roy}},\ and\ \bibinfo {author}
  {\bibfnamefont {A.~S.}\ \bibnamefont {Majumdar}},\ }\bibfield  {title}
  {\bibinfo {title} {Genuine einstein-podolsky-rosen steering of three-qubit
  states by multiple sequential observers},\ }\href
  {https://doi.org/10.1103/PhysRevA.103.022421} {\bibfield  {journal} {\bibinfo
   {journal} {Phys. Rev. A}\ }\textbf {\bibinfo {volume} {103}},\ \bibinfo
  {pages} {022421} (\bibinfo {year} {2021})}\BibitemShut {NoStop}%
\bibitem [{\citenamefont {Yao}\ and\ \citenamefont {Ren}(2021)}]{yr21}%
  \BibitemOpen
  \bibfield  {author} {\bibinfo {author} {\bibfnamefont {D.}~\bibnamefont
  {Yao}}\ and\ \bibinfo {author} {\bibfnamefont {C.}~\bibnamefont {Ren}},\
  }\bibfield  {title} {\bibinfo {title} {Steering sharing for a two-qubit
  system via weak measurements},\ }\href
  {https://doi.org/10.1103/PhysRevA.103.052207} {\bibfield  {journal} {\bibinfo
   {journal} {Phys. Rev. A}\ }\textbf {\bibinfo {volume} {103}},\ \bibinfo
  {pages} {052207} (\bibinfo {year} {2021})}\BibitemShut {NoStop}%
\bibitem [{\citenamefont {Zhu}\ \emph {et~al.}(2022)\citenamefont {Zhu},
  \citenamefont {Hu}, \citenamefont {Li}, \citenamefont {Guo},\ and\
  \citenamefont {Zhang}}]{zh21}%
  \BibitemOpen
  \bibfield  {author} {\bibinfo {author} {\bibfnamefont {J.}~\bibnamefont
  {Zhu}}, \bibinfo {author} {\bibfnamefont {M.-J.}\ \bibnamefont {Hu}},
  \bibinfo {author} {\bibfnamefont {C.-F.}\ \bibnamefont {Li}}, \bibinfo
  {author} {\bibfnamefont {G.-C.}\ \bibnamefont {Guo}},\ and\ \bibinfo {author}
  {\bibfnamefont {Y.-S.}\ \bibnamefont {Zhang}},\ }\bibfield  {title} {\bibinfo
  {title} {Einstein-podolsky-rosen steering in two-sided sequential
  measurements with one entangled pair},\ }\href
  {https://doi.org/10.1103/PhysRevA.105.032211} {\bibfield  {journal} {\bibinfo
   {journal} {Phys. Rev. A}\ }\textbf {\bibinfo {volume} {105}},\ \bibinfo
  {pages} {032211} (\bibinfo {year} {2022})}\BibitemShut {NoStop}%
\bibitem [{\citenamefont {Liu}\ \emph {et~al.}(2021)\citenamefont {Liu},
  \citenamefont {Liu}, \citenamefont {Fang}, \citenamefont {Li},\ and\
  \citenamefont {Wang}}]{ll21}%
  \BibitemOpen
  \bibfield  {author} {\bibinfo {author} {\bibfnamefont {K.}~\bibnamefont
  {Liu}}, \bibinfo {author} {\bibfnamefont {T.}~\bibnamefont {Liu}}, \bibinfo
  {author} {\bibfnamefont {W.}~\bibnamefont {Fang}}, \bibinfo {author}
  {\bibfnamefont {J.}~\bibnamefont {Li}},\ and\ \bibinfo {author}
  {\bibfnamefont {Q.}~\bibnamefont {Wang}},\ }\bibfield  {title} {\bibinfo
  {title} {Both qubits of the singlet state can be steered simultaneously by
  multiple independent observers via sequential measurement},\ }\href@noop {}
  {\bibfield  {journal} {\bibinfo  {journal} {arXiv preprint arXiv:2102.12166}\
  } (\bibinfo {year} {2021})}\BibitemShut {NoStop}%
\end{thebibliography}%

\end{document}